\documentclass[journal]{IEEEtran}
\usepackage{amsmath,amsfonts}
\usepackage{algorithmic}
\usepackage{array}
\usepackage[caption=false,font=normalsize,labelfont=sf,textfont=sf]{subfig}
\usepackage{textcomp}
\usepackage{stfloats}
\usepackage{url}
\usepackage{verbatim}
\usepackage{graphicx}
\def\BibTeX{{\rm B\kern-.05em{\sc i\kern-.025em b}\kern-.08em
		T\kern-.1667em\lower.7ex\hbox{E}\kern-.125emX}}
\usepackage{balance}

\usepackage{amsxtra}\usepackage{latexsym}\usepackage{amscd}\usepackage{amsthm}\usepackage{amsfonts}
\usepackage{amssymb}
\usepackage{algorithm}
\usepackage{url}
\usepackage{verbatim}
\usepackage{graphicx}
\usepackage{cite}
\usepackage{bm}
\usepackage{xr-hyper}
\usepackage{hyperref}
\usepackage{indentfirst}
\usepackage{epstopdf}
\usepackage{varwidth}
\usepackage{multirow}
\usepackage{color}
\usepackage{wasysym}
\usepackage{amssymb}
\usepackage{makecell}
\usepackage[normalem]{ulem}
\usepackage{rotating}
\usepackage{float}

\DeclareRobustCommand{\foreign@language}[1]{%
	\lowercase{\oldforeign@language{#1}}}
\theoremstyle{plain}
\newtheorem{thm}{\theoremname}
\theoremstyle{definition}
\newtheorem{defn}[thm]{\definitionname}
\theoremstyle{plain}
\newtheorem{prop}[thm]{\propositionname}
\theoremstyle{plain}
\newtheorem{cor}[thm]{\corollaryname}
\theoremstyle{remark}
\newtheorem{rem}[thm]{\remarkname}

\newcommand{\carprod}{\ensuremath{\prod}\kern-1.02em{\times}}
\newcommand{\cprod}{\ensuremath{\prod}\kern-0.945em{{\scriptstyle \times}}\kern+0.1em}

\makeatother

\providecommand{\corollaryname}{Corollary}
\providecommand{\definitionname}{Definition}
\providecommand{\propositionname}{Proposition}
\providecommand{\remarkname}{Remark}
\providecommand{\theoremname}{Theorem}

\begin{document}
\markboth{PREPRINT: IEEE TRANSACTIONS ON SIGNAL PROCESSING, VOL. 74, JUNE 2026, PP. 2803-2819, DOI: 10.1109/TSP.2026.3702946}{}
\title{Tractable Approximation of Labeled Multi-Object Posterior Densities}
\author{Thi Hong Thai Nguyen, Ba-Ngu Vo, and Ba-Tuong Vo\thanks{Acknowledgment: This work was supported by the Australian Research
Council under grants LP200301507 and FT210100506.}\thanks{The authors are with the School of Electrical Engineering, Computing
and Mathematical Sciences, Curtin University, Bentley, WA 6102, Australia
(email: t.nguyen346@postgrad.curtin.edu.au, \{ba-ngu.vo, ba-tuong.vo\}@curtin.edu.au).}}
\maketitle
\begin{abstract}
Multi-object estimation in state-space models (SSMs) wherein the system state is represented as a finite set has attracted significant interest in recent years. In Bayesian inference, the posterior density captures all information on the system trajectory since it considers the history of the states. In most multi-object SSM applications, closed-form multi-object posteriors are not available for non-standard multi-object models. Thus, functional approximation is necessary because these posteriors are very high-dimensional. This work provides a tractable multi-scan Generalized Labeled Multi-Bernoulli (GLMB) approximation that matches the trajectory cardinality distribution of the labeled multi-object posterior density. The proposed approximation is also proven to minimize the Kullback-Leibler divergence over a special class of multi-scan GLMB model. Additionally, we develop a tractable algorithm for computing the approximate multi-object posteriors over finite windows. Numerical multi-object tracking experiments, using a simulated social force model with uninformative observations, and a real-world social force pedestrian dataset are presented to validate the proposed approximation method.
\end{abstract}


\section{Introduction}\label{sec:Introduction}

Multi-object estimation is a generalization of estimation for state-space
models (SSMs), where the system state is a finite set. The aim is
to infer the underlying system trajectory, herein called the multi-object
trajectory, consisting of the set of trajectories of individual objects.
Unlike the standard SSM, a sequence of multi-object states (finite
sets) does not necessarily represent the multi-object trajectory.
However, a labeled set representation enables the multi-object trajectory
to be represented by a sequence of multi-object states analogous
to traditional state space models \cite{vo2024}. Multi-object estimation
has a wide range of applications from multi-sensor data fusion \cite{fantacci2016,wang2018,li2018,li2019,moratuwage2022}
to simultaneous mapping and localization \cite{deusch2015,moratuwage2018},
and is far more challenging than the traditional (vector) state estimation
problem due to the unknown and random number of objects, false positives
and negatives in the observations, and data association uncertainty. 

From a Bayesian perspective, given the observation history, all information
on the system trajectory is captured in the posterior density \cite{meditch1973,anderson1979,doucet2009,briers2010,sarkka2013}.
In practice, the current marginal of the posterior--commonly known
as the filtering density--is often used for computational efficiency
since it captures the information on current system state. The filtering
density is adequate for multi-object applications with high signal-to-noise-ratio
(SNR). However, in low SNR or low observability applications including
Track-before-Detect \cite{salmond2001,papi2015,kim2019} or tracking
with superpositional measurements \cite{nannuru2013,papi22015}, multi-object
estimation via the filtering density is not satisfactory. Hence, it
becomes necessary to resort to the posterior although computing the
multi-object posterior is far more difficult than its single-object
counterpart. Under the standard multi-object system model, the multi-object
posterior assumes the form of a Generalized Labeled Multi-Bernoulli
(GLMB) that can be tractably computed. However, in general, posterior
computation is still an active research topic due to its fundamental
importance \cite{bunch2013,papi2014,finke2017,olsson2017,gerber2017,xia2022}. 

\begin{figure*}
	\centering
	\subfloat[]{\includegraphics[bb=0cm 1.5cm 14.5cm 10cm,clip,scale=0.37]{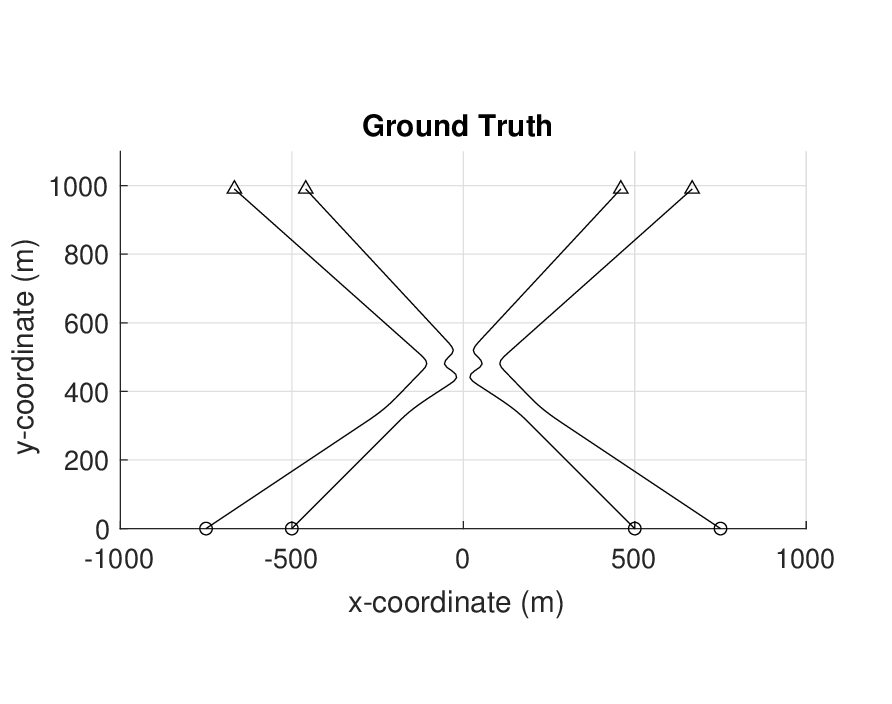}\label{fig:ground_truth}}
	\hfil
	\subfloat[]{\includegraphics[bb=0.5cm 1.5cm 14.5cm 10cm,clip,scale=0.38]{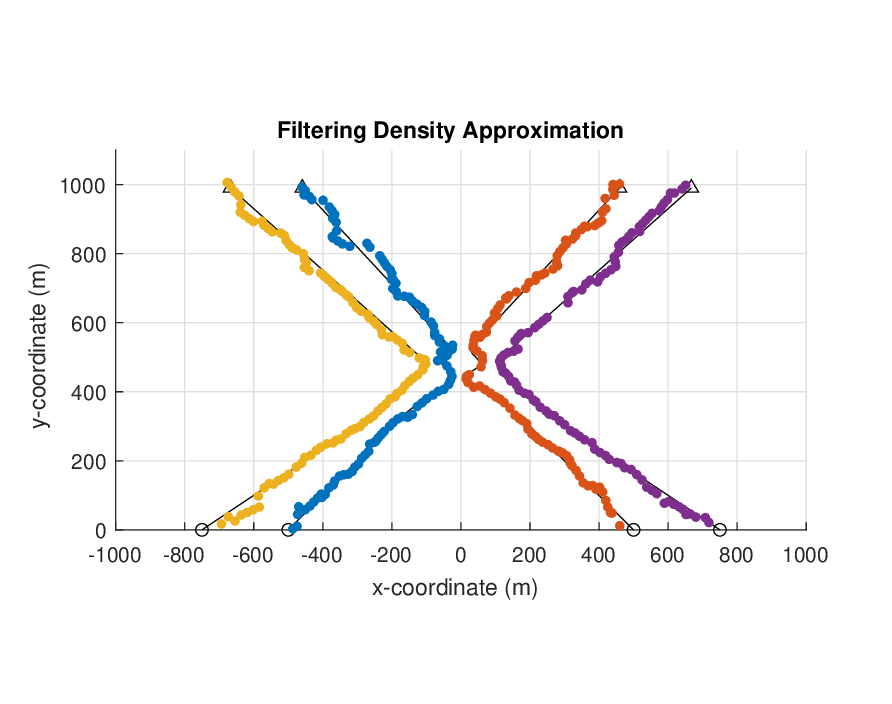}\label{fig:filtering_approx}}
	\hfil
	\subfloat[]{\includegraphics[bb=0.5cm 1.5cm 14.5cm 10cm,clip,scale=0.38]{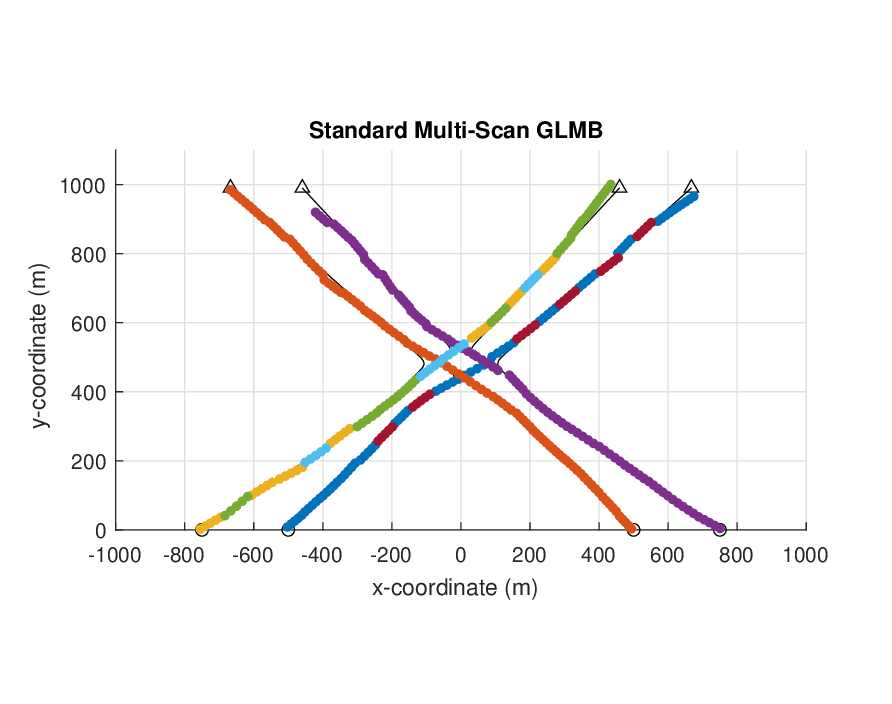}\label{fig:standard_MS_GLMB}}
	\hfil
	\includegraphics[bb=4.8cm 0.8cm 11cm 10.5cm,clip,scale=0.3]{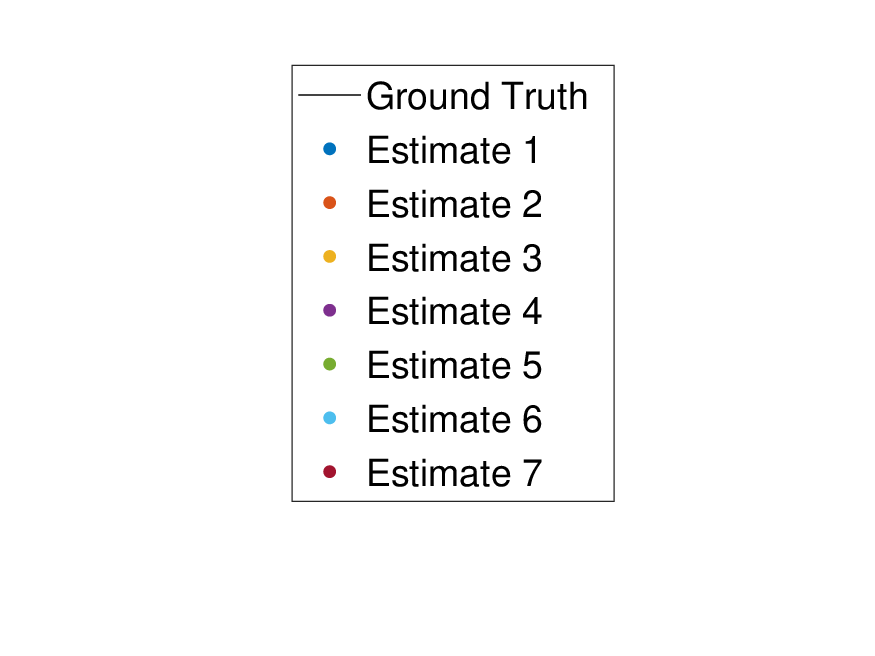}
	\caption{Objects moving and interacting according to the social force model (with details given in Section \ref{sec:numerical_study}). Starting and stopping positions are indicated by $\fullmoon$ and $\APLup$, respectively. (a) Ground truth: the 4 objects approach each other, but change their directions to avoid collision and hence, no crossing. (b) Estimated multi-object trajectory from the multi-object filtering density with social force modeling: no track switching but suffers from significant track fragmentation. (c) Estimated multi-object trajectory from the multi-object posterior under the standard multi-object dynamic model (no interaction modeling): multiple erroneous trajectory crossings and identity switchings.}\label{fig:intro_fig}
\end{figure*}

While most of the contributions in multi-object posterior computation
are based on the standard multi-object SSM \cite{vu2014,vo2019,moratuwage2022},
this is inadequate for many applications. The standard multi-object
dynamic model assumes conditional independence of individual object
motions. However, in many real-world multi-object applications, interaction
between objects in their dynamic systems is critical, see e.g., the
social force model \cite{helbing1995} and its applications in pedestrian
modeling \cite{helbing2005,krishanth2017}, crowd simulation \cite{helbing2000,mehran2009},
multi-object tracking\cite{pellegrini2009}, and references therein.
The importance of capturing inter-object interaction in dynamic modeling
is illustrated in Figure \ref{fig:intro_fig}, highlighting erroneous
posterior multi-object estimation when interactions are ignored. Moreover,
the standard multi-object observation model also considers the conditional
independence of individual object detection probabilities, and that
a detection can only result from at most one object, which are not
valid in general due to occlusions.

In general, the posterior is computationally intractable under
a non-standard multi-object SSM and hence, functional approximation
that preserves relevant multi-object trajectory information is needed.
Inspired by the IID cluster approximation proposed in \cite{mahlerjuly2007},
multi-object filtering solutions have been developed for non-standard
multi-object SSM via an optimal information theoretic GLMB approximation
of the filtering density \cite{papi2015}. This approach has been
used in many non-standard problems including merged and shared measurements
\cite{beard2015}, Track-before-Detect \cite{papi2015,papi22015},
space debris tracking \cite{jones2015}, multi-object tracking with
spawning \cite{bryant2018}, biomedical cell-tracking \cite{nguyen2021},
multi-object tracking with occlusions \cite{ong2022}, resolvable
group target tracking \cite{wu2025}. However, for more complex problems
with inter-object interactions, approximation of the filtering density
is not adequate as demonstrated in Figure \ref{fig:filtering_approx},
while tractable multi-object posterior approximation has not been
addressed. 

In this work, we propose a tractable approximation of the labeled
multi-object posterior density via the multi-scan GLMB family \cite{vo2019}.
This approximation preserves the trajectory cardinality information
of the true posterior and minimizes the Kullback-Leibler divergence
over a special class of multi-scan GLMB thereby generalizing the multi-object
filtering density approximation introduced in \cite{papi2015}. In
addition, we derive a multi-object transition kernel that incorporates
the social force model into the multi-object dynamic systems, and
the resulting multi-object posterior. For numerical implementations,
we develop two GLMB approximation strategies that numerically illustrate
our approximation method. Further, a tractable multi-scan multi-object
approximation is proposed to compute the approximate multi-object
posteriors with complexity per time step that does not grow with time.
For numerical validations, we consider the multi-object tracking scenarios
with social force model and non-standard measurements, as well as
real-world pedestrian tracking benchmarks. 

The remainder of this article is organized into
seven sections. Section \ref{sec:background} provides background
on the labeled random finite set (RFS) and multi-object estimation.
Section \ref{sec:multi_object_density_approx} introduces tractable
and principled approximations via the multi-scan GLMB model, and thereby
providing a practical solution for multi-object posterior modeling.
Section \ref{sec:interactive_modeling} presents the approximate labeled
multi-object posterior recursion for multi-object trajectory estimation.
Section \ref{sec:implementations} details the implementation of the
approximation methods as outlined in Section \ref{sec:interactive_modeling}.
Section \ref{sec:numerical_study} reports the multi-object tracking
experiments for the social force model and uninformative observations,
validating the real-world applicability of the proposed approximation
methods through pedestrian tracking scenarios. Section \ref{sec:conclusion}
concludes the article with a summary of key findings. Mathematical
derivations and additional performance evaluations are provided in
the Supplementary Materials.

\section{Background}\label{sec:background}

This section presents the basic background on labeled RFS and multi-object
estimation. In Subsection \ref{subsec:labeled_rfs}, we outline the
concept of labeled RFS for multi-object state and trajectory modeling.
Subsections \ref{subsec:multi_object_posterior_density} and \ref{subsec:standard_multi_object_ssm}
summarize notions of the multi-object posterior (density) and multi-object
SSM, while Subsection \ref{subsec:posterior_GLMB_recursion} presents
the multi-scan GLMB recursion. Mathematical symbols used throughout
this work are summarized in Table \ref{tab:math_notations}. 
\begin{table}[!t]
	\centering
	\caption{List of common mathematical notations\label{tab:math_notations}}
	\begin{tabular}{|c|p{6.2cm}|}
		\hline 
		\textbf{Notation} & \textbf{Description}\\
		\hline 
		$x_{u:v}$ & $x_{u},x_{u+1},...,x_{v}$\\
		$\langle\phi,\varphi\rangle$ & Inner product $\int\phi(x)\varphi(x)dx$ of function $\phi$ and $\psi$\\
		$\mathbb{X}$ & Finite dimensional state space\\
		$\mathbb{L}$ & (Discrete) space of labels\\
		$\mathbb{B}_{k}$ & Label space of objects born at time $k$\\
		$\mathbb{L}_{k}$ & $\mathbb{B}_{k}\uplus\mathbb{L}_{k-1}$, label space at time $k$\\
		$\ell$ & Label of an object/trajectory\\
		$\mathcal{F}(\mathcal{X})$ & Space of all finite subsets of a set $\mathcal{X}$\\
		$|X|$ & Cardinality (or number of elements) of a set $X$\\
		$\boldsymbol{1}_{S}(\cdot)$ & Indicator function of a finite set $S$\\
		$\delta_{Y}[X]$ & Kronecker-$\delta$, $\delta_{Y}[X]=1$ if $X=Y$, and $0$ otherwise\\
		$\mathbb{X}\times\mathbb{L}$ & Cartesian product of $\mathbb{X}$ and $\mathbb{L}$\\
		$\mathcal{A}$ & Attribute projection $(x,\ell)\mapsto x$\\
		$\mathcal{L}$ & Label projection $(x,\ell)\mapsto\ell$\\
		$\boldsymbol{X}$ & Labeled multi-object state\\
		$\boldsymbol{X}_{k}$ & Multi-object state at time $k$\\
		$\Delta(\boldsymbol{X})$ & Distinct label indicator $\delta_{|\boldsymbol{X}|}(|\mathcal{L}(\boldsymbol{X})|)$
		of $\boldsymbol{X}$\\
		$\boldsymbol{X}_{j:k}$ & Labeled multi-object sequence/trajectory on $\{j:k\}$\\
		$\mathcal{L}(\boldsymbol{X}_{j:k})$ & $(\mathcal{L}(\boldsymbol{X}_{j}),...,\mathcal{L}(\boldsymbol{X}_{k}))$\\
		$\{\mathcal{L}(\boldsymbol{X}_{j:k})\}$ & $\cup_{i=j}^{k}\mathcal{L}(\boldsymbol{X}_{i})$\\
		$T(\ell)$ & Set of time instances that $\boldsymbol{X}_{j:k}$ contains $\ell$\\
		$\boldsymbol{x}_{T(\ell)}$ & $[\boldsymbol{x}_{i}=(x_{i},\ell)\in\boldsymbol{X}_{i}]_{i\in T(\ell)}$,
		trajectory of $\ell$ in $\boldsymbol{X}_{j:k}$\\
		$\cprod_{i=j}^{k}P_{i}$ & $P_{j}\times...\times P_{k}$\\
		$\Delta(\boldsymbol{X}_{j:k})$ & Multi-scan distinct label indicator $\prod_{i=j}^{k}\Delta(\boldsymbol{X}_{i})$\\
		$h^{X}$ & Multi-object exponential $\prod_{x\in X}h(x)$ with $h^{\emptyset}=1$\\
		$h^{\boldsymbol{X}_{j:k}}$ & 
		$\prod_{\ell \in \{\mathcal{L}(\bm{X}_{j:k})\}} h(\bm{x}_{T(\ell)})$, \newline
		where $\bm{X}_{j:k} =\{\bm{x}_{T(\ell)} : \ell \in \{\mathcal{L}(\bm{X}_{j:k})\}\}$\\ 
		$\int f(\boldsymbol{X})\delta\boldsymbol{X}$ & $\sum\limits_{n=0}^{\infty}\frac{1}{n!}\int f(\{\boldsymbol{x}^{(1)},...,\boldsymbol{x}^{(n)}\})d\boldsymbol{x}^{(1)}...d\boldsymbol{x}^{(n)}$\\
		$\boldsymbol{X}_{k}$ & Multi-object state at time $k$\\
		$Z_{k}$ & Multi-object observation of time $k$\\
		$\boldsymbol{g}_{k}(Z_{k}|\boldsymbol{X}_{k})$ & Multi-object likelihood observing $Z_{k}$ given $\boldsymbol{X}_{k}$\\
		$\boldsymbol{f}_{k}(\cdot|\boldsymbol{X}_{k-1})$ & Multi-object Markov transition density given $\boldsymbol{X}_{k-1}$\\
		$\boldsymbol{\pi}_{0:k}(\boldsymbol{X}_{0:k})$ & Multi-object posterior density at $\boldsymbol{X}_{0:k}$\\
		$|\boldsymbol{X}_{j:k}|$ & Trajectory cardinality $|\cup_{i=j}^{k}\mathcal{L}(\boldsymbol{X}_{i})|$
		of $\boldsymbol{X}_{j:k}$\\
		$X \circledast L$ & Finite set $\{(x^{(1)},\ell^{(1)}),...,(x^{(n)},\ell^{(n)})\}$ where $X=\{x^{(1)},...,x^{(n)}\}$ and $L=\{\ell^{(1)},...,\ell^{(n)}\}$ is a set of $n$ distinct labels\\
		\hline 
	\end{tabular}
\end{table}

\subsection{Labeled RFS}\label{subsec:labeled_rfs}

An RFS of $\mathcal{X}$ is a simple finite point
process, or a random variable on the class $\mathcal{F}(\mathcal{X})$
of finite subsets of $\mathcal{X}$ \cite{mahler2014}. A \textit{labeled
RFS} is a special class of RFSs used for modeling multi-object states/trajectories
\cite{vo2013}. Formally, a labeled RFS $\boldsymbol{X}\in\mathcal{F}(\mathbb{X}\times\mathbb{L})$
with attribute space $\mathbb{X}$ and label (or mark) space $\mathbb{L}$
is a simple finite marked point process of $\mathbb{X}\times\mathbb{L}$,
where each realization of $\boldsymbol{X}$ has distinct labels. 

\begin{comment}
Define a ``stitch'' operator $\otimes$ such that each attribute
$x\in\mathcal{A}(\mathbf{X})$ stitches with a distinct (unique) label
$\ell\in\mathcal{L}(\mathbf{X})$, i.e. $x\otimes\ell=(x,\ell)=\boldsymbol{x}\in\mathbf{X}$.
Without generality, the labeled RFS $\mathbf{X}$ can be rewritten
as a result of stitching its sets of attributes $\mathcal{A}(\mathbf{X})=X$
and labels $\mathcal{L}(\mathbf{X})=L$ as follows
\begin{equation}
\mathbf{X}=\mathcal{A}(\mathbf{X})\otimes\mathcal{L}(\mathbf{X})=\{(X,L)\}.
\end{equation}
\vspace{-0.5cm}

Note that since $\mathbf{X}$ has distinct labels, the stitch operator
between its sets of attributes $\mathcal{A}(\mathbf{X})$ and labels
$\mathcal{L}(\mathbf{X})$ is unique.
\end{comment}
Let $\mathcal{A}\colon(x,\ell)\mapsto x$, and $\mathcal{L}\colon(x,\ell)\mapsto\ell$
denote, respectively, the attribute and label projections. Then $\mathcal{A}(\boldsymbol{X})$
and $\mathcal{L}(\boldsymbol{X})$ are, respectively, the sets of
attributes and labels of $\boldsymbol{X}$. We say $\boldsymbol{X}$
has distinct labels, if and only if $\boldsymbol{X}$ has the same
cardinality as \textbf{$\mathcal{L}(\boldsymbol{X})$}, and define
the distinct label indicator function as $\delta_{|\boldsymbol{X}|}[|\mathcal{L}(\boldsymbol{X})|]$,
where $\delta_{Y}[X]=1$ if $X=Y$, and $0$ otherwise. Given the
arrays of attributes and labels, i.e. $X=(x^{(1)},...,x^{(n)})$ and
$L=(\ell^{(1)},...,\ell^{(n)})$, respectively, we abbreviate the
finite set $\{(x^{(1)},\ell^{(1)}),...,(x^{(n)},\ell^{(n)})\}$ as
$X\circledast L$, if the labels are distinct. 

\subsection{Multi-Object Posterior Density}\label{subsec:multi_object_posterior_density}

A multi-object state is modeled as a labeled RFS that evolves in time,
such that the label of each element remains unchanged \cite{vo2013}.
Hence, the multi-object trajectory can be represented by a time sequence
of multi-object states \cite{vo2019,vo2024}. 

Consider a time sequence $\boldsymbol{X}_{j:k}$ of multi-object states
from times $j:k$, and let $T(\ell)\subseteq\{j:k\}$ denote the set
of time instances that $\boldsymbol{X}_{j:k}$ contains a state with
label $\ell$. The trajectory of (an object with) label $\ell$ in
$\{\mathcal{L}(\boldsymbol{X}_{j:k})\}\triangleq\cup_{i=j}^{k}\mathcal{L}(\boldsymbol{X}_{i})$,
is the time sequence $\boldsymbol{x}_{T(\ell)}=[\boldsymbol{x}_{i}\in\boldsymbol{X}_{i}]_{i\in T(\ell)}$
consisting of $\boldsymbol{x}_{i}\in\boldsymbol{X}_{i}$, $i\in T(\ell)$,
such that $\mathcal{L}(\boldsymbol{x}_{i})=\ell$. Moreover, we can
equivalently represent $\boldsymbol{X}_{j:k}$ as the set of trajectories
\cite{vo2019}, i.e. 
\[
\boldsymbol{X}_{j:k}\equiv\left\{ \boldsymbol{x}_{T(\ell)}:\,\ell\in\{\mathcal{L}(\boldsymbol{X}_{j:k})\}\right\} .
\]
Note that for an unfragmented trajectory, we have $T(\ell)=\{s(\ell):t(\ell)\}$,
where $s(\ell)\triangleq\min T(\ell)$ and $t(\ell)\triangleq\max T(\ell)$.
In this article, we only consider unfragmented trajectories. Given
a multi-object trajectory from $\{j:k\}$, all information on the
multi-object trajectory is captured in the \textit{multi-object trajectory
density}, i.e. \textit{multi-scan multi-object density}.

From a Bayesian perspective, the labeled multi-object posterior density
captures all information on the multi-object trajectory from time
$0$ to $k$, given the observation history $Z_{0:k}$. The multi-object
posterior $\boldsymbol{\pi}_{0:k}(\boldsymbol{X}_{0:k}|Z_{0:k})$,
or simply $\boldsymbol{\pi}_{0:k}(\boldsymbol{X}_{0:k})$, can be
propagated forward by the following (posterior) Bayes recursion \cite{vo2019}%
\begin{comment}
\begin{align}
 & \boldsymbol{\pi}_{0:k}(\mathbf{X}_{0:k})=\\
 & \frac{\boldsymbol{g}_{k}(Z_{k}|\mathbf{X}_{k})\boldsymbol{f}_{k|k-1}(\mathbf{X}_{k}|\mathbf{X}_{k-1})\boldsymbol{\pi}_{0:k-1}(\mathbf{X}_{0:k-1})}{{\displaystyle \int}\boldsymbol{g}_{k}(Z_{k}|\mathbf{Y}_{k})\boldsymbol{f}_{k|k-1}(\mathbf{X}_{k}|\mathbf{Y}_{k-1})\boldsymbol{\pi}_{0:k-1}(\mathbf{Y}_{0:k-1})\delta\mathbf{Y}_{0:k}},\nonumber 
\end{align}
\end{comment}
{} 
\[
\boldsymbol{\pi}_{0:k}(\boldsymbol{X}_{0:k})\propto\boldsymbol{g}_{k}(Z_{k}|\boldsymbol{X}_{k})\boldsymbol{f}_{k}(\boldsymbol{X}_{k}|\boldsymbol{X}_{k-1})\boldsymbol{\pi}_{0:k-1}(\boldsymbol{X}{}_{0:k-1}),
\]
where $\boldsymbol{g}_{k}(\cdot|\cdot)$ is the multi-object observation
likelihood function, and $\boldsymbol{f}_{k}(\cdot|\cdot)$ is the
multi-object Markov transition density, to be presented in the next
subsection. 

\subsection{Standard Multi-Object State-Space Model}\label{subsec:standard_multi_object_ssm}

The multi-object likelihood function $\boldsymbol{g}_{k}(Z_{k}|\boldsymbol{X}_{k})$
is the probability density of the observation $Z_{k}$ given the multi-object
state $\boldsymbol{X}_{k}$ at time $k$. In the \textit{standard
multi-object observation model}, given a multi-object state $\boldsymbol{X}{}_{k}$,
each element $\boldsymbol{x}_{k}\in\boldsymbol{X}_{k}$ is detected
with probability $P_{k,D}(\boldsymbol{x}_{k})$ and generates a measurement
$z_{k}$ with likelihood $g_{k}(z_{k}|\boldsymbol{x}_{k})$, or missed
with probability $Q_{k,D}(\boldsymbol{x}_{k})=1-P_{k,D}(\boldsymbol{x}_{k}).$
The multi-object observation $Z_{k}$ is the superposition of detections
and clutter (or false observations/alarms) modeled as an Poisson RFS
with intensity $\kappa_{k}(\cdot)$. Assuming that conditioned on
$\boldsymbol{X}_{k}$, detections are independent of each other and
clutter \cite{vo2013}
\begin{equation}
\boldsymbol{g}_{k}(Z_{k}|\boldsymbol{X}_{k})\propto\sum\limits_{\theta_{k}\in\Theta_{k}(\mathcal{L}(\boldsymbol{X}_{k}))}\left[\psi_{k,Z_{k}}^{(\theta_{k}\circ\mathcal{L}(\cdot))}\right]^{\boldsymbol{X}_{k}},\label{eq:likelihood_func}
\end{equation}
where $(\theta_{k}\circ\mathcal{L})(\boldsymbol{x}_{k})=\theta_{k}(\mathcal{L}(\boldsymbol{x}_{k}))$,
and 
\[
\psi_{k,Z_{k}}^{(j)}(\boldsymbol{x}_{k})=\begin{cases}
\frac{P_{k,D}(\boldsymbol{x}_{k})g_{k}(z_{j}|\boldsymbol{x}_{k})}{\kappa_{k}(z_{j})}, & j\in\left\{ 1:|Z_{k}|\right\} ,\\
Q_{k,D}(\boldsymbol{x}_{k}), & j=0,
\end{cases}
\]
and $\Theta_{k}$ is the set of association maps $\theta_{k}:\mathbb{L}_{k}\rightarrow\{0:|Z_{k}|\}$
such that $\theta_{k}$ is a positive $1-1$ mapping, i.e. $\theta_{k}(i)=\theta_{k}(j)>0$
implies $i=j$. Each detected label $\ell$ is assigned a measurement
$z_{\theta_{k}(\ell)}\in Z_{k},$ whereas $\theta_{k}(\ell)=0$ for
an undetected label.

The multi-object Markov transition density $\boldsymbol{f}_{k}(\cdot|\boldsymbol{X}_{k-1})$
is the probability density of the multi-object state at time $k$
given the multi-object state $\boldsymbol{X}_{k-1}$ at time $k-1$.
In the \textit{standard multi-object dynamic model}, at time $k$,
an object with state $\boldsymbol{x}_{k}=(x_{k},\ell)$, $\ell\in\mathbb{B}_{k}$
is either born with birth probability $P_{k,B}(\ell)$ and birth density
$p_{k,B}(x_{k},\ell)$, or not born with probability $Q_{k,B}(\ell)=1-P_{k,B}(\ell)$,
where $\mathbb{B}_{k}$ denotes the discrete space of birth labels.
At time $k$, the label space $\mathbb{L}_{k}$ is a disjoint union
of the birth label space $\mathbb{B}_{k}$ and the label space $\mathbb{L}_{k-1}$,
i.e. $\mathbb{L}_{k}=\mathbb{B}_{k}\uplus\mathbb{L}_{k-1}$. Given
a multi-object state $\boldsymbol{X}_{k-1}$ at time $k-1$, each
element/object with state $\boldsymbol{x}_{k-1}=(x_{k-1},\ell)\in\boldsymbol{X}_{k-1}$
either survives with survival probability $P_{k,S}(\boldsymbol{x}_{k-1})$
and transitions to a new state $\boldsymbol{x}_{k}=(x_{k},\ell)$
with survival density $f_{k,S}(x_{k}|x_{k-1},\ell)$, or dies with
probability $Q_{k,S}(\boldsymbol{x}_{k-1})=1-P_{k,S}(\boldsymbol{x}_{k-1}).$
The multi-object state at time $k$ is the superposition of new born
states and surviving states. Assuming that conditioned on $\boldsymbol{X}_{k-1}$,
the objects evolve and born independently of each other, then $\boldsymbol{f}_{k}(\boldsymbol{X}_{k}|\boldsymbol{X}_{k-1})$
is given by equation (6) of \cite{vo2014}.

For simplicity we often omit the time subscript $k$, and use the
subscript ``$-$'' to indicate time $k-1$. We also adopt the following
abbrevations of commonly used terms that involve the standard multi-object
model
\begin{align*}
\chi_{B}^{(j)}(x,\ell) & \triangleq\psi_{Z}^{(j)}(x,\ell)p_{B}(x,\ell)P_{B}(\ell),\\
\chi_{S}^{(j)}(x|\nu,\ell) & \triangleq\psi_{Z}^{(j)}(x,\ell)f_{S}(x|\nu,\ell)P_{S}(\nu,\ell),\\
\bar{\chi}_{B}^{(j)}(\ell) & \triangleq\langle\chi_{B}^{(j)}(\cdot,\ell),1\rangle,\\
\bar{\chi}_{S}^{(\xi,j)}(\ell) & \triangleq\int\langle\chi_{S}^{(j)}(x|\cdot,\ell),p_{-}^{(\xi)}(\cdot,\ell)\rangle dx.
\end{align*}
where $\langle\phi,\varphi\rangle$ denotes the inner product $\int\phi(x)\varphi(x)dx$
of the functions $\phi$, $\varphi$, and $p_{-}^{(\xi)}(\cdot,\ell)$
is a given probability density (on the attribute space $\mathbb{X}$)
of the attribute of $\ell$ at time $k-1$. 

\subsection{Posterior GLMB Recursion}\label{subsec:posterior_GLMB_recursion}

Under the standard multi-object SSM, the posterior recursion admits
a close form solution known as the multi-scan GLMB \cite{vo2019}.
Since the standard multi-object model does not permit fragmented trajectories,
we only consider the multi-scan GLMB for contiguous trajectories. 

\subsubsection{Multi-scan GLMB}

We start with a basic building block of the multi-scan GLMB, called
the \textit{multi-scan multi-object exponential} or \textit{multi-object
trajectory exponential}, defined for a multi-object trajectory $\boldsymbol{X}_{j:k}$
and a suitable function $h$ by
\begin{align}
h^{\boldsymbol{X}_{j:k}} & \triangleq h^{\{\boldsymbol{x}_{T(\ell)}:\,\ell\in\{\mathcal{L}(\boldsymbol{X}_{j:k})\}\}},\nonumber \\
 & =\prod\limits_{\ell\in\{\mathcal{L}(\boldsymbol{X}_{j:k})\}}h(\boldsymbol{x}_{T(\ell)}).\label{eq:multiscanexponential}
\end{align}
Since the function $h$ takes trajectories on different sub-intervals
of $\{j:k\}$ as its argument, its domain is the disjoint union of
trajectory spaces on these sub-intervals. More concisely, let $\cprod_{i=j}^{k}P_{i}$
denote the Cartesian product $P_{j}\times...\times P_{k}$. Then the
space of trajectories on the sub-interval $\{t_{1},...,t_{n}\}$ is
$\mathbb{T}_{\{t_{1},...,t_{n}\}}\triangleq\cprod_{i=t_{1}}^{t_{n}}(\mathbb{X}\times\mathbb{L}_{i})$,
with $\mathbb{T}_{\emptyset}=\emptyset$, and the domain of $h$ is
$\uplus_{I\subseteq\{j:k\}}\mathbb{T}_{I}$. Note that if $j=k$,
equation \eqref{eq:multiscanexponential} becomes $h^{\mathbf{X}_{j}}$,
i.e. the single-scan multi-object exponential \cite{vo2013}. The
\textit{exponential-like properties} of multi-object trajectory exponential
are summarized in the following \cite{vo2019}.

Suppose that $\boldsymbol{X}_{j:k}$ and $\boldsymbol{Y}_{j:k}$ are
multi-object trajectories with disjoint label sets on the interval
$\{j:k\}$ and $g,h$ are two functions on $\uplus_{I\subseteq\{j:k\}}\mathbb{T}_{I}$.
Then,
\begin{itemize}
\item $[g\,h]^{\boldsymbol{X}_{j:k}}=g^{\boldsymbol{X}_{j:k}}h^{\boldsymbol{X}_{j:k}}$,
\item $h^{\boldsymbol{X}_{j:k}\uplus\boldsymbol{Y}_{j:k}}=h^{\boldsymbol{X}_{j:k}}h^{\boldsymbol{Y}_{j:k}}$.
\end{itemize}
Formally, a multi-scan GLMB on an interval $\left\{ j:k\right\} $
is a joint multi-object density on $\cprod_{i=j}^{k}\mathcal{F}(\mathbb{X}\times\mathbb{L}_{i})$
of the form 
\begin{equation}
\boldsymbol{\pi}_{j:k}(\boldsymbol{X}_{j:k})=\Delta(\boldsymbol{X}_{j:k})\sum\limits_{\xi\in\Xi}w^{(\xi)}(\mathcal{L}(\boldsymbol{X}_{j:k}))[p^{(\xi)}]^{\boldsymbol{X}_{j:k}},\label{eq:multi_scan_GLMB}
\end{equation}
where: $\Delta(\boldsymbol{X}_{j:k})\triangleq\prod_{i=j}^{k}\Delta(\boldsymbol{X}_{i})$;
$\Xi$ is a set of indices; $\mathcal{L}(\boldsymbol{X}_{j:k})\triangleq(\mathcal{L}(\boldsymbol{X}_{j}),...,\mathcal{L}(\boldsymbol{X}_{k}))$;
$w^{(\xi)}(I_{j:k})$ is non-negative such that $\sum_{\xi,I_{j:k}}w^{(\xi)}(I_{j:k})=1$,
with the sum is taken over $\xi\in\Xi$ and $I_{j:k}\in\cprod_{i=j}^{k}\mathcal{F}(\mathbb{L}_{i})$;
and $p^{(\xi)}(x_{s(\ell):t(\ell)},\ell)$ is the joint density of
the attribute sequence $x_{s(\ell):t(\ell)}$, with $\ell\in\{I_{j:k}$\}
and $\int p^{(\xi)}(x_{s(\ell):t(\ell)},\ell)dx_{s(\ell):t(\ell)}=1$.
Note that $s(\ell)$ and $t(\ell)$ implicitly depend on $(\xi,I_{j:k})$. 

For numerical implementations, the multi-scan GLMB \eqref{eq:multi_scan_GLMB}
can be rewritten in the following \textit{$\delta$-GLMB} form 
\[
\boldsymbol{\pi}_{j:k}(\boldsymbol{X}_{j:k})=\Delta(\boldsymbol{X}_{j:k})\sum\limits_{\xi,I_{j:k}}w^{(\xi,I_{j:k})}\delta_{I_{j:k}}[\mathcal{L}(\boldsymbol{X}_{j:k})][p^{(\xi)}]^{\boldsymbol{X}_{j:k}}.
\]
For simplicity, we denote the multi-scan GLMB $\boldsymbol{\pi}_{j:k}$
as
\[
\boldsymbol{\pi}_{j:k}=\{(w^{(\xi)}(I_{j:k}),p^{(\xi)}):(\xi,I_{j:k})\},
\]
where it is understood that $\xi\in\Xi$ and $I_{j:k}\in\cprod_{i=j}^{k}\mathcal{F}(\mathbb{L}_{i})$.

Analogous to the GLMB, the multi-scan GLMB is a sum of weighted components,
each consisting of a probability/weight $w^{(\xi)}(I_{j:k})$ of hypothesis
$(\xi,I_{j:k})$, and the joint probability density $p^{(\xi)}(\cdot,\ell)$
for the states of trajectory $\ell$ conditioned on hypothesis $(\xi,I_{j:k})$. 

\subsubsection{Multi-scan GLMB recursion}

Under the standard multi-object system model, the multi-scan GLMB
is closed under the Bayes posterior recursion and hence admits an
analytic solution \cite{vo2019}. Specifically, given a multi-scan
GLMB 
\[
\boldsymbol{\pi}_{0:k-1}=\{(w_{-}^{(\xi)}(I_{0:k-1}),p_{-}^{(\xi)}):(\xi,I_{0:k-1})\},
\]
 at time $k-1$ and the measurement $Z_{k}$, the multi-scan GLMB
posterior density at time $k$ is 
\[
\boldsymbol{\pi}_{0:k}(\cdot|Z_{k})\propto\{(w_{Z}^{(\xi,\theta_{k})}(I_{0:k}),p_{Z}^{(\xi,\theta_{k})}):(\xi,I_{0:k},\theta_{k})\},
\]
where $\xi\in\Xi,\,I_{0:k}=(I_{0:k-1},I_{k})\in\cprod_{i=0}^{k}\mathcal{F}(\mathbb{L}_{i}),\,\theta_{k}\in\Theta_{k},$ 
\[
w_{Z}^{(\xi,\theta_{k})}(I_{0:k})=w_{Z}^{(\xi,I_{k-1},\theta_{k},I_{k})}w_{-}^{(\xi)}(I_{0:k-1}),
\]
\[
w_{Z}^{(\xi,I_{k-1},\theta_{k},I_{k})}=\boldsymbol{1}_{\Theta_{k}(I_{k})}^{\mathcal{F}(\mathbb{B}_{k}\uplus I_{k-1})}(\theta_{k})\prod_{\ell\in\mathbb{B}_{k}\uplus I_{k-1}}\omega_{Z}^{(\xi,I_{k},\ell)}(\theta_{k}(\ell)),
\] 
\[
\boldsymbol{1}_{\Theta_{k}(I_{k})}^{\mathcal{F}(\mathbb{B}_{k}\uplus I_{k-1})}(\theta_{k})=\boldsymbol{1}_{\mathcal{F}(\mathbb{B}_{k}\uplus I_{k-1})}(I_{k})\boldsymbol{1}_{\Theta_{k}(I_{k})}(\theta_{k}),
\]
\[
\omega_{Z}^{(\xi,I_{k},\ell)}(j)=\begin{cases}
1-\langle P_{k,S}p_{-}^{(\xi)}\rangle(\ell), & \ell\in\mathbb{L}_{k-1}-I_{k},\\
\overline{\chi}_{S}^{(\xi,j)}(\ell), & \ell\in\mathbb{L}_{k-1}\cap I_{k},\\
1-P_{k,B}(\ell), & \ell\in\mathbb{B}_{k}-I_{k},\\
\overline{\chi}_{B}^{(j)}(\ell), & \ell\in I_{k}\cap\mathbb{B}_{k},
\end{cases}
\]
\begin{align*}
 & p_{Z}^{(\xi,\theta_{k})}(x_{s(\ell):t(\ell)},\ell)\propto\\
 & \quad\begin{cases}
p_{-}^{(\xi)}(x_{s(\ell):t(\ell)},\ell), & t(\ell)<k-1,\\
(1-P_{k,S}(x_{t(\ell)},\ell))p_{-}^{(\xi)}(x_{s(\ell):t(\ell)},\ell), & t(\ell)=k-1,\\
\chi_{S}^{(\theta_{k}(\ell))}(x_{k}|x_{k-1},\ell)p_{-}^{(\xi)}(x_{s(\ell):k-1},\ell), & s(\ell)<t(\ell)=k,\\
\chi_{B}^{(\theta_{k}(\ell))}(x_{k},\ell), & s(\ell)=t(\ell)=k.
\end{cases}
\end{align*}

The multi-scan GLMB contains all information on new born trajectories,
surviving trajectories, terminating trajectories, and previously terminated
trajectories. New trajectories are initiated and surviving trajectories
are updated similarly to those of the GLMB recursion (apart from marginalization
of the past attribute states). However, the multi-scan GLMB recursion
retains and manages information on disappearing and disappeared trajectories,
which are otherwise discarded in the GLMB recursion.

\section{Multi-Object Trajectory Density Approximation}\label{sec:multi_object_density_approx}

This section presents the approximation of multi-object trajectory
densities. Subsection \ref{subsec:multi_object_trajectory_density}
introduces the general form of the multi-object trajectory density,
followed by tractable multi-scan GLMB approximations that minimize
the multi-object information divergence in Subsection \ref{subsec:multi_object_posterior_approx}.
Additionally, Subsection \ref{subsec:SW_approx} provides a tractable
multi-scan multi-object approximation over finite windows.

\subsection{Multi-Object Trajectory Density}\label{subsec:multi_object_trajectory_density}

In many applications involving non-standard multi-object system models,
the multi-object trajectory density of interest is not a GLMB due
to inter-object correlations present in the dynamic or observation
models. Therefore, it is necessary to consider a general form for
the multi-object trajectory density on the interval $\{j:k\}$ that
can capture inter-object dependencies and the multi-modality arising
from data association: 
\begin{equation}
\boldsymbol{\pi}_{j:k}(\boldsymbol{X}_{j:k})=\Delta(\boldsymbol{X}_{j:k})\sum\limits_{\xi\in\Xi}w^{(\xi)}(\mathcal{L}(\boldsymbol{X}_{j:k}))p^{(\xi)}(\boldsymbol{X}_{j:k}),\label{eq:labeledMS_general_form}
\end{equation}
where $\sum_{\xi,I_{j:k}}w^{(\xi)}(I_{j:k})=1$ and 
\[
\int p^{(\xi)}(\boldsymbol{X}_{j:k})\delta\boldsymbol{X}_{j:k}={\displaystyle {\displaystyle \int}...{\displaystyle \int}p^{(\xi)}}(\boldsymbol{X}_{j:k})\delta\boldsymbol{X}_{j}...\delta\boldsymbol{X}_{k}=1.
\]
For notational convenience we denote \eqref{eq:labeledMS_general_form}
as
\[
\boldsymbol{\pi}_{j:k}=\{(w_{j:k}^{(\xi)}(I_{j:k}),p_{j:k}^{(\xi)}):(\xi,I_{j:k})\},
\]
where it is understood that $\xi\in\Xi$ and $I_{j:k}\in\cprod_{i=j}^{k}\mathcal{F}(\mathbb{L}_{i})$.

Unlike the multi-scan GLMB \eqref{eq:multi_scan_GLMB}, each $p_{j:k}^{(\xi)}(\cdot)$
operates on multi-object trajectory $\boldsymbol{X}_{j:k}$ and can
jointly capture all the dependencies between the trajectories.

\subsection{Multi-Object Trajectory Density Approximation}\label{subsec:multi_object_posterior_approx}

Computing the multi-object trajectory density \eqref{eq:labeledMS_general_form}
is numerically intractable due to the exponentially growing number
of hypotheses/components as well as the high-dimensional densities
of the components. This subsection presents a tractable multi-scan
GLMB approximation of arbitrary multi-object trajectory density, which
matches the trajectory cardinality distribution. The proposed approximation
also minimizes the Kullback-Leibler divergence over a special class
of multi-scan GLMB densities.
\begin{defn}
Given a function $f$ on $\cprod_{i=j}^{k}\mathcal{F}(\mathbb{X}\times\mathbb{L}_{i})$,
and $L_{j:k}\in\cprod_{i=j}^{k}\mathcal{F}(\mathbb{L}_{i})$, we define
the \textit{joint label-marginal} of $f$ at $L_{j:k}$, as
\begin{equation}
\langle f\rangle(L_{j:k})=\int f(X_{j}\circledast L_{j},...,X_{k}\circledast L_{k})\delta X_{j:k}.
\end{equation}
Note that:
\begin{equation}
{\displaystyle {\displaystyle \int}f}(\boldsymbol{X}_{j:k})\delta\boldsymbol{X}_{j:k}=\sum_{L_{j}\subseteq\mathbb{L}_{j}}...\sum_{L_{k}\subseteq\mathbb{L}_{k}}\langle f\rangle(L_{j:k}).
\end{equation}
\end{defn}
Following \cite{papi2015}, consider a multi-object trajectory density
$\boldsymbol{\pi}_{j:k}$ on the interval $\{j:k\}$. The \textit{joint
existence probability} of a sequence of label sets $L_{j:k}$ is given
by
\begin{equation}
w(L_{j:k})\triangleq\langle\boldsymbol{\pi}_{j:k}\rangle(L_{j:k}).
\end{equation}
where $w(\emptyset_{j:k})\triangleq\langle\boldsymbol{\pi}_{j:k}\rangle(\emptyset_{j:k})=1$
is a convention. Further, given the label sets $L_{i}$ and $\boldsymbol{X}_{i}=X_{i}\circledast L_{i}$
for $i\in\{j:k\}$, the \textit{joint probability density} of the
multi-object attributes conditioned on $L_{j:k}$ is given by 
\begin{equation}
p_{L_{j:k}}(X_{j:k})\triangleq\frac{\boldsymbol{\pi}_{j:k}(X_{j}\circledast L_{j},...,X_{k}\circledast L_{k})}{w(L_{j:k})}.
\end{equation}
If $w(\mathcal{L}(\boldsymbol{X}_{j:k}))$ is zero, then $p_{\mathcal{L}(\boldsymbol{X}_{j:k})}(\boldsymbol{X}_{j:k})$
is implicitly zero. Thus, the multi-object trajectory density can
be rewritten as 
\begin{equation}
\boldsymbol{\pi}_{j:k}(\boldsymbol{X}_{j:k})=w(\mathcal{L}(\boldsymbol{X}_{j:k}))p_{\mathcal{L}(\boldsymbol{X}_{j:k})}(\boldsymbol{X}_{j:k}).\label{eq:labeled_posterior}
\end{equation}

Following \cite{vo2024}, the \textit{trajectory cardinality} of
$\boldsymbol{X}_{j:k}$ is the number of trajectories in $\{\mathcal{L}(\boldsymbol{X}_{j:k})\}$,
i.e. $|\boldsymbol{X}_{j:k}|\triangleq|\cup_{i=j}^{k}\mathcal{L}(\boldsymbol{X}_{i})|$,
and the \textit{trajectory cardinality distribution} is given by 
\begin{align}
\rho_{\boldsymbol{X}_{j:k}}(n) & =\mathbb{P}_{\boldsymbol{\pi}_{j:k}}(|\boldsymbol{X}_{j:k}|=n)=\mathbb{E}_{\boldsymbol{\pi}_{j:k}}\left[\delta_{n}[|\boldsymbol{X}_{j:k}|]\right],\label{eq:cardinality_distribution}
\end{align}
where $\delta_{n}[|\boldsymbol{X}_{j:k}|]=\delta_{n}[|\cup_{i=j}^{k}\mathcal{L}(\boldsymbol{X}_{i})|]$.
For simplicity, hereon the subscript ``$\boldsymbol{X}_{j:k}$''
is omitted.

The strategy of preserving the trajectory cardinality distribution
and minimizing the Kullback-Leibler divergence over a certain class
of multi-scan GLMB in the approximation bellow is a generalization
of the filtering counterpart in \cite{papi2015}. A multi-scan generalization
of \textit{Marginalized GLMB (M-GLMB)} \cite{fantacci2018,vo2024}
is a sub-class of multi-scan GLMB with density of the following form:
\[
\mathring{\boldsymbol{\pi}}_{j:k}(\boldsymbol{X}_{j:k})=\Delta(\boldsymbol{X}_{j:k})\sum_{I_{j:k}}\mathring{w}^{(I_{j:k})}\delta_{I_{j:k}}[\mathcal{L}(\boldsymbol{X}_{j:k})][\mathring{p}^{(I_{j:k})}]^{\boldsymbol{X}_{j:k}}.
\]
For simplicity, we denote the multi-scan M-GLMB $\mathring{\boldsymbol{\pi}}_{j:k}$
as 
\[
\mathring{\boldsymbol{\pi}}_{j:k}=\{(\mathring{w}^{(I_{j:k})},\mathring{p}^{(I_{j:k})}):I_{j:k}\},
\]
 where it is understood that $I_{j:k}\in\cprod_{i=j}^{k}\mathcal{F}(\mathbb{L}_{i})$.
\begin{prop}
	\label{thm:MSapproximation}Given any multi-object trajectory density $\boldsymbol{\pi}_{j:k}$, the multi-scan GLMB density $\hat{\boldsymbol{\pi}}_{j:k}$ that matches the trajectory cardinality distribution has hypothesis weights satisfying 
	\begin{align*}
	\sum_{\xi\in\Xi}\hat{w}^{(\xi)}(I_{j:k}) & =\langle\boldsymbol{\pi}_{j:k}\rangle(I_{j:k}).
	\end{align*}

	Further, the multi-scan M-GLMB density $\mathring{\boldsymbol{\pi}}_{j:k}$
	that matches the trajectory cardinality distribution and minimizes
	the Kullback-Leibler divergence from $\boldsymbol{\pi}_{j:k}$ is
	\begin{align*}
	\mathring{\boldsymbol{\pi}}_{j:k} & =\{(\mathring{w}^{(I_{j:k})},\mathring{p}^{(I_{j:k})}):I_{j:k}\},\\
	\mathring{w}^{(I_{j:k})} & =\langle\boldsymbol{\pi}_{j:k}\rangle(I_{j:k}),\\
	\mathring{p}^{(I_{j:k})}(\boldsymbol{x}_{T(\ell)}) & =\boldsymbol{1}_{\{I_{j:k}\}}(\ell)\langle p_{I_{j:k}}(\{\boldsymbol{x}_{T(\ell)}\}\uplus\cdot)\rangle(\{I_{j:k}\}-\{\ell\}).
	\end{align*}
\end{prop}
The above result establishes that, a multi-scan GLMB density can
be used to approximate the multi-object trajectory density with matching
trajectory cardinality distribution. Further, if we restrict ourselves
to the class of multi-scan M-GLMB, the approximation that also minimizes
the Kullback-Leibler divergence can be obtained by replacing its label
trajectory-conditioned attribute probability densities $p_{I_{j:k}}(\boldsymbol{X}_{j:k})$
by the product of their trajectory marginals $\mathring{p}^{(I_{j:k})}(\boldsymbol{x}_{T(\ell)})$.
The proof is given in Subsection \ref{subsec:proof_approx} of the
Supplementary Materials. 

\subsection{Approximation with Finite Windows}\label{subsec:SW_approx}

In practice, a growing time window is infeasible. A practical alternative
is to approximate the multi-scan multi-object density from $\{j:k\}$
using shorter windows \cite{vo2019,moratuwage2022}. Let $N_{S}$
be the number of disjoint sub-windows of $\{j:k\}$ such that $\{j:k\}=\uplus_{i=1}^{N_{S}}\{j^{(i)}:k^{(i)}\}$,
where $\{j^{(i)}:k^{(i)}\}$ is a smoothing window from $j^{(i)}$
to $k^{(i)}$ with $j^{(i)}\geq j$ and $k^{(i)}\leq k$, for all
$i\in\{1:N_{S}\}$. For notational convenience, we use $\{\bar{j}^{(i)}:\bar{k}^{(i)}\}$
to denote $\{j:k\}\setminus\{j^{(i)}:k^{(i)}\}$.  

Analogous to single-object smoothing \cite{meditch1973,anderson1979,doucet2009},
a finite-window approximation of the multi-scan multi-object density
with minimal Kullback-Leibler divergence is summarized in Proposition
\ref{prop:smoothing_window_approx}.
\begin{prop}
\label{prop:smoothing_window_approx}Given the multi-scan multi-object
density $\boldsymbol{\pi}_{j:k}$ on the window $\{j:k\}$, the minimal
Kullback-Leibler divergence approximation of $\boldsymbol{\pi}_{j:k}$
by multi-densities on the sub-windows $\{j^{(i)}:k^{(i)}\}_{i\in\{1:N_{S}\}}$
is given by 
\begin{align*}
\check{\boldsymbol{\pi}}_{j:k}(\boldsymbol{X}_{j:k}) & =\prod_{i=1}^{N_{S}}\check{\boldsymbol{\pi}}_{j^{(i)}:k^{(i)}}(\boldsymbol{X}_{j^{(i)}:k^{(i)}}),\\
\check{\boldsymbol{\pi}}_{j^{(i)}:k^{(i)}}(\boldsymbol{X}_{j^{(i)}:k^{(i)}}) & =\int\boldsymbol{\pi}_{j:k}(\boldsymbol{X}_{j:k})\delta\boldsymbol{X}_{\bar{j}^{(i)}:\bar{k}^{(i)}}.
\end{align*}
\end{prop}
Using a set of $N_{S}$ disjoint shorter windows, the multi-scan multi-object
density on $\{j:k\}$ can be approximated by the product of multi-scan
multi-object densities on $\{j^{(i)}:k^{(i)}\}$, $i=1:N_{S}$ with
minimal Kullback-Leibler divergence (see Subsection \ref{subsec:proof_moving_window}
of the Supplementary Materials for proof).

\section{Approximate Multi-Object Posterior Recursion}\label{sec:interactive_modeling}

This section presents a tractable approximation of the labeled multi-object
posterior recursion for interacting trajectories. Subsection \ref{subsec:interacting_recursion}
specifies the multi-object transition model for interacting objects
and the corresponding posterior Bayes recursion. In Subsection \ref{subsec:interacting_multi_object_approx},
the approximation method from Section \ref{sec:multi_object_density_approx}
is applied to develop two strategies for approximating this labeled
multi-object posterior.

\subsection{Interacting Multi-Object Posterior Recursion}\label{subsec:interacting_recursion}

At time $k$, the multi-object state $\boldsymbol{X}_{k}=\boldsymbol{B}_{k}\uplus\boldsymbol{S}_{k}$
is the disjoint union of newly born objects $\boldsymbol{B}_{k}=\boldsymbol{X}_{k}\cap(\mathbb{X}\times\mathbb{B}_{k})$
and surviving objects $\boldsymbol{S}_{k}=\boldsymbol{X}_{k}-(\mathbb{X}\times\mathbb{B}_{k})$.
Assuming births and survivals occur independently, and that correlations
exist only among surviving objects. Let $\boldsymbol{f}_{k,S}(\boldsymbol{S}_{k}|\boldsymbol{X}_{k-1})$
be the general form of the surviving multi-object density. The multi-object
Markov transition density is given by  
\begin{align}
\boldsymbol{f}_{k}(\boldsymbol{X}_{k}|\boldsymbol{X}_{k-1}) & =\Delta(\boldsymbol{X}_{k})\boldsymbol{f}_{k,B}(\boldsymbol{B}_{k})\boldsymbol{\Phi}_{k,S}(\boldsymbol{S}_{k}|\boldsymbol{X}_{k-1}),\label{eq:Markov_transition}\\
\boldsymbol{f}_{k,B}(\boldsymbol{B}_{k}) & =w_{k,B}(\mathcal{L}(\boldsymbol{B}_{k}))[p_{k,B}]^{\boldsymbol{B}_{k}},\nonumber \\
w_{k,B}(\mathcal{L}(\boldsymbol{B}_{k})) & =[Q_{k,B}]^{\mathbb{B}_{k}-\mathcal{L}(\boldsymbol{B}_{k})}[P_{k,B}]^{\mathcal{L}(\boldsymbol{B}_{k})},\nonumber \\
\boldsymbol{\Phi}_{k,S}(\boldsymbol{S}_{k}|\boldsymbol{X}_{k-1}) & =\boldsymbol{1}_{\mathcal{L}(\boldsymbol{X}_{k-1})}^{\mathcal{L}(\boldsymbol{S}_{k})}w_{k,S}(\mathcal{L}(\boldsymbol{S}_{k}))\boldsymbol{f}_{k,S}(\boldsymbol{S}_{k}|\boldsymbol{X}_{k-1}),\nonumber \\
\boldsymbol{1}_{\mathcal{L}(\boldsymbol{X}_{k-1})}^{\mathcal{L}(\boldsymbol{S}_{k})} & =\prod_{\ell\in\mathcal{L}(\boldsymbol{S}_{k})}\boldsymbol{1}_{\mathcal{L}(\boldsymbol{X}_{k-1})}(\ell),\nonumber \\
w_{k,S}(\mathcal{L}(\boldsymbol{S}_{k})) & =[Q_{k,S}]^{\mathcal{L}(\boldsymbol{X}_{k-1})-\mathcal{L}(\boldsymbol{S}_{k})}[P_{k,S}]^{\mathcal{L}(\boldsymbol{S}_{k})}.\nonumber 
\end{align}

Given the multi-object trajectory $\boldsymbol{X}_{0:k-1}$ at time
$k-1$, the multi-object trajectory $\boldsymbol{X}_{0:k}$ at time
$k$ can be decomposed as $\boldsymbol{B}_{k}\uplus\boldsymbol{S}_{0:k}\uplus\boldsymbol{D}_{0:k-1}$,
where $\boldsymbol{B}_{k}$ is the set of new births, $\boldsymbol{S}_{0:k}=\{\boldsymbol{x}_{T(\ell)}\in\boldsymbol{X}_{0:k}:\ell\in\mathcal{L}(\boldsymbol{X}_{0:k-1})\cap\mathcal{L}(\boldsymbol{X}_{k})\}$
is the set of surviving trajectories, and $\boldsymbol{D}_{0:k-1}=\{\boldsymbol{x}_{T(\ell)}\in\boldsymbol{X}_{0:k}:\ell\in\mathcal{L}(\boldsymbol{X}_{0:k-1})-\mathcal{L}(\boldsymbol{X}_{k})\}$
is the set of trajectories that have either just disappeared or were
previously terminated. Proposition \ref{prop:jointpredictionprop}
and Proposition \ref{prop:jointupdateprop}, describe the propagation
of the labeled multi-object posterior through prediction and update
steps, respectively (see Subsections \ref{subsec:proof_pred} and
\ref{subsec:proof_update} of the Supplementary Materials for proofs).
\begin{prop}
\label{prop:jointpredictionprop}Given the labeled multi-object posterior
density $\boldsymbol{\pi}_{0:k-1}=\{(w_{-}^{(\xi)}(I_{0:k-1}),p_{-}^{(\xi)}):(\xi,I_{0:k-1})\}$
at time $k-1$, the multi-object prediction density at time $k$,
under the multi-object dynamic model described by \eqref{eq:Markov_transition},
is given by %
\begin{comment}
\begin{align*}
 & \boldsymbol{\pi}_{0:k,+}(\mathbf{X}_{0:k})=\\
 & \quad\Delta(\mathbf{X}_{0:k})\sum\limits_{\xi,I_{0:k}}w_{+}^{(\xi,I_{0:k})}\delta_{I_{0:k}}[\mathcal{L}(\mathbf{X}_{0:k})]p_{+}^{(\xi)}(\mathbf{X}_{0:k}),
\end{align*}
\end{comment}
{} 
\begin{equation}
\boldsymbol{\pi}_{0:k}=\{(w_{0:k}^{(\xi)}(I_{0:k}),p_{0:k}^{(\xi)}):(\xi,I_{0:k})\},\label{eq:predictionMSlabeled}
\end{equation}
where $\xi\in\Xi,I_{0:k}\in\cprod_{i=0}^{k}\mathcal{F}(\mathbb{L}_{i}),$
and \allowdisplaybreaks
\begin{comment}
\begin{align}
w_{+}^{(\xi,I_{0:k})} & =[Q_{B,k}]^{\mathbb{B}_{k}-(\mathbb{B}_{k}\cap I_{k})}[P_{B,k}]^{\mathbb{B}_{k}\cap I_{k}},\\
 & \times[Q_{S,k}]^{I_{k-1}-I_{k}-\mathbb{B}_{k}}[P_{S,k}]^{I_{k-1}\cap I_{k}},\nonumber \\
 & \times1_{\mathcal{F}(\mathbb{B}_{k}\uplus\mathcal{L}(\mathbf{X}_{k-1}))}(I_{k})w_{-}^{(\xi)}(I_{0:k-1}),\nonumber 
\end{align}
\begin{align}
p_{+}^{(\xi)}(\mathbf{X}_{0:k}) & =[p_{B,+}]^{\mathbf{B}_{k}}p_{S,+}^{(\xi)}(\mathbf{S}_{0:k})\\
 & \times p_{-}^{(\xi)}(\mathbf{D}_{0:k-1})p_{-}^{(\xi)}(\mathbf{T}_{0:k-1}),\nonumber 
\end{align}
\begin{equation}
p_{S,+}^{(\xi)}(\mathbf{S}_{0:k})=\boldsymbol{f}_{S,+}(\mathbf{S}_{k}|\mathbf{X}_{k-1})p_{-}^{(\xi)}(\mathbf{S}_{0:k-1}).
\end{equation}
\end{comment}
\begin{align*}
w_{0:k}^{(\xi)}(I_{0:k}) & =\boldsymbol{1}_{I_{k-1}}^{I_{k}-\mathbb{B}_{k}}\eta_{k}^{(I_{k-1},I_{k})}w_{-}^{(\xi)}(I_{0:k-1}),\\
\boldsymbol{1}_{I_{k-1}}^{I_{k}-\mathbb{B}_{k}} & =\prod_{\ell\in I_{k}-\mathbb{B}_{k}}\boldsymbol{1}_{I_{k-1}}(\ell),\\
\eta_{k}^{(I_{k-1},I_{k})} & =w_{k,B}(\mathbb{B}_{k}\cap I_{k})w_{k,S}(I_{k-1}\cap I_{k}),\\
p_{0:k}^{(\xi)}(\boldsymbol{X}_{0:k}) & =[p_{k,B}]^{\boldsymbol{B}_{k}}p_{k,S}^{(\xi)}(\boldsymbol{S}_{0:k}\uplus\boldsymbol{D}_{0:k-1}),\\
p_{k,S}^{(\xi)}(\boldsymbol{S}_{0:k}\uplus\boldsymbol{D}_{0:k-1}) & =p_{k,S}^{(\xi)}(\boldsymbol{S}_{0:k}|\boldsymbol{D}_{0:k-1})p_{-}^{(\xi)}(\boldsymbol{D}_{0:k-1}),\\
p_{k,S}^{(\xi)}(\boldsymbol{S}_{0:k}|\boldsymbol{D}_{0:k-1}) & =\boldsymbol{f}_{k,S}(\boldsymbol{S}_{k}|\boldsymbol{X}_{k-1})p_{-}^{(\xi)}(\boldsymbol{S}_{0:k-1}|\boldsymbol{D}_{0:k-1}).
\end{align*}
\end{prop}
The hypothesis $(\xi,I_{0:k-1})$ of the previous labeled multi-object
posterior generates the set of children hypotheses $(\xi,I_{0:k})$
of the multi-object prediction density. The weight $w_{0:k}^{(\xi)}(I_{0:k})$
of the predictive hypothesis $(\xi,I_{0:k})$ is given by the product
of: the probabilities of unborn and newly born labels, the predictive
survival probabilities of disappearing and surviving labels, and the
previous hypothesis weight. Similarly, for each predictive hypothesis
$(\xi,I_{0:k})$, its corresponding probability density $p_{0:k}^{(\xi)}$
is given by the product of: the prediction density of newly born trajectories,
the prediction density of surviving trajectories, and the density
of trajectories that have just disappeared or were previously terminated. 
\begin{prop}
\label{prop:jointupdateprop}Given, at time $k$, the multi-object
prediction density \eqref{eq:predictionMSlabeled} and the measurement
set $Z_{k}$, under the standard multi-object observation model described
by \eqref{eq:likelihood_func}, the multi-object posterior density
is%
\begin{comment}
\begin{align*}
 & \boldsymbol{\pi}_{0:k,+}(\mathbf{X}_{0:k}|Z_{0:k})\propto\\
 & \Delta(\mathbf{X}_{0:k})\sum\limits_{\xi,I_{0:k},\theta_{k}}w_{Z_{+}}^{(\xi,I_{0:k},\theta_{k})}\delta_{I_{0:k}}[\mathcal{L}(\mathbf{X}_{0:k})]p_{Z_{+}}^{(\xi,\theta_{k})}(\mathbf{X}_{0:k}),
\end{align*}
\end{comment}
{} 
\begin{equation}
\boldsymbol{\pi}_{0:k}(\cdot|Z_{k})\propto\{(w_{0:k,Z}^{(\xi,\theta_{k})}(I_{0:k}),p_{0:k,Z}^{(\xi,\theta_{k})}):(\xi,I_{0:k},\theta_{k})\},\label{eq:updateMSlabeled}
\end{equation}
where $\xi\in\Xi,I_{0:k}\in\cprod_{i=0}^{k}\mathcal{F}(\mathbb{L}_{i}),\theta_{k}\in\Theta_{k},$
and %
\begin{comment}
\begin{align*}
w_{Z_{+}}^{(\xi,\theta_{k})}(I_{0:k}) & =\boldsymbol{1}_{\Theta_{k}(I_{k})}(\theta_{k})w_{+}^{(\xi)}(I_{0:k}),\\
 & \times[\mu_{B,Z_{+}}^{(\theta_{k})}]^{I_{k}\cap\mathbb{B}_{k}}\mu_{S,Z_{+}}^{(\xi,\theta_{k})}(I_{k}-\mathbb{B}_{k}),
\end{align*}
\begin{equation}
\mu_{B,Z_{+}}^{(\theta_{k})}(\ell)=\left\langle p_{B,+}(\cdot,\ell),\psi_{k,Z_{+}}^{(\theta_{k}(\mathcal{L}(\cdot)))}(\cdot,\ell)\right\rangle ,
\end{equation}
\begin{equation}
\mu_{S,Z_{+}}^{(\xi,\theta_{k})}(L)=\left\langle p_{S,+}^{(\xi,\theta_{k})}(\cdot),\left[\psi_{k,Z_{+}}^{(\theta_{k}(\mathcal{L}(\cdot)))}\right]^{(\cdot)}\right\rangle ,
\end{equation}
\begin{align*}
p_{Z_{+}}^{(\xi,\theta_{k})}(\mathbf{X}_{0:k}) & \propto\left[p_{B,Z_{+}}^{(\theta_{k})}\right]^{\mathbf{B}_{k}}p_{S,Z_{+}}^{(\xi,\theta_{k})}(\mathbf{S}_{0:k})\\
 & \times p_{-}^{(\xi)}(\mathbf{D}_{0:k-1})p_{-}^{(\xi)}(\mathbf{T}_{0:k-1}),
\end{align*}
\begin{equation}
\left[p_{B,Z_{+}}^{(\theta_{k})}\right]^{\mathbf{B}_{k}}=\left[p_{B,+}\psi_{k,Z_{+}}^{(\theta_{k}(\mathcal{L}(\cdot)))}\right]^{\mathbf{B}_{k}},
\end{equation}
\begin{equation}
p_{S,Z_{+}}^{(\xi,\theta_{k})}(\mathbf{S}_{0:k})=p_{S,+}^{(\xi)}(\mathbf{S}_{0:k})\left[\psi_{k,Z_{+}}^{(\theta_{k}(\mathcal{L}(\cdot)))}\right]^{\mathbf{S}_{0:k}}.
\end{equation}
\end{comment}
\begin{align*}
w_{0:k,Z}^{(\xi,\theta_{k})}(I_{0:k}) & =\boldsymbol{1}_{\Theta_{k}(I_{k})}(\theta_{k})\mu_{Z}^{(\xi,I_{k},\theta_{k})}w_{0:k}^{(\xi)}(I_{0:k}),\\
\mu_{Z}^{(\xi,I_{k},\theta_{k})} & =[\mu_{B,Z}^{(\theta_{k})}]^{I_{k}\cap\mathbb{B}_{k}}\mu_{S,Z}^{(\xi,\theta_{k})}(I_{k}-\mathbb{B}_{k}),\\
\mu_{B,Z}^{(\theta_{k})}(\ell) & =\left\langle p_{k,B}(\cdot,\ell),\psi_{k,Z}^{(\theta_{k}(\mathcal{L}(\cdot)))}(\cdot,\ell)\right\rangle ,\\
\mu_{S,Z}^{(\xi,\theta_{k})}(L) & =\left\langle p_{k,S}^{(\xi)}(\cdot),\left[\psi_{k,Z}^{(\theta_{k}(\mathcal{L}(\cdot)))}\right]^{(\cdot)}\right\rangle (L),\\
p_{0:k,Z}^{(\xi,\theta_{k})}(\boldsymbol{X}_{0:k}) & \propto[p_{B,Z}^{(\theta_{k})}]^{\boldsymbol{B}_{k}}p_{S,Z}^{(\xi,\theta_{k})}(\boldsymbol{S}_{0:k}\uplus\boldsymbol{D}_{0:k-1}),\\{}
[p_{B,Z}^{(\theta_{k})}]^{\boldsymbol{B}_{k}} & =[p_{k,B}\psi_{k,Z}^{(\theta_{k}(\mathcal{L}(\cdot)))}]^{\boldsymbol{B}_{k}},\\
p_{S,Z}^{(\xi,\theta_{k})}(\boldsymbol{S}_{0:k}\uplus\boldsymbol{D}_{0:k-1}) & =p_{S,Z}^{(\xi,\theta_{k})}(\boldsymbol{S}_{0:k}|\boldsymbol{D}_{0:k-1})p_{-}^{(\xi)}(\boldsymbol{D}_{0:k-1}),\\
p_{S,Z}^{(\xi,\theta_{k})}(\boldsymbol{S}_{0:k}|\boldsymbol{D}_{0:k-1}) & =p_{k,S}^{(\xi)}(\boldsymbol{S}_{0:k}|\boldsymbol{D}_{0:k-1})[\psi_{k,Z}^{(\theta_{k}(\mathcal{L}(\cdot)))}]^{\boldsymbol{S}_{0:k}}.
\end{align*}
\end{prop}
The predictive hypothesis $(\xi,I_{0:k})$ generates the set of children
hypotheses $(\xi,I_{0:k},\theta_{k})$ (assuming $\theta_{k}$ is
a valid association map i.e. $\boldsymbol{1}_{\Theta_{k}(I_{k})}(\theta_{k})=1$)
for the labeled multi-object posterior density. The weight $w_{0:k,Z}^{(\xi,\theta_{k})}(I_{0:k})$
of the updated hypothesis $(\xi,I_{0:k},\theta_{k})$ is the product
of: the predictive weight, the weight of updated new born labels,
and the weight of updated surviving labels. Since the disappearing
trajectories are terminated, its predictive multi-object density is
not updated with measurements. Further, for each updated hypothesis
$(\xi,I_{0:k},\theta_{k}),$ its probability density $p_{0:k,Z}^{(\xi,\theta_{k})}$
is given by the product of: the updated density of new born trajectories,
the updated density of surviving trajectories, and the density of
trajectories that have just disappeared or were previously terminated.

\subsection{Approximate Multi-Object Posterior Recursion}\label{subsec:interacting_multi_object_approx}

In certain estimation problems with data association uncertainty (e.g.,
\cite{beard2015}, \cite{nguyen2021}, \cite{ong2022}), the multi-scan
M-GLMB density in Proposition \ref{thm:MSapproximation} cannot approximate
the labeled multi-object density in \eqref{eq:labeledMS_general_form}
whilst capturing its modes and associated information.\textcolor{red}{{}
}However, we can use Proposition \ref{thm:MSapproximation} to approximate
\eqref{eq:labeledMS_general_form} with matching trajectory cardinality
whilst still capturing the modes by
\begin{equation}
\hat{\boldsymbol{\pi}}_{j:k}=\{\hat{w}^{(\xi,I_{j:k})},\hat{p}^{(\xi,I_{j:k})}:(\xi,I_{j:k})\},\label{eq:multi-scan-approx}
\end{equation}
where $\hat{w}^{(\xi,I_{j:k})}=w^{(\xi)}(I_{j:k})$, and 
\[
\hat{p}^{(\xi,I_{j:k})}(\boldsymbol{x}_{T(\ell)})=\boldsymbol{1}_{\{I_{j:k}\}}(\ell)\langle p^{(\xi)}(\{\boldsymbol{x}_{T(\ell)}\}\uplus\cdot)\rangle(\{I_{j:k}\}-\{\ell\}).
\]
The rationale of approximating $p^{(\xi)}(\boldsymbol{X}_{j:k})$
at each of the modes of \eqref{eq:labeledMS_general_form} by the
product of its trajectory marginals $\hat{p}^{(\xi,I_{j:k})}(\boldsymbol{x}_{T(\ell)})$
is to preserve the information contained in the modes. This approximation
requires more components than the multi-scan M-GLMB approximation,
but intuitively incurs less information loss \cite{vo2024}. However,
deriving an analytical expression for the Kullback-Leibler divergence
to formally verify the approximation in \eqref{eq:multi-scan-approx}
remains intractable due to its mixture structure \footnote{Even for the much simpler case of Gaussian mixtures, there is no analytic
expressions for the Kullback Leibler divergence.}.\textcolor{red}{}

\subsubsection{Prediction Approximation}

This strategy, summarized in Corollary \ref{cor:predictionapproximation},
approximates the multi-object prediction density \eqref{eq:predictionMSlabeled}
by a multi-scan GLMB in Proposition \ref{thm:MSapproximation} with
matching trajectory cardinality distribution, which is then updated
with the new measurements to yield a multi-scan GLMB posterior.
\begin{cor}
\label{cor:predictionapproximation}A multi-scan GLMB matching the
multi-object prediction density \eqref{eq:predictionMSlabeled} in
trajectory cardinality distribution is
\begin{equation}
\hat{\boldsymbol{\pi}}_{0:k}=\{(\hat{w}^{(\xi,I_{0:k})},\hat{p}^{(\xi,I_{0:k})}):(\xi,I_{0:k})\},\label{eq:predictionMSapprox}
\end{equation}
where: $\xi\in\Xi$, $I_{0:k}=(I_{0:k-1},I_{k})$, $\hat{w}^{(\xi,I_{0:k})}=w_{0:k}^{(\xi)}(I_{0:k})$,
and
\begin{align*}
 & \hat{p}^{(\xi,I_{0:k})}(x_{s(\ell):t(\ell)},\ell)=\\
 & \quad\begin{cases}
p_{-}^{(\xi)}(x_{s(\ell):t(\ell)},\ell), & t(\ell)<k-1,\\
\hat{p}_{-}^{(\xi,I_{k-1}-I_{k}-\mathbb{B}_{k})}(x_{s(\ell):t(\ell)},\ell), & t(\ell)=k-1,\\
\hat{p}_{k,S}^{(\xi,I_{k}-\mathbb{B}_{k})}(x_{s(\ell):t(\ell)},\ell), & s(\ell)<t(\ell)=k,\\
p_{k,B}(x_{k},\ell), & s(\ell)=t(\ell)=k,
\end{cases}\\
 & \hat{p}_{-}^{(\xi,L_{k-1})}(x_{s(\ell):t(\ell)},\ell)=\\
 & \quad\boldsymbol{1}_{L_{k-1}}(\ell)\left\langle p_{-}^{(\xi)}(\{(x_{s(\ell):t(\ell)},\ell)\}\uplus\cdot)\right\rangle (L_{k-1}-\{\ell\}),\\
 & \hat{p}_{k,S}^{(\xi,L_{k})}(x_{s(\ell):t(\ell)},\ell)=\\
 & \quad\boldsymbol{1}_{L_{k}}(\ell)\left\langle p_{k,S}^{(\xi)}(\{(x_{s(\ell):t(\ell)},\ell)\}\uplus\cdot)\right\rangle (L_{k}-\{\ell\}).
\end{align*}
In addition, given the multi-object measurement $Z_{k}$ and the approximate
multi-scan GLMB prediction density \eqref{eq:predictionMSapprox},
under the standard multi-object observation model \eqref{eq:likelihood_func},
the multi-scan GLMB posterior density is 
\begin{equation}
\bar{\boldsymbol{\pi}}_{0:k}(\cdot|Z_{k})\propto\{(\bar{w}_{Z}^{(\xi,I_{0:k},\theta_{k})},\bar{p}_{Z}^{(\xi,I_{0:k},\theta_{k})}):(\xi,I_{0:k},\theta_{k})\},\label{eq:updateafterpredictionMS}
\end{equation}
where $\theta_{k}\in\Theta_{k}$, and 
\begin{align}
\bar{w}_{Z}^{(\xi,I_{0:k},\theta_{k})} & =1_{\Theta_{k}(I_{k})}(\theta_{k})\hat{w}^{(\xi,I_{0:k})}\left[\bar{\psi}_{k,Z}^{(\xi,\theta_{k})}\right]^{I_{0:k}},\label{eq:weightpredictionMSapprox}\\
\bar{p}_{Z}^{(\xi,I_{0:k},\theta_{k})}(\cdot,\ell) & =\begin{cases}
\frac{\hat{p}^{(\xi,I_{0:k})}(\cdot,\ell)\psi_{k,Z}^{(\theta_{k}(\ell))}(\cdot,\ell)}{\bar{\psi}_{k,Z}^{(\xi,\theta_{k})}(\ell)}, & t(\ell)=k,\\
\hat{p}^{(\xi,I_{0:k})}(\cdot,\ell), & t(\ell)<k,
\end{cases}\nonumber \\
\bar{\psi}_{k,Z}^{(\xi,\theta_{k})}(\ell) & =\left\langle \hat{p}^{(\xi,I_{0:k})}(\cdot,\ell),\psi_{k,Z}^{(\theta_{k}(\ell))}(\cdot,\ell)\right\rangle .\nonumber 
\end{align}
\end{cor}

\subsubsection{Update approximation}

This strategy, summarized in Corollary \ref{cor:updateapproximation},
first performs a posterior Bayes recursion to obtain the labeled multi-object
posterior \eqref{eq:updateMSlabeled}, which is then approximated
by a multi-scan GLMB in Proposition \ref{thm:MSapproximation} with
matching trajectory cardinality distribution.
\begin{cor}
\label{cor:updateapproximation}A multi-scan GLMB matching the labeled
multi-object posterior \eqref{eq:updateMSlabeled} in trajectory cardinality
distribution is %
\begin{comment}
\begin{align*}
 & \widetilde{\boldsymbol{\pi}}_{0:k,+}(\mathbf{X}_{0:k}|Z_{0:k})\propto\\
 & \quad\Delta(\mathbf{X}_{0:k})\sum\limits_{\xi,I_{0:k},\theta_{k}}w_{Z_{+}}^{(\xi,I_{0:k},\theta_{k})}\delta_{I_{0:k}}[\mathcal{L}(\mathbf{X}_{0:k})][p_{Z_{+}}^{(\xi,I_{0:k},\theta_{k})}]^{\mathbf{X}_{0:k}},
\end{align*}
\end{comment}
{} 
\begin{equation}
\hat{\boldsymbol{\pi}}_{0:k}(\cdot|Z_{k})\propto\{(\hat{w}_{Z}^{(\xi,I_{0:k},\theta_{k})},\hat{p}_{Z}^{(\xi,I_{0:k},\theta_{k})}):(\xi,I_{0:k},\theta_{k})\},\label{eq:MSupdateapprox}
\end{equation}
where: $\xi\in\Xi$, $I_{0:k}=(I_{0:k-1},I_{k})$, $\theta_{k}\in\Theta_{k}$,
$\hat{w}_{Z}^{(\xi,I_{0:k},\theta_{k})}=w_{0:k,Z}^{(\xi,\theta_{k})}(I_{0:k})$,
and 
\begin{align*}
 & \hat{p}_{Z}^{(\xi,I_{0:k},\theta_{k})}(x_{s(\ell):t(\ell)},\ell)=\\
 & \quad\begin{cases}
p_{-}^{(\xi)}(x_{s(\ell):t(\ell)},\ell), & t(\ell)<k-1,\\
\hat{p}_{-}^{(\xi,I_{k-1}-I_{k}-\mathbb{B}_{k})}(x_{s(\ell):t(\ell)},\ell), & t(\ell)=k-1,\\
\hat{p}_{S,Z}^{(\xi,I_{k}-\mathbb{B}_{k},\theta_{k})}(x_{s(\ell):t(\ell)},\ell), & s(\ell)<t(\ell)=k,\\
p_{B,Z}^{(\theta_{k})}(x_{k},\ell), & s(\ell)=t(\ell)=k,
\end{cases}\\
 & \hat{p}_{-}^{(\xi,L_{k-1})}(x_{s(\ell):t(\ell)},\ell)=\\
 & \quad\boldsymbol{1}_{L_{k-1}}(\ell)\left\langle p_{-}^{(\xi)}(\{(x_{s(\ell):t(\ell)},\ell)\}\uplus\cdot)\right\rangle (L_{k-1}-\{\ell\}),\\
 & \hat{p}_{S,Z}^{(\xi,L_{k},\theta_{k})}(x_{s(\ell):t(\ell)},\ell)=\\
 & \quad\boldsymbol{1}_{L_{k}}(\ell)\left\langle p_{S,Z}^{(\xi,\theta_{k})}(\{(x_{s(\ell):t(\ell)},\ell)\}\uplus\cdot)\right\rangle (L_{k}-\{\ell\}).
\end{align*}
\end{cor}
In principle, performing the approximation in \eqref{eq:predictionMSapprox}
is computationally less demanding than propagating \eqref{eq:updateMSlabeled}
due to the dimension of joint densities. However, the approximation
via \eqref{eq:MSupdateapprox} is expected to be more accurate than
\eqref{eq:updateafterpredictionMS}, at the price of the highly expensive
computation of the joint densities \eqref{eq:updateMSlabeled}. 
\begin{rem}
Approximating the joint density as the product of independent marginals
may yield overconfident covariance estimates. In the single-object
setting, such overconfidence can be assessed using the Normalized
Estimation Error Squared (NEES) metric \cite{barshalom2001}. Extending
NEES to enable covariance calibration analysis in the multi-object
case remains an open problem and warrants further investigation.
\end{rem}

\section{Implementations}\label{sec:implementations}

This section outlines the implementation of the approximate multi-object
posterior recursions described in Section \ref{sec:interactive_modeling}.
Due to the super-exponential growth in the number of posterior GLMB
components, truncation is essential to maintain computational tractability.
However, the scale of the problem and the presence of inter-object
dependencies render traditional ranked assignment methods intractable.
To address this, we employ Gibbs sampling techniques \cite{vo2019}
to truncate the multi-scan GLMB while effectively capturing inter-object
interactions in the approximation. 

\subsection{Multi-Dimensional Ranked Assignment Problem}\label{subsec:ranked_assignment_problem}

Truncating the multi-scan GLMB by discarding components with the
smallest weights minimizes the $L_{1}$-norm of the approximation
error \cite{vo2019}. To formulate the truncation problem, we represent
each association map $\theta_{k}\in\Theta_{k}$ of the multi-scan
GLMB by an extended association map $\gamma_{k}$ that inherits the
positive 1-1 property, defined by
\begin{equation}
\gamma_{k}(\ell)=\begin{cases}
\theta_{k}(\ell), & \textrm{if \ensuremath{\ell\in\mathcal{D}(\theta_{k})},}\\
-1, & \textrm{otherwise.}
\end{cases}\label{eq:MS-extended-assoc-map}
\end{equation}
Denote the set of all $\gamma_{k}:\mathbb{L}_{k}\rightarrow\{-1:|Z_{k}|\}$
by $\Gamma_{k}$ and the live labels of $\gamma_{k}$ by $\mathcal{L}(\gamma_{k})\triangleq\{\ell\in\mathbb{L}_{k}:\gamma_{k}(\ell)\geq0\}$.
For $\gamma_{k}\in\Gamma_{k}$, each $\theta_{k}\in\Theta_{k}$ is
recovered by $\theta_{k}(\ell)=\gamma_{k}(\ell)$, $\ell\in\mathcal{L}(\gamma_{k})$.
Hence there exists a bijection between $\Gamma_{k}$ and $\Theta_{k}$
\cite{vo2017}.

\subsubsection{Prediction Approximation}\label{subsec:PA}

Enumerating $I_{k-1}\uplus\mathbb{B}_{k}=\{\ell_{1:P_{k}}\},$ for
any $\ell_{i}$, $i\in\{1:P_{k}\}$, $u=\gamma_{k}(\ell_{i})$, define
\begin{align*}
\eta_{k}^{(i)}(u) & =\begin{cases}
\bar{Q}_{S}^{(\xi)}(\ell_{i}), & \ell_{i}\in\mathbb{L}_{k-1},u<0,\\
\bar{\Lambda}_{S}^{(\xi,u)}(\ell_{i}), & \ell_{i}\in\mathbb{L}_{k-1},u\geq0,\\
Q_{B}(\ell_{i}), & \ell_{i}\in\mathbb{B}_{k},u<0,\\
\bar{\chi}_{B}^{(u)}(\ell_{i}), & \ell_{i}\in\mathbb{B}_{k},u\geq0,
\end{cases}\\
\bar{Q}_{S}^{(\xi)}(\ell) & =\langle Q_{k,S}(\cdot,\ell)\hat{p}_{-}^{(\xi,I_{k-1}-I_{k}-\mathbb{B}_{k})}(\cdot,\ell)\rangle,\\
\bar{\Lambda}_{S}^{(\xi,u)}(\ell) & =\langle P_{k,S}(\cdot,\ell)\hat{p}_{k,S}^{(\xi,I_{k}-\mathbb{B}_{k})}(\cdot,\ell)\psi_{k,Z}^{(u)}(\cdot,\ell)\rangle.
\end{align*}
By iteratively propagating the initial multi-scan GLMB $\boldsymbol{\pi}_{0}=\{(\omega_{0}(\gamma_{0}),p^{(\gamma_{0})}):\gamma_{0}\in\Gamma_{0}\}$,
the weight \eqref{eq:weightpredictionMSapprox} can be explicitly
rewritten as a function of $\gamma_{0:k}$ as follows 
\begin{equation}
\omega_{0:k}(\gamma_{0:k})=\prod_{j=1}^{k}\left[\boldsymbol{1}_{\Gamma_{j}}^{(\gamma_{j-1})}(\gamma_{j})\prod_{i=1}^{P_{j}}\eta_{j}^{(i)}(\gamma_{j}(\ell_{i}))\right]\omega_{0}(\gamma_{0}),\label{eq:weightGibbsMS}
\end{equation}
where $\boldsymbol{1}_{\Gamma_{j}}^{(\gamma_{j-1})}(\gamma_{j})=\boldsymbol{1}_{\Gamma_{j}}(\gamma_{j})\boldsymbol{1}_{\mathcal{F}(\mathbb{B}_{j}\uplus\mathcal{L}(\gamma_{j-1}))}(\mathcal{L}(\gamma_{j})),$
and $\eta_{j}^{(i)}(\gamma_{j}(\ell_{i}))$ implicitly depends on
$\gamma_{0:j-1}(\ell_{i})$.

\subsubsection{Update Approximation}

Following Corollary \ref{cor:updateapproximation},
to truncate the GLMB update approximation \eqref{eq:MSupdateapprox},
we can construct a similar multi-dimensional ranked assignment as
in \eqref{eq:weightGibbsMS} using the extended association map $\gamma_{k}$
to express the updated weight $\hat{w}_{Z}^{(\xi,I_{0:k},\theta_{k})}$
in terms of $\gamma_{0:k}$. Since computing the multi-object posterior
density \eqref{eq:MSupdateapprox} involves the joint attribute densities
of multiple trajectories, it is highly expensive and is intractable
to solve the resulting multi-dimensional ranked assignment problem,
especially with a large number of measurements. A more tractable alternative
is exploiting the prediction approximation strategy to sample the
significant GLMB hypotheses/components in terms of $\gamma_{0:k}$,
then recompute their updated weights via \eqref{eq:updateMSlabeled}
and its approximate multi-scan GLMB posterior densities via \eqref{eq:MSupdateapprox}.

\subsection{Algorithm Description}\label{subsec:MS_gibbs} 

This subsection describes the implementation of multi-scan GLMB approximation
strategies presented in Subsection \ref{subsec:ranked_assignment_problem}.
Due to more accurate approximation results, we only perform the update
approximation strategy. To truncate the multi-scan GLMB approximation,
we use the multi-scan Gibbs sampler presented in \cite{vo2019} for
solving the multi-dimensional ranked assignment \eqref{eq:weightGibbsMS}.
Specifically, a discrete probability distribution $\nu$ is used to
sample extended association maps $\gamma_{0:k}$ such that hypotheses
with higher weights are more likely to be chosen. 

Choosing 
\begin{equation}
\nu(\gamma_{0:k})\triangleq\prod_{j=1}^{k}\nu^{(j)}(\gamma_{j}|\gamma_{0:j-1})\nu_{0}(\gamma_{0}),\label{eq:target_distribution}
\end{equation}
where $\nu_{0}=\omega_{0}$, for $j\in\{1:k\}$:
\[
\nu{}^{(j)}(\gamma_{j}|\gamma_{0:j-1})\propto\boldsymbol{1}_{\Gamma_{j}}^{(\gamma_{j-1})}(\gamma_{j})\prod_{i=1}^{P_{j}}\eta_{j}^{(i)}(\gamma_{j}(\ell_{i})),
\]
and using \eqref{eq:weightGibbsMS}, we obtain $\nu(\gamma_{0:k})\propto\omega_{0:k}(\gamma_{0:k})$.
Following \cite{vo2019,vo2024}, to sample $\gamma_{0:k}$ from \eqref{eq:target_distribution},
two Gibbs sampling techniques, i.e. sampling from the factors and
the multi-scan Gibbs sampler (MS-Gibbs), are implemented. For numerical
implementations, we often denote the multi-scan GLMB posterior by
$G_{0:k}\triangleq(\gamma_{0:k},w_{0:k},p_{0:k})$.

The Gibbs sampler is a computationally efficient MCMC technique in
which the proposed samples are always accepted \cite{geman1984,casella1992}.
However, depending on the initialization, MCMC algorithms may require
a significant duration for the chain to converge. Further, there are
no universal bounds on the burn-in period, nor are there reliable
sampling techniques to determine when the convergence
has occurred in the algorithm (see e.g. \cite{eladlouni2006} and
references therein). 

To implement the update approximation (UA) strategy, it is necessary
to reconstruct the exact posterior from $\{0:k\}$. Given the most
significant hypotheses $\gamma_{0:k}$, the joint updated weights
$w_{0:k}$ and the joint updated densities $p_{0:k}$ are computed
by recursively propagating the posterior Bayes recursion from $\{0:k\}$.
We then apply Corollary \ref{cor:updateapproximation} to obtain
the multi-scan GLMB update approximation $\hat{G}_{0:k}=(\gamma_{0:k},w_{0:k},\hat{p}_{0:k})$. 

Algorithm \ref{alg:MSGibbs-then-UA} shows the implementation steps
of UA strategy following the significant hypotheses $\gamma_{0:k}$
from MS-Gibbs of the entire smoothing period. Since reconstructing
the exact posterior from $\{0:k\}$ is expensive, a cheaper alternative
is computing its approximate density at each time scan as shown in
Algorithm \ref{alg:Joint-MSGibbs-UA}. In this algorithm, the UA strategy
is jointly implemented with MS-Gibbs at each time scan from $\{0:k\}$,
which yields the multi-scan GLMB update approximation at every time
step with less computational cost. For $j\in\{1:k\}$, let $P_{j}=|\mathbb{B}_{j}\uplus\mathcal{L}(\gamma_{j-1})|$
and $M_{j}=|Z_{j}|$, setting $\bar{P}=\max_{j\in\{1:k\}}P_{j}$ and
$\bar{M}=\max_{j\in\{1:k\}}M_{j}$, both Algorithms \ref{alg:MSGibbs-then-UA}
and \ref{alg:Joint-MSGibbs-UA} have complexities of $\mathcal{O}(kT\bar{P}^{2}\bar{M})$
\cite{vo2019}. 

%------- First Algorithm
\begin{algorithm}
	\caption{\textbf{MSGibbs-then-UA}\label{alg:MSGibbs-then-UA} \\
		{\footnotesize$\bullet$ Input: $G_{0:k}=(\gamma_{0:k},w_{0:k},p_{0:k})$,
			$T$ (no. samples);} \\
		{\footnotesize$\bullet$ Output: $[\hat{G}_{0:k}^{(t)}]_{t=1}^{T}$;}}
	\begin{algorithmic}
		\STATE {\footnotesize Initialize $\hat{G}_{0:k}^{(0)}:=G_{0:k}$;}
		\STATE {\footnotesize \textbf{for} $t=1:T$}
		\STATE {\footnotesize$\quad$}{\footnotesize \textbf{for} $j=1:k$}
		\STATE {\footnotesize$\quad$$\quad$$\gamma_{0:j}^{(t)}:=\text{\textrm{MSGibbs}}(\hat{G}_{0:k}^{(t-1)})$;}
		\STATE {\footnotesize$\quad$}{\footnotesize \textbf{end}}
		\STATE {\footnotesize$\quad$Compute $w_{0:k}^{(t)}$, $p_{0:k}^{(t)}$ from $\gamma_{0:k}^{(t)}$ and $G_{0}^{(t)}$ via Bayes recursion;}
		\STATE {\footnotesize$\quad$Compute the approximation $\hat{p}_{0:k}^{(t)}$ of $p_{0:k}^{(t)}$ via Corollary \ref{cor:updateapproximation};}
		\STATE {\footnotesize$\quad$$\hat{G}_{0:k}^{(t)}:=(\gamma_{0:k}^{(t)},w_{0:k}^{(t)},\hat{p}_{0:k}^{(t)})$;}
		\STATE {\footnotesize\textbf{end}}
	\end{algorithmic}
\end{algorithm}
%---------Second Algorithm----
\begin{algorithm}
	\caption{\textbf{JointMSGibbs-UA}\label{alg:Joint-MSGibbs-UA}\\
		{\footnotesize$\bullet$ Input: $G_{0:k}=(\gamma_{0:k},w_{0:k},p_{0:k})$, $T$ (no. samples);} \\
		{\footnotesize$\bullet$ Output: $[\hat{G}_{0:k}^{(t)}]_{t=1}^{T}$;}}
	\begin{algorithmic}
		\STATE {\footnotesize Initialize $\hat{G}_{0:k}^{(0)}:=G_{0:k}$;}
		\STATE {\footnotesize\textbf{for} $t=1:T$}
		\STATE {\footnotesize$\quad$}{\footnotesize\textbf{for} $j=1:k$}
		\STATE {\footnotesize$\quad$$\quad$$\gamma_{0:j}^{(t)}:=\text{\textrm{MSGibbs}}(\hat{G}_{0:k}^{(t-1)})$;}
		\STATE {\footnotesize$\quad$$\quad$Compute $w_{0:j}^{(t)}$, $p_{0:j}^{(t)}$ from $\gamma_{0:j}^{(t)}$ and $\hat{G}_{0:j-1}^{(t)}$ via Bayes recursion;}
		\STATE {\footnotesize$\quad$$\quad$Compute the approximation $\hat{p}_{0:j}^{(t)}$ of $p_{0:j}^{(t)}$ via Corollary \ref{cor:updateapproximation};}
		\STATE {\footnotesize$\quad$$\quad$$\hat{G}_{0:j}^{(t)}:=(\gamma_{0:j}^{(t)},w_{0:j}^{(t)},\hat{p}_{0:j}^{(t)})$;}
		\STATE {\footnotesize$\quad$ \textbf{end}}
		\STATE {\footnotesize \textbf{end}}
	\end{algorithmic}
\end{algorithm}

\begin{rem}
In practice, the choice between Algorithm \ref{alg:MSGibbs-then-UA}
and Algorithm \ref{alg:Joint-MSGibbs-UA} is governed by trade-offs
between estimation accuracy and computational efficiency. Specifically,
Algorithm \ref{alg:Joint-MSGibbs-UA} is suitable for scenarios with
high object density or limited computational resources. While computationally
cheaper due to the approximation of the joint density
at each time scan, it may incur a slightly higher approximation error
compared to Algorithm \ref{alg:MSGibbs-then-UA}. Conversely, Algorithm
\ref{alg:MSGibbs-then-UA} is well-suited for scenarios where accurate
estimation results are prioritized and computational resources are
unconstrained. However, this comes at a greater computational expense
due to the evaluation over multiple time scans.
\end{rem}
Overall, to compute the approximate multi-scan GLMB posterior, we
adopt the smoothing-while-filtering algorithm(Algorithm
4, \cite{vo2019}) to our implementation. This algorithm is therefore
named the \textit{smoothing-while-filtering approximation (SFA)}.
Algorithm \ref{alg:SFA} shows the implementation of SFA where the
multi-scan GLMB approximation $\hat{G}_{0:k}=(\gamma_{0:k},w_{0:k},\hat{p}_{0:k})$
is recursively propagated using factor sampling to generate $\gamma_{k}$
on-the-fly and multi-scan Gibbs sampling to generate the new significant
$\gamma_{0:k}$. Due to parallelization of the for loops, Algorithm
\ref{alg:SFA} has complexity of $\mathcal{O}(kT\bar{P}^{2}\bar{M})$
\cite{vo2019}. For simplicity, the SFA algorithm using Algorithm
\ref{alg:MSGibbs-then-UA} is named \textit{SFA-then-UA}, and similarly
\textit{JointSFA-UA} is the SFA algorithm using Algorithm \ref{alg:Joint-MSGibbs-UA}.

\begin{algorithm}
	\caption{\textbf{SFA} \label{alg:SFA}\\
		{\footnotesize$\bullet$ Input: $[G_{0:k-1}^{(h)}]_{h=1}^{H_{k-1}}$,
			$[R^{(h)}]_{h=1}^{H_{k-1}}$, $T$;} \\
		{\footnotesize$\bullet$ Output: $[\hat{G}_{0:k}^{(h)}]_{h=1}^{H_{k}}$;}}
	\begin{algorithmic}
		\STATE {\footnotesize\textbf{for} $h=1:H_{k-1}$}
		\STATE {\footnotesize$\quad$$[G_{0:k}^{(h,r)}]_{r=1}^{\bar{R}^{(h)}}:=\textrm{Unique}(\textrm{FactorSampling}(G_{0:k-1}^{(h)},R^{(h)}))$;}
		\STATE {\footnotesize\textbf{end}}
		\STATE {\footnotesize Keep $\bar{H}_{k}$ best $[G_{0:k}^{(h)}]_{h=1}^{\bar{H}_{k}}$;}
		\STATE {\footnotesize\textbf{for} $h=1:\bar{H}_{k}$}
		\STATE {\footnotesize$\quad$Compute $[\hat{G}_{0:k}^{(h,t)}]_{t=1}^{T}$ via Algorithm \ref{alg:MSGibbs-then-UA} or Algorithm \ref{alg:Joint-MSGibbs-UA};}
		\STATE {\footnotesize\textbf{end}}
		\STATE {\footnotesize$[\hat{G}_{0:k}^{(h)}]_{h=1}^{\hat{H}_{k}}:=\textrm{Unique}([\hat{G}_{0:k}^{(h,t)}]_{h,t=(1,1)}^{(\bar{H}_{k},T)})$; }
		\STATE {\footnotesize Keep $H_{k}$ best $[\hat{G}_{0:k}^{(h)}]_{h=1}^{H_{k}}$;}
		\STATE {\footnotesize Normalize weights $[w_{0:k}^{(h)}]_{h=1}^{H_{k}}$;}
	\end{algorithmic}
\end{algorithm}

\subsection{Windowing Approximation Technique}\label{subsec:MSW_implementation}

This subsection describes the implementation of multi-scan multi-object
approximation in Subsection \ref{subsec:SW_approx}. The multi-object
posterior density can be approximated by multi-densities on a set
of finite sub-windows using Proposition \ref{prop:smoothing_window_approx}.
This allows us to model the multi-object posterior with lower computational
cost due to shorter windows.

Given a multi-object posterior density $G_{0:k}$ on the window $\{0:k\}$
with the initial condition $G_{0}$, we divide the smoothing window
$\{1:k\}$ into $N_{O}$ number of overlapping sub-windows, such that
$\{1:k\}=\cup_{i=1}^{N_{O}}\{j^{(i)}:k^{(i)}\}$, where $\{j^{(i)}:k^{(i)}\}$
is a smoothing window from $j^{(i)}$ to $k^{(i)}$, and $\{j^{(i)}:k^{(i)}\}\cap\{j^{(i-1)}:k^{(i-1)}\}\neq\emptyset$.
For each $i\in\{1:N_{O}\}$, let $m^{(i)}\in\{j^{(i)}:k^{(i)}\}$
be the marginalization time for multi-scan multi-object approximation,
then each sub-window can be decomposed as $\{j^{(i)}:k^{(i)}\}=\{j^{(i)}:m^{(i)}\}\uplus\{m^{(i)}+1:k^{(i)}\}$.
We assume the current window overlaps with the previous window if
$\{j^{(i)}:m^{(i)}\}=\{m^{(i-1)}+1:k^{(i-1)}\}$. For numerical implementations,
we often denote the sub-window $\{j^{(i)}:k^{(i)}\}$ by three values
$(j^{(i)},m^{(i)},k^{(i)})$, for all $i\in\{1:N_{O}\}$. Algorithm
\ref{alg:MSW} shows the implementation steps of overlapping smoothing
window (SW) technique, where Proposition \ref{prop:smoothing_window_approx}
is applied to compute multi-scan multi-object approximations over
sub-windows.

\begin{algorithm}
	\caption{\textbf{Overlapping-SW} \label{alg:MSW}\\
		{\footnotesize$\bullet$ Input: $N_{O}$ (no. overlapped sub-windows);} \\
		{\footnotesize$\bullet$ Output: SW approximation $[G_{0:k}^{(h)}]_{h=1}^{H}$;}}
	\begin{algorithmic}
		\STATE {\footnotesize Compute $\{(j^{(i)},m^{(i)},k^{(i)})\}_{i\in\{1:N_{O}\}}$;}
		\STATE {\footnotesize Initialize $G_{0}^{(1)}:=(\gamma_{0}^{(1)},w_{0}^{(1)},p_{0}^{(1)})$;}
		\STATE {\footnotesize \textbf{for} $i=1:N_{O}$ }
		\STATE {\footnotesize$\quad$1. Initialize $[G_{j^{(i)}:m^{(i)}}^{(h)}]_{h=1}^{H}:=[G_{m^{(i-1)}+1:k^{(i-1)}}^{(h)}]_{h=1}^{H}$;}
		\STATE {\footnotesize$\quad$2. Compute $[G_{j^{(i)}:k^{(i)}}^{(h)}]_{h=1}^{H}$ via SFA-then-UA or JointSFA-UA;}
		\STATE {\footnotesize$\quad$3. Compute $[G_{j^{(i)}:m^{(i)}}^{(h)}]_{h=1}^{H}$, $[G_{m^{(i)}+1:k^{(i)}}^{(h)}]_{h=1}^{H}$ via Proposition \ref{prop:smoothing_window_approx};}
		%\STATE {\footnotesize$\quad$4. Compute $[G_{0:m^{(i)}}^{(h)}]_{h=1}^{H}$ from $[G_{0:m^{(i-1)}}^{(h)}]_{h=1}^{H}$ and $[G_{j^{(i)}:m^{(i)}}^{(h)}]_{h=1}^{H}$;}
		\STATE {\footnotesize$\quad$4. Compute $[G_{0:m^{(i)}}^{(h)}]_{h=1}^{H}$ from $[G_{0:m^{(i-1)}}^{(h)}]_{h=1}^{H}$ and $[G_{j^{(i)}:m^{(i)}}^{(h)}]_{h=1}^{H}$ via Proposition \ref{prop:smoothing_window_approx};}
		\STATE {\footnotesize\textbf{end }}
	\end{algorithmic}
\end{algorithm}

The computational complexity of Algorithm \ref{alg:MSW} is dominated
by the computation of the multi-object posterior $[G_{j^{(i)}:k^{(i)}}^{(h)}]_{h=1}^{H}$
over sub-windows in Step 2. Specifically, for any overlapping sub-window
indexed by $i\in\{1:N_{O}\}$, let $L^{(i)}=k^{(i)}-m^{(i)}$ denote
its length, while $\bar{P}_{i}=\max_{j\in\{m^{(i)}+1:k^{(i)}\}}P_{j}$
and $\bar{M}_{i}=\max_{j\in\{m^{(i)}+1:k^{(i)}\}}M_{j}$ represent
the maximum number of objects and measurements. The asymptotic complexity
of Step 2 is $\mathcal{O}(L^{(i)}T\bar{P}_{i}^{2}\bar{M}_{i})$ since
it uses Algorithm \ref{alg:SFA} to compute $[G_{m^{(i)}+1:k^{(i)}}^{(h)}]_{h=1}^{H}$.
Thus, the complexity of Algorithm \ref{alg:MSW} is determined by
the worst-case scenario computed across all local windows, i.e. 
\[
\mathcal{O}\left(\max_{i\in\{1:N_{O}\}}L^{(i)}T\bar{P}_{i}^{2}\bar{M}_{i}\right),
\]
where the maximization is taken over all overlapping sub-windows $i\in\{1:N_{O}\}$.
Note that the window parameters are bounded and independent of the
total duration $k$, hence offering a significant reduction compared
to Algorithm \ref{alg:SFA}, which entails a complexity of $\mathcal{O}(kT\bar{P}^{2}\bar{M})$
due to consideration of the entire duration $\{0:k\}$.  
\begin{rem}
When multi-object posterior inference becomes computationally prohibitive,
the non-overlapping SW technique (the special case with marginalization
time $m^{(i)}=k^{(i)}$ and the initial condition $G_{j^{(i)}}=G_{m^{(i-1)}}$,
$i\in\{1:N_{O}\}$) may be needed to further reduce the computational
cost.
\end{rem}

\section{Numerical Studies}\label{sec:numerical_study}

This section evaluates the proposed multi-scan GLMB approximation
across three numerical studies with the repulsive social force model
\cite{helbing1995,johansson2007,pellegrini2009}, where the correlated
object motion is described by a system of nonlinear differential equations
(DEs). In Subsection \ref{subsec:repulsive_social_force}, we present
the multi-object transition density that incorporates the repulsive
social force model. Using the simulated tracking scenario introduced
in Section \ref{sec:Introduction}, Subsection \ref{subsec:standard_meas}
verifies the proposed method's ability to capture social force interactions
under a standard multi-object observation model wherein detections
are conditionally independent. To further demonstrate the effectiveness
of our method in handling more complex interactions, Subsection \ref{subsec:merged_meas}
presents a more challenging simulated scenario featuring the social
force model coupled with merged measurements \cite{beard2015}, in
which object detections are also correlated. Subsection \ref{subsec:pedestrian_tracking}
validates the real-world applicability of the proposed approximation
using the BIWI Walking Pedestrian dataset from ETHZ CVL in \cite{pellegrini2009}.

\subsection{Multi-Object Transition with Social Force Model}\label{subsec:repulsive_social_force}

This subsection specifies the labeled RFS transition presented in
Subsection \ref{subsec:interacting_recursion} using the repulsive
social force model \cite{helbing1995}. A default discrete sample
period $\Delta T=1s$ is used. An object with label $\ell$ has a
4D kinematic state $x_{k}^{(\ell)}=\left[p_{k,x}^{(\ell)},v_{k,x}^{(\ell)},p_{k,y}^{(\ell)},v_{k,y}^{(\ell)}\right]^{T}$
of 2D position and velocity. The survival probability is set to $P_{k,S}=0.99$
for each object. Given the multi-object state $\boldsymbol{X}_{k-1}$
at time $k-1$, the surviving multi-object density contained in \eqref{eq:Markov_transition}
can be written as a product of each single-object transition density:
\begin{equation}
\boldsymbol{f}_{k,S}(\boldsymbol{S}_{k}|\boldsymbol{X}_{k-1})=\prod_{x_{k}^{(\ell)}\in\boldsymbol{S}_{k}}f_{k,S}(x_{k}^{(\ell)}|x_{k-1}^{(\ell)};\boldsymbol{X}_{k-1}),\label{eq:surviving_multi_object_density}
\end{equation}
where for each surviving object $x_{k}^{(\ell)}\in\boldsymbol{S}_{k}$,
its transition density is given by $f_{k,S}(x_{k}^{(\ell)}|x_{k-1}^{(\ell)};\boldsymbol{X}_{k-1})=\mathcal{N}(x_{k}^{(\ell)};\hat{m}_{k}(x_{k-1}^{(\ell)},\boldsymbol{X}_{k-1}),I_{2})$.
Given the initial condition $x_{k-1}^{(\ell)}$ at time $t=k-1$,
the mean $\hat{m}_{k}(x_{k-1}^{(\ell)},\boldsymbol{X}_{k-1})$ is
obtained by solving the system of nonlinear DEs in \eqref{eq:DEs}
that describes the repulsive force on each object using standard numerical
approximation techniques. 

For notational convenience, we denote $\vec{p}^{(\ell)}=[p_{x}^{(\ell)},p_{y}^{(\ell)}]$
and $\vec{v}^{(\ell)}=[v_{x}^{(\ell)},v_{y}^{(\ell)}]$. The kinematic
state of object $\ell$ is governed by the following system of nonlinear
DEs \cite{helbing1995} 
\begin{equation}
\begin{cases}
\dot{\vec{p}}^{(\ell)}(t) & =\vec{v}^{(\ell)}(t),\\
\dot{\vec{v}}^{(\ell)}(t) & =\vec{F}^{(\ell)}(t),
\end{cases}\label{eq:DEs}
\end{equation}
where the repulsive social force on object $\ell$ is given by
\begin{align*}
\vec{F}^{(\ell)}(t) & =\sum_{\ell^{'}\in\mathcal{L}(\boldsymbol{X}_{k-1}),\,\ell'\neq\ell}\vec{F}^{(\ell,\ell')}\left(\vec{p}^{(\ell,\ell^{'})}\right),\\
 & =\frac{V}{\alpha^{2}}\sum_{\ell'\neq\ell}\left(\vec{p}^{(\ell,\ell^{'})}-\Delta T\vec{v}^{(\ell^{'})}\right)\exp\left(-\frac{b(\vec{p}^{(\ell,\ell^{'})})}{2\alpha^{2}}\right);
\end{align*}
$\vec{p}^{(\ell,\ell^{'})}=\vec{p}^{(\ell)}-\vec{p}^{(\ell^{'})}$;
$b(\vec{p}^{(\ell,\ell^{'})})=\lVert\vec{p}^{(\ell,\ell^{'})}-\Delta T\vec{v}^{(\ell^{'})}\rVert^{2}$;
$V=550m^{2}s^{-2}$; and $\alpha=30m$.

Due to the inherent non-linearity of the social force model, the prediction
step cannot be performed analytically. For each component in the multi-object
posterior, the surviving multi-object density in \eqref{eq:surviving_multi_object_density}
is approximated by a joint Gaussian using the Unscented Kalman Filter
(UKF). The sigma points are propagated through the system of nonlinear
DEs in \eqref{eq:DEs}, allowing the predicted mean and covariance
to explicitly capture object interactions. A static LMB birth model
with 4 components is used, with birth parameters are $\{P_{k,B}(\ell_{i}),p_{k,B}(x_{k},\ell_{i})\}_{i=1}^{4}$,
where: $\ell_{i}=(k,i);\,P_{k,B}(\ell_{i})=0.01$; and $p_{k,B}(x_{k},\ell_{i})=\mathcal{N}(x_{k};m_{B}^{(\ell_{i})},P_{B})$
with 
\begin{align*}
m_{B}^{(\ell_{1})}=[-500,10,0,10]^{T}, & \quad m_{B}^{(\ell_{2})}=[500,-10,0,10]^{T},\\
m_{B}^{(\ell_{3})}=[-750,15,0,10]^{T}, & \quad m_{B}^{(\ell_{4})}=[750,-15,0,10]^{T},\\
P_{B}=\textrm{diag}([10,10,10,10]^{T})^{2}.
\end{align*}

\subsection{Validation with Social Force and Standard Measurements}\label{subsec:standard_meas}

This subsection evaluates performance of the proposed multi-scan
GLMB approximation in tracking interacting objects under the repulsive
social force model of Subsection \ref{subsec:repulsive_social_force}.
Direct comparisons with the standard multi-object dynamic model are
shown to highlight the improvement in tracking performance due to
modeling of multi-object interactions and the effectiveness of our
approximation. 

The ground truth for this scenario shown in Figure \ref{fig:intro_fig}
involves four objects over 100 time steps. At time $k=1$, four objects
are born at widely separated positions and interact according to the
social force model of Subsection \ref{subsec:repulsive_social_force}
when they enter a circular region of radius $50m$; outside this region,
their movements are independent. The repulsive interactions between
the objects prevent them from crossing paths and result in evasive
turns in the middle of the scenario. For simplicity all objects remain
present for the entire duration.

Measurements are generated according to the standard multi-object
measurement model \eqref{eq:likelihood_func}. Each measurement is
a noisy 2D bearing-range detection $z=[\theta,r]^{T}$ generated by
a static sensor at the origin $[0,0]^{T}$ with likelihood $g_{k}(z|x_{k}^{(\ell)})=\mathcal{N}(z;h(x_{k}^{(\ell)}),R)$;
\[
h(x_{k}^{(\ell)})=\left[\arctan\left(\frac{p_{k,x}^{(\ell)}}{p_{k,y}^{(\ell)}}\right),\sqrt{\left(p_{k,x}^{(\ell)}\right)^{2}+\left(p_{k,y}^{(\ell)}\right)^{2}}\right]^{T};
\]
and $R=\mathrm{diag}([\sigma_{\theta}^{2},\sigma_{r}^{2}])^{T}$ with
$\sigma_{\theta}=\frac{2\pi}{180}\mathrm{rad}$ and $\sigma_{r}=10m$.
The detection probability is $P_{k,D}=0.7$, and the Poisson clutter
rate is $10$ per scan. The measurement updates are conducted using
the UKF.

\begin{figure*}
	\centering
	\subfloat[]{\includegraphics[bb=0.5cm 1.5cm 14.5cm 10cm,clip,scale=0.38]{figures/est_overlap_corr_0.7_10_10ms_4}\label{fig:estimates_bearing_range_standard_MS}}
	\hfil 
	\subfloat[]{\includegraphics[bb=0.5cm 1.5cm 14.5cm 10cm,clip,scale=0.38]{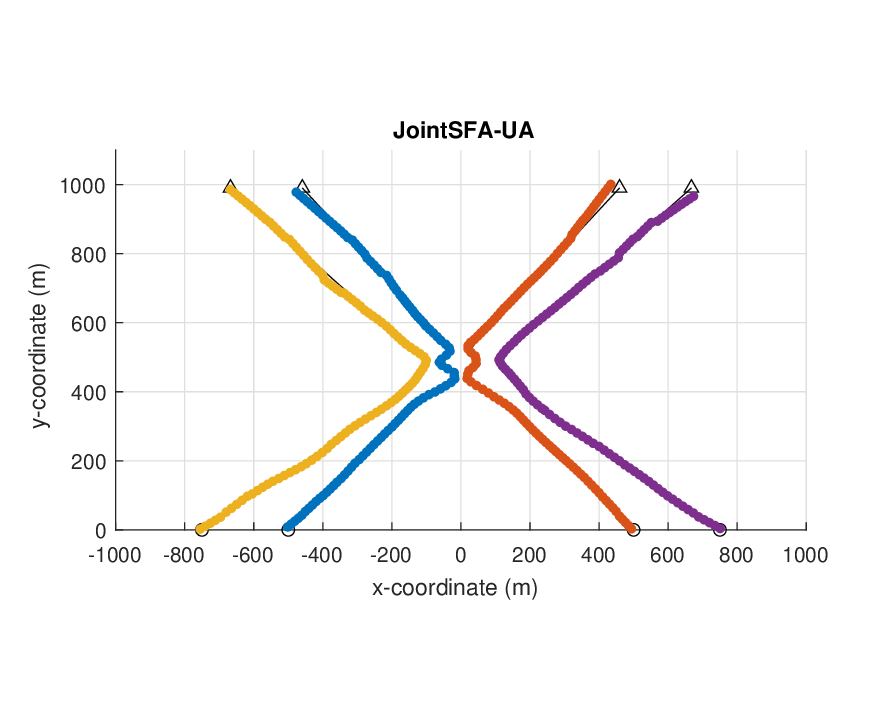}\label{fig:estimates_bearing_range_joint_SFA_UA}}
	\hfil
	\subfloat[]{\includegraphics[bb=0.5cm 1.5cm 14.5cm 10cm,clip,scale=0.38]{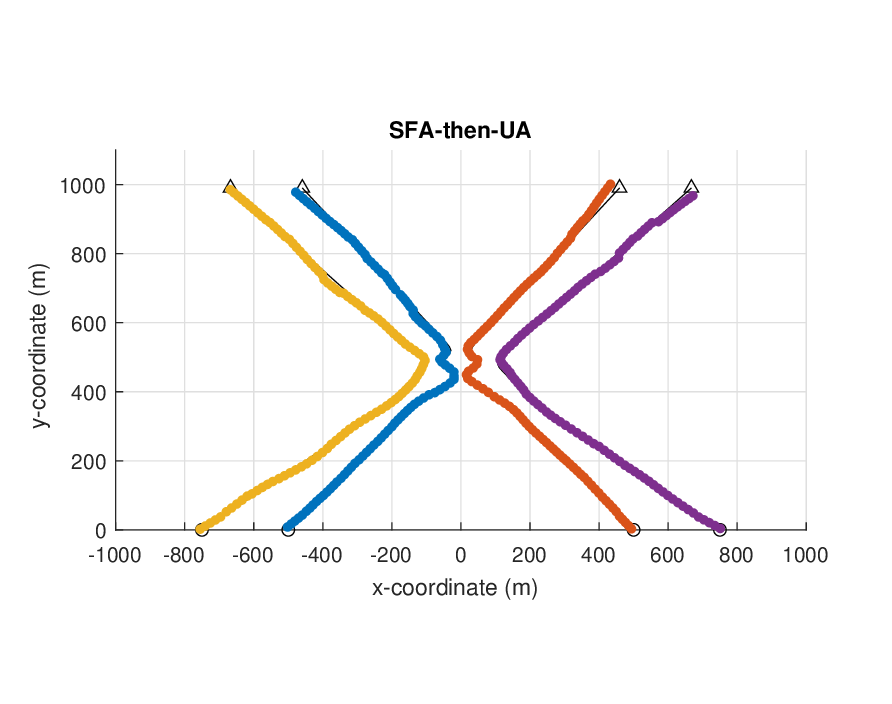}\label{fig:estimates_bearing_range_SFA_then_UA}}
	\hfil
	\includegraphics[bb=4.8cm 0.75cm 10.5cm 10.5cm,clip,scale=0.3]{figures/legends_4}
	
	\caption{(a) Estimates from the standard multi-scan GLMB suffer from label switchings and object crossings around intersection circles. (b) and (c) Estimates from JointSFA-UA and SFA-then-UA where the social force is accounted for, show no crossings as well as good track continuity.}\label{fig:estimates_bearing_range}
\end{figure*}

\begin{figure}
	\centering
	\includegraphics[bb=0.5cm 5.75cm 14cm 11.5cm,clip,scale=0.55]{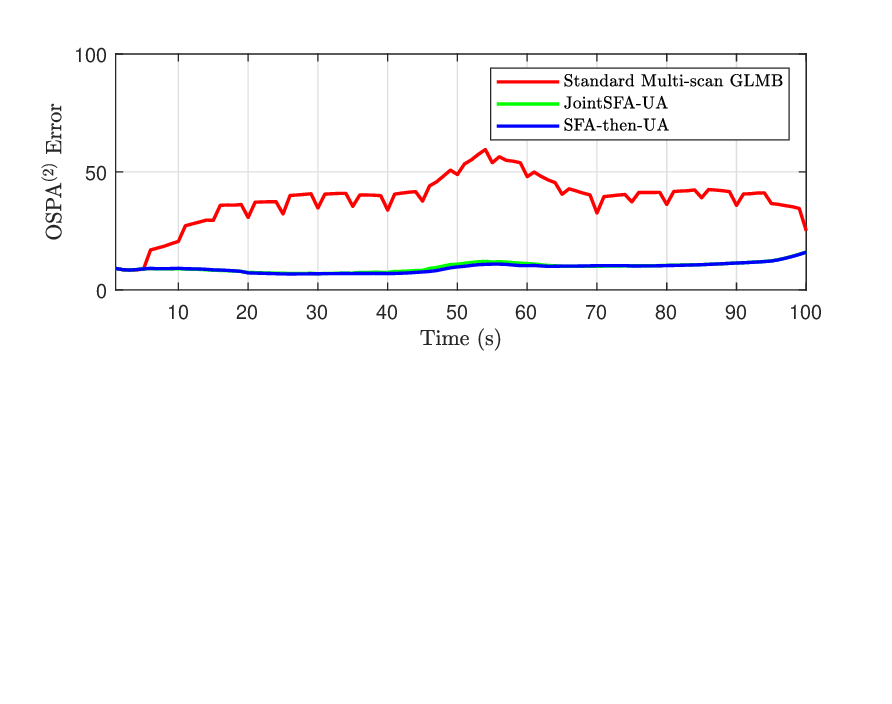}
	
	\caption{$\textrm{OSPA}^{(2)}$ (cutoff $c=100m$; order $p=1$; over 10-scan window length) errors \cite{schuhmacher2008,beard2020} of final estimates from the standard multi-scan GLMB, JointSFA-UA, and SFA-then-UA over 100 Monte Carlo runs.}\label{fig:error_bearing_range}
\end{figure}

All multi-object posterior densities are computed with a maximum
of $1000$ components, $T=10$ iterations of the Markov chain, and
component weight threshold of $10^{-5}$. Each Markov chain is initialized
by sampling from the factors \cite{vo2019}. The windowing approximation
technique is implemented on a set of 10-scan sub-windows with an overlap
of length 5. In this experiment we only implement the update approximation
strategies. Figures \ref{fig:estimates_bearing_range} and \ref{fig:error_bearing_range},
respectively, show the multi-object trajectory estimates and the $\textrm{OSPA}^{(2)}$
errors \cite{schuhmacher2008,beard2020} (over 100 Monte Carlo runs)
obtained from the standard multi-scan GLMB, JointSFA-UA, and SFA-then-UA
filters. 

The standard multi-scan GLMB filter assumes independent object motion,
whereas JointSFA-UA and SFA-then-UA incorporate the repulsive social
force model. Figure \ref{fig:estimates_bearing_range} highlights
the differences in multi-object trajectory estimates produced by these
filters. From the same set of measurements, Figure \ref{fig:estimates_bearing_range_standard_MS}
shows the failure of standard multi-scan GLMB filter when the objects
come into close proximity with erroneous object crossings during the
time interval $\{45:55\}$, due to the independent motion assumption.
In contrast, Figures \ref{fig:estimates_bearing_range_joint_SFA_UA}
and \ref{fig:estimates_bearing_range_SFA_then_UA} show that both
JointSFA-UA and SFA-then-UA successfully track all objects by accounting
for the repulsive social forces and thus successfully mitigate label
switching.

Figure \ref{fig:error_bearing_range} shows the $\textrm{OSPA}^{(2)}$
errors \cite{schuhmacher2008,beard2020} of the final estimates over
100 Monte Carlo runs. Due to low detectability, there is a significant
performance difference between both the JointSFA-UA/SFA-then-UA from
the standard multi-scan GLMB filter, as indicated by their substantially
lower $\textrm{OSPA}^{(2)}$ errors, albeit at the cost of increased
computations. These results are consistent with the trajectory comparisons
in Figure \ref{fig:estimates_bearing_range}, supporting the conclusion
that the proposed functional approximations are effective for capturing
object interactions with the repulsive social force model. 

It is worth noting that SFA-then-UA performs slightly better than
JointSFA-UA when objects exhibit strong interactions during the interval
$\{45:55\}$, due to the different ways in which the two methods handle
posterior density approximation. JointSFA-UA approximates the joint
posterior densities using a multi-scan GLMB at every time scan, whereas
SFA-then-UA defers the multi-scan GLMB marginalization until the end
of the smoothing period. 

\subsection{Validation with Social Force and Merged Measurements}\label{subsec:merged_meas}

This subsection presents a similar but progressively more challenging
scenario in which merged measurements introduce additional inter-object
correlations. Specifically, the multi-object likelihood function is
not of the standard form \eqref{eq:likelihood_func} but takes a more
general form to accommodate measurement merging \cite{beard2015}
by considering all partitions of the set of objects. Each element
or group of a given partition generates at most one merged measurement.

Given a set of objects $\boldsymbol{X}_{k}$ at time $k$, a \textit{partition}
$\mathcal{U}(\boldsymbol{X}_{k})$ of $\boldsymbol{X}_{k}$ is a disjoint
collection of subsets of $\boldsymbol{X}_{k}$, whose union is equal
to $\boldsymbol{X}_{k}$. Denote $\mathcal{P}(\boldsymbol{X}_{k})$
the set of all partitions of $\boldsymbol{X}_{k}$, the multi-object
likelihood function is given by \cite{beard2015}
\[
\tilde{\boldsymbol{g}}_{k}(Z_{k}|\boldsymbol{X}_{k})\propto\sum_{\begin{smallmatrix}{\scriptsize {\footnotesize \mathcal{U}(\boldsymbol{X}_{k})\in\mathcal{P}(\boldsymbol{X}_{k}),}}\\
{\scriptsize {\footnotesize \tilde{\theta}_{k}\in\tilde{\Theta}_{k}(\mathcal{U}(\mathcal{L}(\boldsymbol{X}_{k})))}}
\end{smallmatrix}}\left[\tilde{\psi}_{k,Z_{k}}^{(\tilde{\theta}_{k}\circ\mathcal{L}(\cdot))}\right]^{\mathcal{U}(\boldsymbol{X}_{k})},
\]
where $\tilde{\Theta}_{k}(\mathcal{U}(\mathcal{L}(\boldsymbol{X}_{k})))$
is the set of positive $1-1$ mappings $\tilde{\theta}_{k}:\mathcal{U}(\mathcal{L}(\boldsymbol{X}_{k}))\rightarrow\{0:|Z_{k}|\}$,
i.e. $\tilde{\theta}_{k}(I)=\tilde{\theta}_{k}(J)>0$ implies $I=J$.
The likelihood $\tilde{\psi}_{k,Z_{k}}^{(j)}(\boldsymbol{Y}_{k})$
is a generalization of standard measurement likelihood whose arguments
are groups of objects, i.e.
\begin{equation}
\tilde{\psi}_{k,Z_{k}}^{(j)}(\boldsymbol{Y}_{k})=\begin{cases}
\frac{\tilde{P}_{k,D}(\boldsymbol{Y}_{k})\tilde{g}_{k}(z_{j}|\boldsymbol{Y}_{k})}{\kappa_{k}(z_{j})}, & j\in\{1:|Z_{k}|\},\\
\tilde{Q}_{k,D}(\boldsymbol{Y}_{k}), & j=0,
\end{cases}\label{eq:merged_meas_likelihood}
\end{equation}
where: $\tilde{P}_{k,D}(\boldsymbol{Y}_{k})$ is the detection probability
for group of objects $\boldsymbol{Y}_{k}$; $\tilde{Q}_{k,D}(\boldsymbol{Y}_{k})=1-\tilde{P}_{k,D}(\boldsymbol{Y}_{k})$
is the misdetection probability for $\boldsymbol{Y}_{k}$; and $\tilde{g}_{k}(z_{j}|\boldsymbol{Y}_{k})$
is the likelihood of measurement $z_{j}$, with $j=\tilde{\theta}_{k}(\mathcal{L}(\boldsymbol{Y}_{k}))$
given $\boldsymbol{Y}_{k}$.

Detections are noisy bearings in $\mathbb{Z}=[0;2\pi]$. The merged
measurements are simulated based on the detection-level model in \cite{beard2015},
where the group of objects in \eqref{eq:merged_meas_likelihood} corresponds
to the objects falling within a geometric cell. Specifically, let
$C=\{c_{1},c_{2},...,c_{N}\}$ be the set of $N$ disjoint cells.
Let $\boldsymbol{T}_{k}^{(i)}$ be the set of objects in $\boldsymbol{X}_{k}$
whose detections are located in cell $c_{i}$ at time $k$, defined
as
\[
\boldsymbol{T}_{k}^{(i)}=\{x_{k}^{(\ell)}\in\boldsymbol{X}_{k}:\:h(x_{k}^{(\ell)},s_{k})\in c_{i}\},
\]
where the measurement function is
\begin{equation}
h\left(x_{k}^{(\ell)},s_{k}\right)=\arctan\left(\frac{p_{k,x}^{(\ell)}-s_{k,x}}{p_{k,y}^{(\ell)}-s_{k,y}}\right),\label{eq:measurement_func}
\end{equation}
and $s_{k}=\left[s_{k,x},s_{k,y}\right]^{T}$ is a moving sensor at
time $k$ given by \allowdisplaybreaks
\begin{align*}
s_{k,x} & =\begin{cases}
1000\cos\left(\mathrm{floor}\left(\frac{k}{2}\right)\frac{\pi}{4}\right), & \text{if mode\ensuremath{(k,2)=1,}}\\
800\cos\left((k-1)\frac{\pi}{8}\right), & \text{otherwise,}
\end{cases}\\
s_{k,y} & =\begin{cases}
1000\sin\left(\mathrm{floor}\left(\frac{k}{2}\right)\frac{\pi}{4}\right), & \text{if mode\ensuremath{(k,2)=1,}}\\
800\sin\left((k-1)\frac{\pi}{8}\right), & \text{otherwise.}
\end{cases}
\end{align*}

Let $w_{k}$ be the measurement noise vector, then cell $c_{i}$
generates the following measurement 
\[
z_{k}^{(i)}=\begin{cases}
\frac{1}{|\boldsymbol{T}_{k}^{(i)}|}\underset{x_{k}^{(\ell)}\in\boldsymbol{T}_{k}^{(i)}}{\sum}h(x_{k}^{(\ell)},s_{k})+w_{k}, & |\boldsymbol{T}_{k}^{(i)}|>0,\\
\emptyset, & |\boldsymbol{T}_{k}^{(i)}|=0,
\end{cases}
\]
with probability $P_{k,D}(\boldsymbol{T}_{k}^{(i)})$, and $z_{k}^{(i)}=\emptyset$
with probability $Q_{k,D}(\boldsymbol{T}_{k}^{(i)})=1-P_{k,D}(\boldsymbol{T}_{k}^{(i)})$.
The set of generated measurements or detections at time $k$ is $D_{k}=\uplus_{i=1}^{N}z_{k}^{(i)}$.
Thus, the overall measurement set is $Z_{k}=D_{k}\uplus K_{k}$, where
$K_{k}$ is the false alarms. 

\begin{figure}
\centering
\includegraphics[bb=0.75cm 5.75cm 14.5cm 11.5cm,clip,scale=0.55]{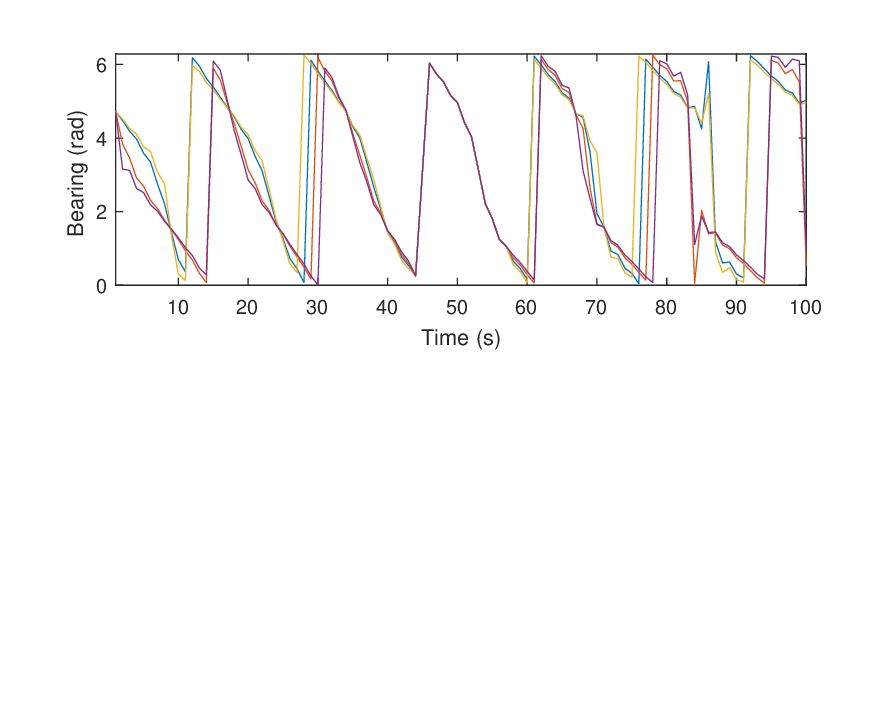}

\caption{True bearing measurements of 4 objects
in 100 time scans, where merged measurements occur from $k=45:55$.}\label{fig:true_bearings}
\end{figure}

In this experiment, the likelihood function conditioned on the group
of objects $\boldsymbol{T}_{k}^{(i)}$ is a Gaussian distribution
$\tilde{g}_{k}(z_{k}^{(i)}|\boldsymbol{T}_{k}^{(i)})=\mathcal{N}(z_{k}^{(i)};m_{k}(\boldsymbol{T}_{k}^{(i)}),R)$,
where $m_{k}(\boldsymbol{T}_{k}^{(i)})$ is the measurement update,
and $R=\sigma_{\theta}^{2}$ with $\sigma_{\theta}=\frac{\pi}{180}\mathrm{rad}$
is the measurement noise covariance. The detection probability is
$P_{k,D}=0.7$, the Poisson clutter is $0.3$ per scan. Bearing cell
widths are fixed at 2 degrees (approximately 0.035 rad). The measurement
updates are conducted using the UKF. Merged measurements are more
likely in the middle of the scenario when the 4 objects are close
together, i.e. from time $k=45:55$ (as depicted in Figure \ref{fig:true_bearings}). 

\begin{figure*}
	\centering
	\subfloat[]{\includegraphics[bb=1cm 0bp 13cm 338bp,clip,scale=0.4]{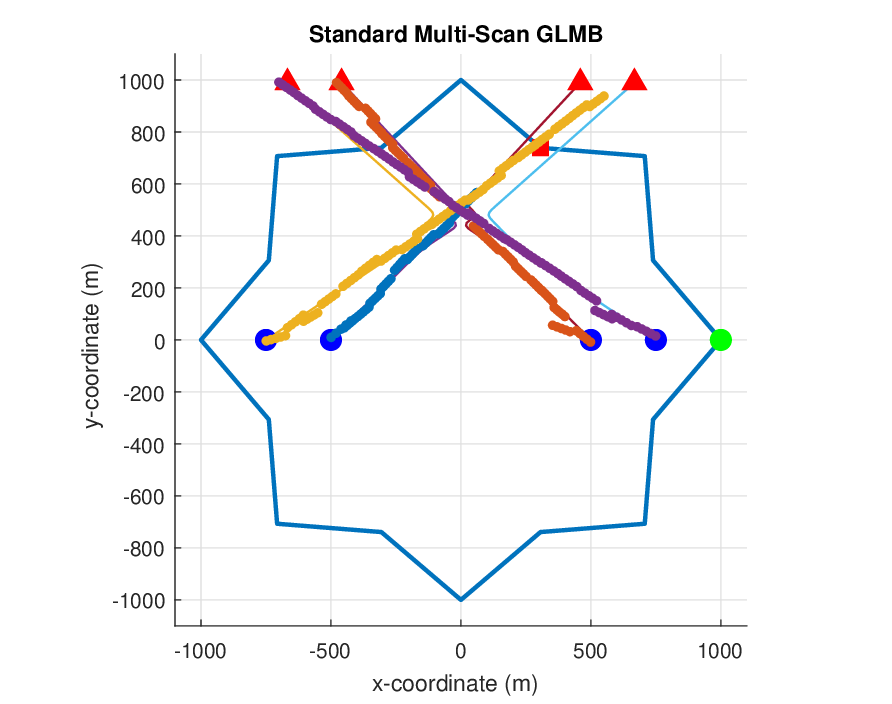}\label{fig:estimates_merged_standard_MS}}
	\hfil
	\subfloat[]{\includegraphics[bb=1cm 0bp 13cm 338bp,clip,scale=0.4]{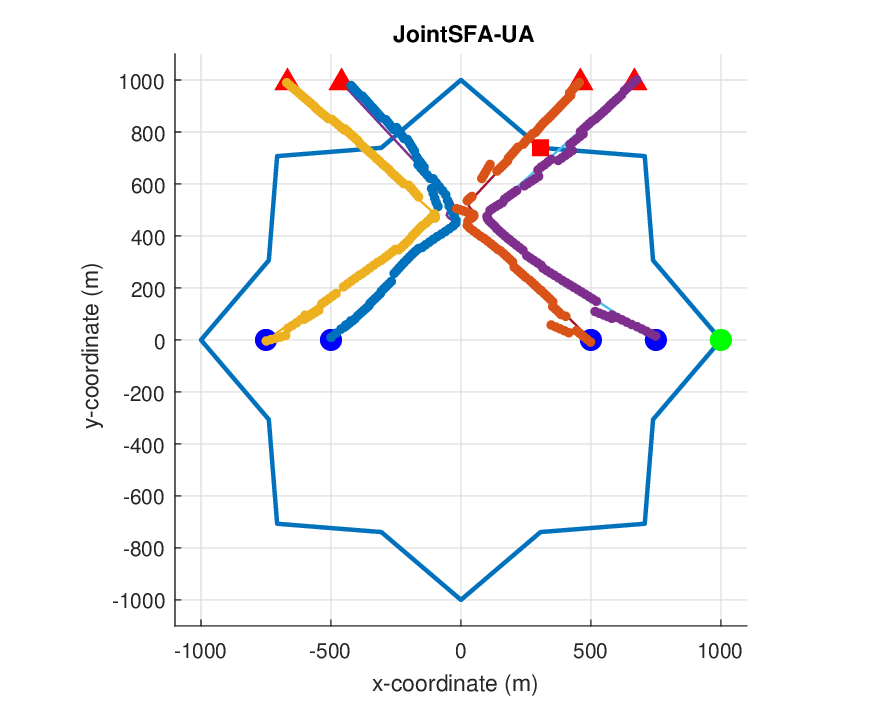}\label{fig:estimates_merged_jointSFA_UA}}
	\hfil
	\subfloat[]{\includegraphics[bb=1cm 0bp 13cm 338bp,clip,scale=0.4]{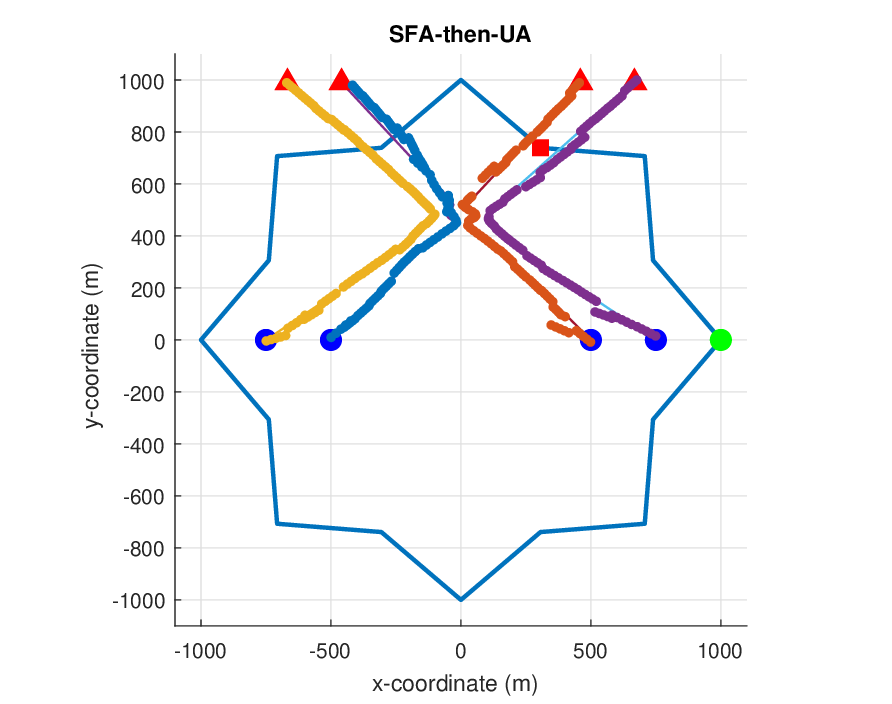}\label{fig:estimates_merged_SFA_then_UA}}
	\hfil
	\includegraphics[bb=3.5cm 0.5cm 10cm 8cm,clip,scale=0.4]{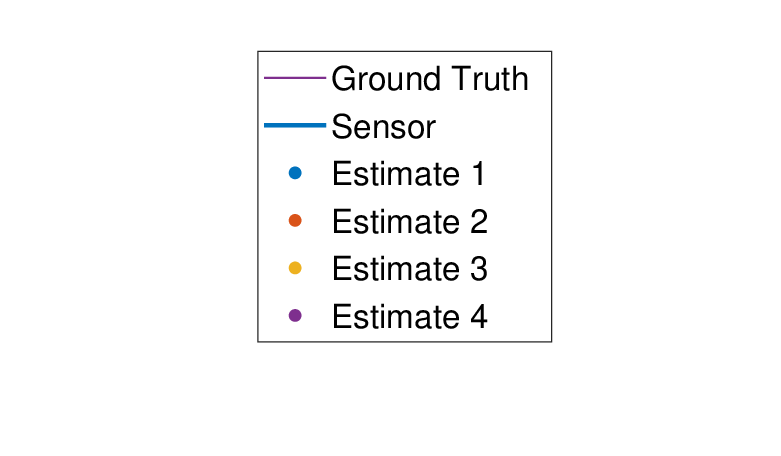}
	
	\caption{(a) Estimates from the standard multi-scan GLMB filter suffer from object crossings and track droppings around intersection circles. (b) and (c) Estimates from JointSFA-UA and SFA-then-UA where the social force is accounted for, show no crossing as well as good track continuity.}\label{fig:estimates_merged}
\end{figure*}

\begin{figure}
	\centering
	\includegraphics[bb=0.5cm 6cm 14.5cm 11.5cm,clip,scale=0.55]{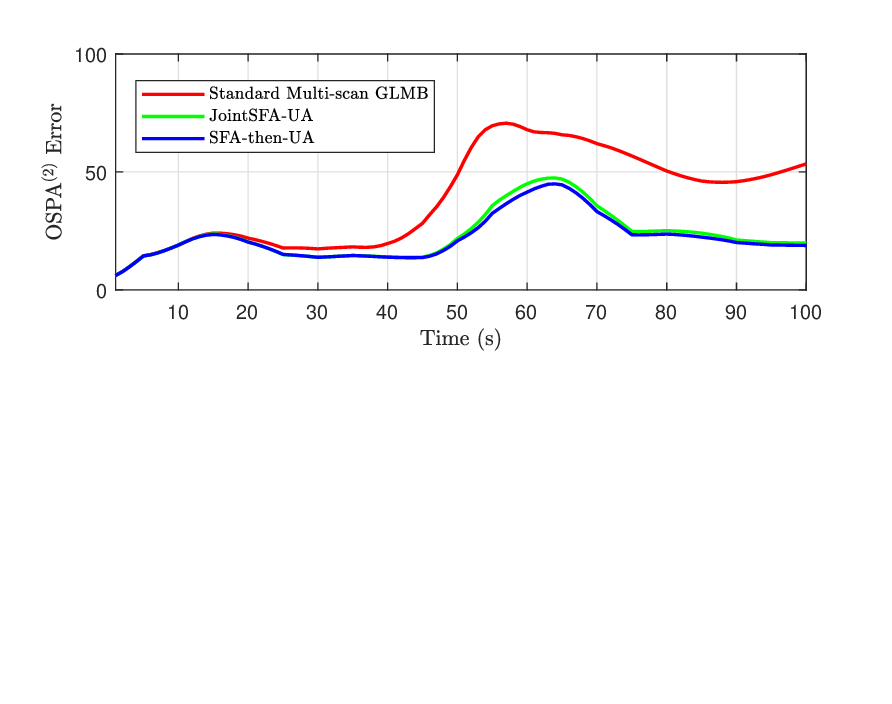}
	
	\caption{$\textrm{OSPA}^{(2)}$ (cutoff $c=100m$ and order $p=1$ over 10-scan window length) errors \cite{schuhmacher2008,beard2020} of final estimates from the standard multi-scan GLMB, JointSFA-UA, and SFA-then-UA over 100 Monte Carlo runs. }\label{fig:error_merged}
\end{figure}

All multi-object posterior densities are computed with a maximum of
$10000$ components, $T=10$ iterations of the Markov chain, and component
weight threshold of $10^{-5}$. Each Markov chain is initialized by
sampling from the factors \cite{vo2019}. In this scenario, the LMB
birth model is active only at time $k=1$. Due to social force with
merged measurements, the windowing approximation is implemented on
a set of 5-scan non-overlapping sub-windows to reduce computation.
Similar to the first experiment, the update approximation strategies
are implemented. Figures \ref{fig:estimates_merged} and \ref{fig:error_merged},
respectively, show the multi-object trajectory estimates and the $\textrm{OSPA}^{(2)}$
errors \cite{schuhmacher2008,beard2020} (over 100 Monte Carlo runs)
obtained from the standard multi-scan GLMB, JointSFA-UA, and SFA-then-UA
filters. 

Trajectory estimates in Figure \ref{fig:estimates_merged} demonstrate
the effectiveness of the proposed functional approximation in capturing
object and measurement interactions. Similar to the results in Subsection
\ref{subsec:standard_meas}, Figure \ref{fig:estimates_merged_standard_MS}
shows that the standard multi-scan GLMB filter exhibits dropped and
switched tracks in the middle of the scenario when objects come into
close proximity, due to the assumption of independent object motion
and detection. In contrast, Figures \ref{fig:estimates_merged_jointSFA_UA}
and \ref{fig:estimates_merged_SFA_then_UA} show that both JointSFA-UA
and SFA-then-UA successfully achieve accurate multi-object tracking
without dropped or switched tracks, as a direct result of incorporating
the social force model and merged measurement model. These results
are verified by the $\textrm{OSPA}^{(2)}$ (over a 10-scan moving
window) errors \cite{schuhmacher2008,beard2020} in Figure \ref{fig:error_merged},
demonstrating that modeling object correlations from the interactive
motion and measurement coupling yields significantly improved trajectory
estimates compared to the standard multi-scan GLMB filter which assumes
conditionally independent dynamics and detections.

Further, compared to the tracking results in Subsection \ref{subsec:standard_meas},
these errors show a noticeable gap of tracking performance between
SFA-then-UA and JointSFA-UA after objects exhibit strong interactions
and merged measurements occur, since SFA-then-UA handles the posterior
density approximation more better than JointSFA-UA. Thus, performing
SFA-then-UA achieves more accurate tracking results than JointSFA-UA
in non-standard multi-object estimation. 

\subsection{Pedestrian Tracking with Social Force Model}\label{subsec:pedestrian_tracking}

In this subsection, we validate the proposed multi-object estimation
solution with the social force model on the BIWI Walking Pedestrian
dataset from ETHZ CVL in \cite{pellegrini2009}. To benchmark the
tracking performance against the standard motion model, we utilize
the ETHZ annotated ground truth\footnote{Available at https://vision.ee.ethz.ch/datasets.html (folder $\texttt{eth\_seq}$)}.
This annotated dataset is recorded at 2.5 fps, i.e 0.4s between frames,
and includes manual annotations for time frames, pedestrian IDs, full
3D positions and velocities. Note that positions and velocities have
been projected from image pixels onto 3D plane via a homography matrix
by the dataset creators. For simplicity, we only focus on the 2D
plane. All positions and velocities are expressed in meters (m) and
meters per seconds (m/s), respectively.

Due to the extensive size of the full ETHZ dataset, a subset is selected
to demonstrate the effectiveness of social force modeling, comprising
6 pedestrians over 65 time frames. This scenario is chosen to satisfy
two key criteria: (i) the presence of strong interactions among pedestrians,
and (ii) sufficiently long trajectories, with each pedestrian observed
for at least 10 frames. Consequently, the subset captures meaningful
interactions while providing adequate track lengths for these effects
to be reflected in the posterior. For clarity, the original dataset
IDs (53, 54, 58, 59, 60, 62) are re-indexed as Pedestrians 1 through
6, respectively. The number of active pedestrians varies due to appearances
(births) and disappearances (deaths). Specifically, births occur at
time frames $1,9,25,41$ (with respectively $2,1,2,1$ births) for
Pedestrians 1-2, 3, 4-5, and 6, respectively. Deaths occur at time
frames $11,26,47$ (with respectively $2,1,2$ deaths), marking the
exits of Pedestrians 1-2, 3, and 4-5, respectively. Further, the scenario
features two distinct social groups A (Pedestrians 1 and 2) and B
(Pedestrians 4 and 5), in which members maintain safe interpersonal
distances and exhibit cohesive motion without trajectory crossings.

\subsubsection*{Repulsive Social Force Model for Pedestrian Dynamics}

We follow the repulsive social force model presented in \cite{pellegrini2009}
for tracking interacting pedestrians. Specifically, a discrete sample
period of $\Delta T=0.4s$ is used. The survival probability is set
to $P_{k,S}=0.99$ for each pedestrian. For each surviving pedestrian
$x_{k}^{(\ell)}\in\boldsymbol{S}_{k}$, the transition density in
\eqref{eq:surviving_multi_object_density} is given by $f_{k,S}(x_{k}^{(\ell)}|x_{k-1}^{(\ell)};\boldsymbol{X}_{k-1})=\mathcal{N}(x_{k}^{(\ell)};\hat{m}_{k}(x_{k-1}^{(\ell)},\boldsymbol{X}_{k-1}),Q)$,
where $Q=\mathrm{diag}([\sigma_{v}^{2},\sigma_{v}^{2}]^{T})$ with
$\sigma_{v}=0.5m/s^{2}$ is the process noise covariance. Given the
initial condition $x_{k-1}^{(\ell)}$ at time $k-1$, the mean $\hat{m}_{k}(x_{k-1}^{(\ell)},\boldsymbol{X}_{k-1})$
is obtained by solving the system of nonlinear DEs in \eqref{eq:DEs_peds}
that describes the repulsive social force on each pedestrian using
standard numerical approximation techniques.

The kinematic state of pedestrian $\ell$ is characterized by the
following system of nonlinear DEs 
\begin{equation}
\begin{cases}
\dot{\vec{p}}^{(\ell)}(t) & =\vec{v}^{(\ell)}(t),\\
\dot{\vec{v}}^{(\ell)}(t) & =\vec{F}_{\text{weighted}}^{(\ell)}(t),
\end{cases}\label{eq:DEs_peds}
\end{equation}
where the repulsive social force on pedestrian $\ell$ is the sum
of weighted effects from other pedestrians $\ell'$, i.e.
\[
\vec{F}_{\text{weighted}}^{(\ell)}(t)=\sum_{\ell'\in\mathcal{L}(\boldsymbol{X}_{k-1}),\ell'\neq\ell}w_{\ell}(\ell')\vec{F}^{(\ell,\ell')}\left(\vec{p}^{(\ell,\ell')}\right);
\]
the weight $w_{\ell}(\ell')$ is assigned to each pedestrian $\ell$
based on the current distance $d$ and angular displacement $\varphi$
from $\ell'$ 
\begin{align*}
w_{\ell}(\ell') & =w_{\ell}^{d}(\ell')w_{\ell}^{\varphi}(\ell');\\
w_{\ell}^{d}(\ell') & =\exp\left(-\frac{\lVert\vec{p}^{(\ell')}-\vec{p}^{(\ell)}\rVert^{2}}{2\sigma_{w}^{2}}\right);\\
w_{\ell}^{\varphi}(\ell') & =\left(\frac{1+\cos\varphi}{2}\right)^{\beta};
\end{align*}
$\sigma_{w}=2.088m$; $\beta=1.462$; the repulsive effect $\vec{F}^{(\ell,\ell')}\left(\vec{p}^{(\ell,\ell')}\right)$
is given by Subsection \ref{subsec:repulsive_social_force} with $V=1m^{2}s^{-2}$
and $\alpha=0.361m$. For simplicity, the field-of-view is restricted
to $\pm\frac{\pi}{2}$rad (i.e. $w_{\ell}^{\varphi}(\ell')=0$ if
$\lvert\varphi\rvert>\frac{\pi}{2}$).

The initial states of the 6 pedestrians are modeled as a static LMB
birth with 6 components, with birth parameters are $\{P_{k,B}(\ell_{i}),p_{k,B}(x_{k},\ell_{i})\}_{i=1}^{6}$,
where $\ell_{i}=(k,i)$; $P_{k,B}(\ell_{i})=0.01;$ and $p_{k,B}(x_{k},\ell_{i})=\mathcal{N}(x_{k},m_{B}^{(\ell_{i})},P_{B})$,
$m_{B}^{(\ell_{i})}$ is the initial state of pedestrian $\ell_{i}$
extracted from the ETHZ annotated ground truth, and 
\[
P_{B}=\mathrm{diag}([0.1,0.1,0.1,0.1]^{T})^{2}.
\]
Each measurement is a 2D position vector $z=[z_{x},z_{y}]^{T}$ on
the region $[-5,15]m\times[-5,15]m$ with likelihood $g(z|x_{k}^{(\ell)})=\mathcal{N}(z;Hx_{k}^{(\ell)},R)$;
\[
H=\left[\begin{array}{cccc}
1 & 0 & 0 & 0\\
0 & 0 & 1 & 0
\end{array}\right];R=\mathrm{diag}([\sigma_{\epsilon}^{2},\sigma_{\epsilon}^{2}]^{T});
\]
and $\sigma_{\epsilon}=0.25m$ is the measurement noise standard deviation.
The detection probability is set to $P_{k,D}=0.99$ and the average
Poisson clutter rate is $\lambda_{c}=0.5$ per scan. 

\subsubsection*{Performance Evaluation}

\begin{figure*}
	\centering
	\subfloat[]{\includegraphics[bb=1cm 2cm 14cm 9cm,clip,scale=0.42]{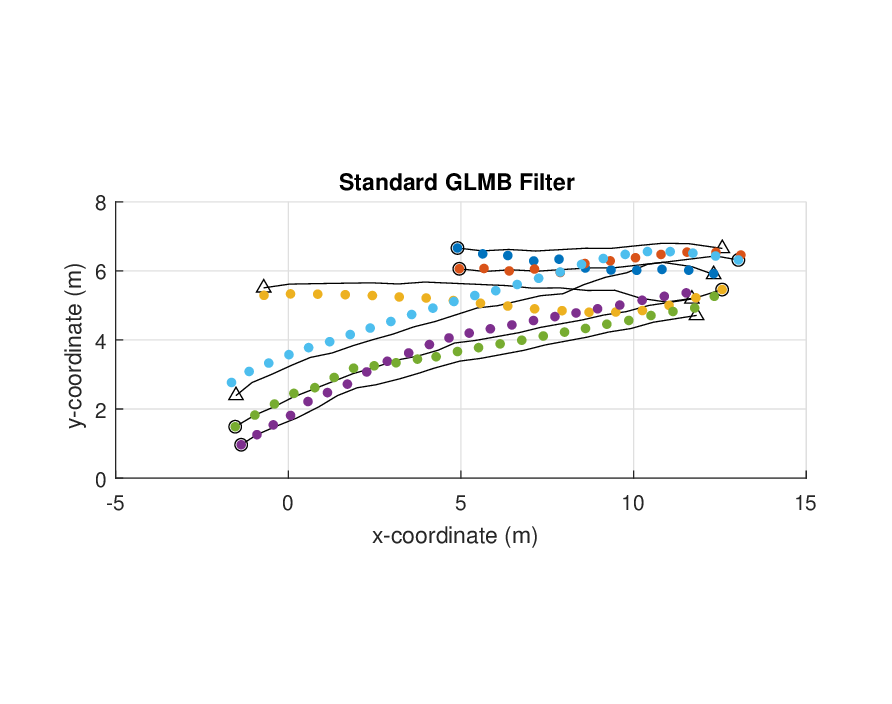}\label{fig:estimates_peds_standard_GLMB_filter}}
	\hfil
	\subfloat[]{\includegraphics[bb=1cm 2cm 14cm 9cm,clip,scale=0.42]{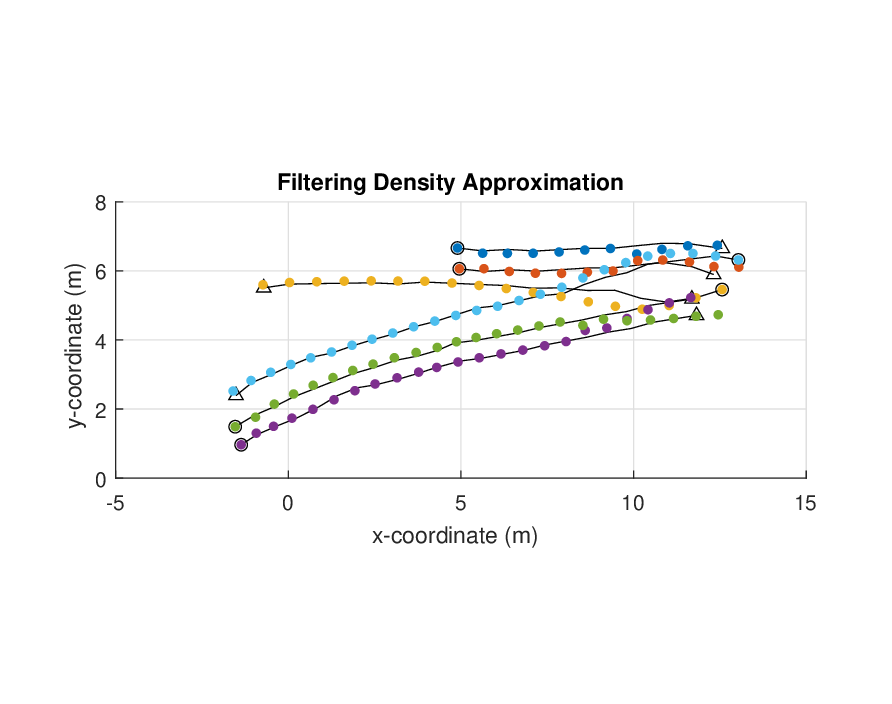}\label{fig:estimates_peds_filtering_approx}}
	\hfil 
	\subfloat[]{\includegraphics[bb=1cm 2cm 14cm 9cm,clip,scale=0.42]{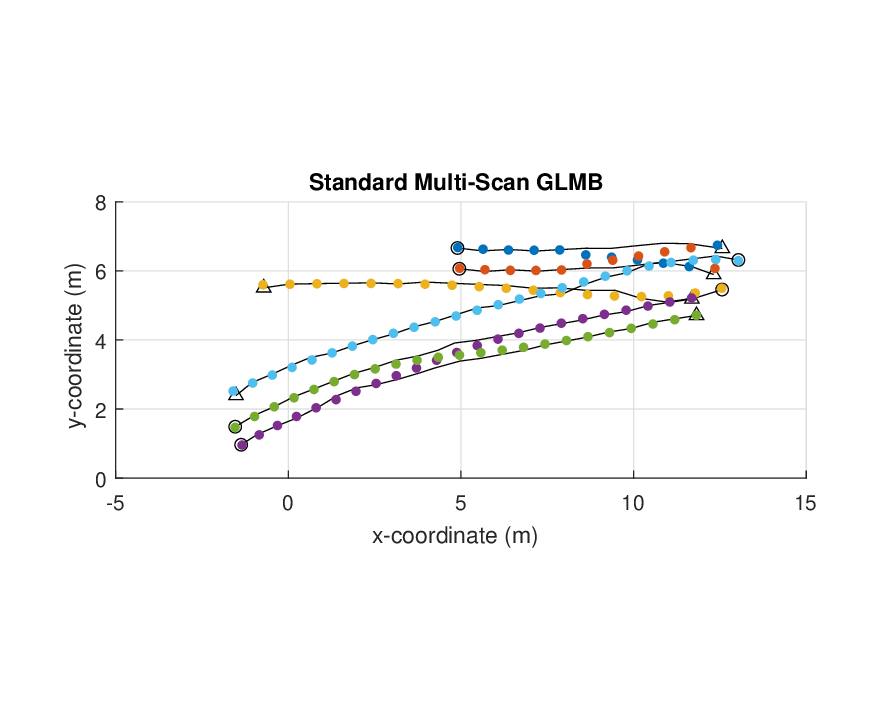}\label{fig:estimates_peds_standard_MS_GLMB}}
	\hfil
	\subfloat[]{\includegraphics[bb=1cm 2cm 14cm 9cm,clip,scale=0.42]{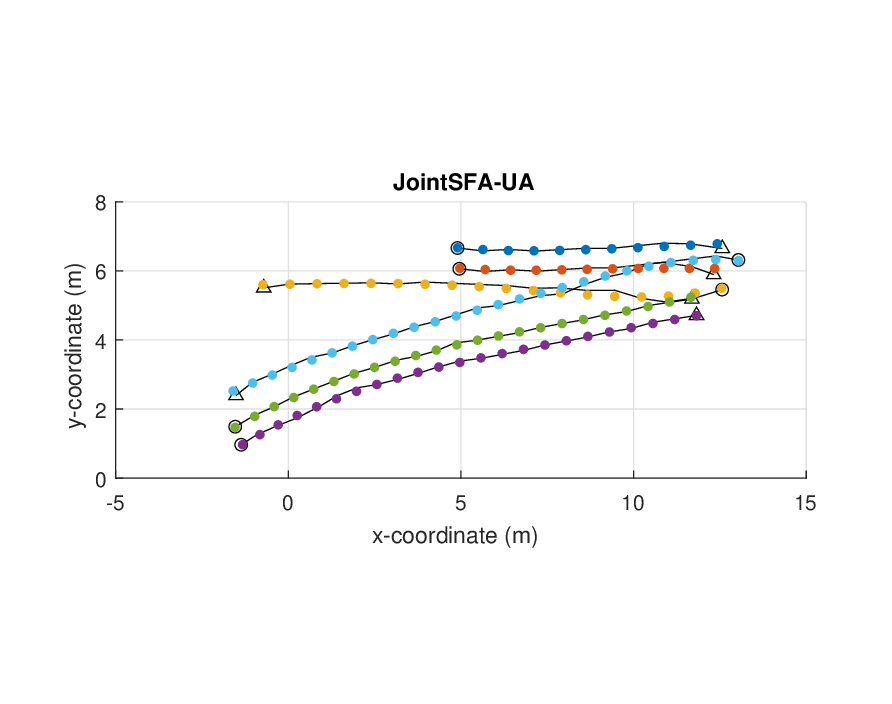}\label{fig:estimates_peds_jointSFA_UA}}
	\hfil
	\subfloat[]{\includegraphics[bb=1cm 2cm 14cm 9cm,clip,scale=0.42]{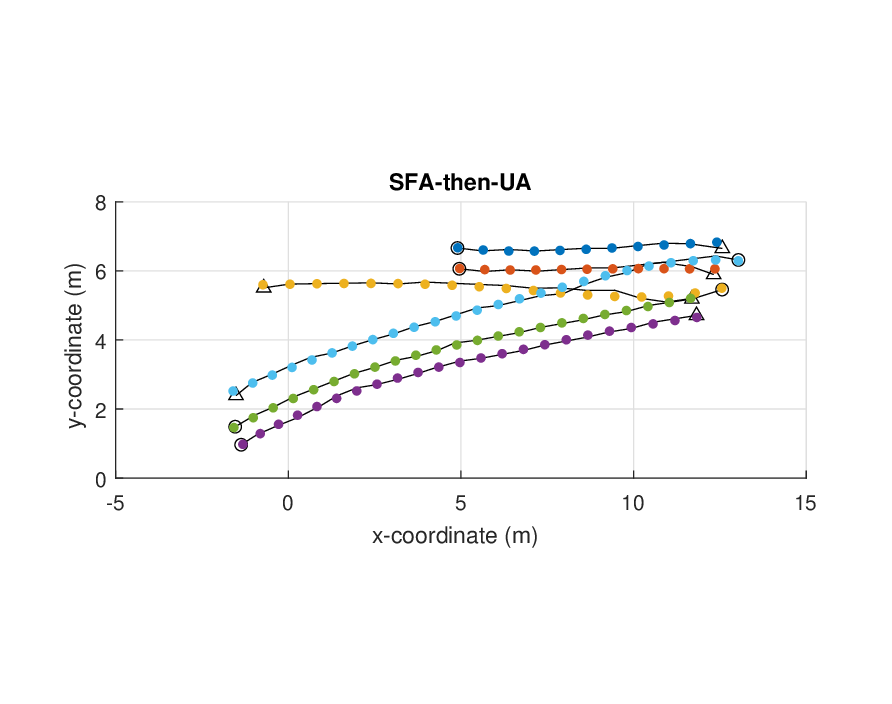}\label{fig:estimates_peds_SFA_then_UA}}
	\hfil 
	\includegraphics[bb=1cm 0.3cm 13cm 6cm,clip,scale=0.46]{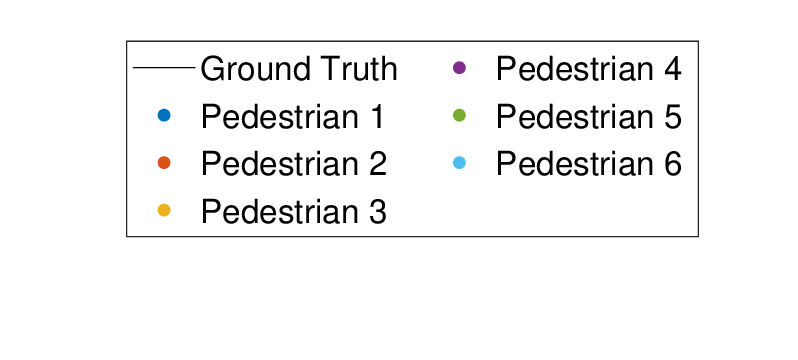}
	
	\caption{Trajectory estimates for six pedestrians, highlighting social force interactions in Group A (blue, orange) and Group B (purple, green). (a) Standard GLMB filter: significant fragmentation and erroneous trajectory crossings due to the independent motion assumption. (b) Filtering density approximation with social force model: no trajectory crossings in Group A, but the lack of full posterior information leads to fragmentations and collisions in Group B. (c) Standard multi-scan GLMB: reduces fragmentation, but the independent motion causes erroneous trajectory crossings. (d) JointSFA-UA and (e) SFA-then-UA: integrating the social force model with the posterior approximation assumption prevents all trajectory crossings and yields continuous trajectories consistent with ground-truth social distances for all pedestrians.}\label{fig:trajectory_ped_tracking}
\end{figure*}

\begin{figure}
	\centering
	\includegraphics[bb=0.5cm 5cm 14cm 12cm,clip,scale=0.55]{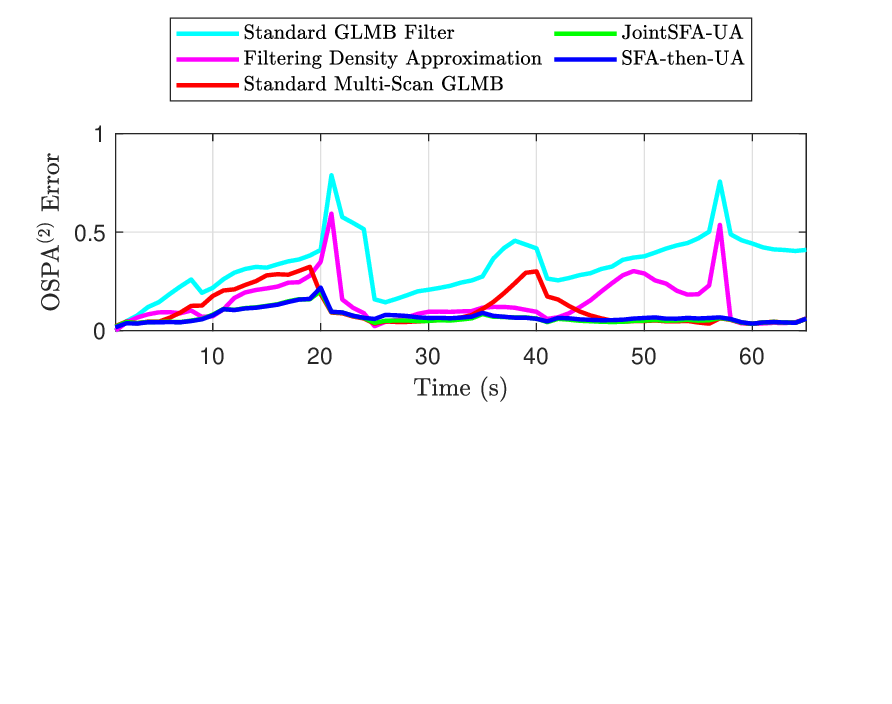}
	
	\caption{$\text{OSPA}^{(2)}$ (cutoff $c=1m$;order $p=1$; over 10-scan window length) errors \cite{schuhmacher2008,beard2020} of the estimates from the standard GLMB filter, filtering density approximation, standard multi-scan GLMB, JointSFA-UA, and SFA-then-UA filters.}\label{fig:ospa_real_data}
\end{figure}

All multi-object posterior densities are computed with a maximum of
$1000$ components, $T=10$ iterations of the Markov chain, and component
weight threshold of $10^{-5}$. Each Markov chain is initialized by
sampling from the factors \cite{vo2019}. The windowing approximation
technique is conducted on a set of $10-$scan sub-windows with an
overlap of length 5. In this experiment we only implement the update
approximation strategies. Figure \ref{fig:trajectory_ped_tracking}
shows the trajectory estimates from five tracking methods: standard
GLMB filter, filtering density approximation, standard multi-scan
GLMB, and the proposed JointSFA-UA and SFA-then-UA. To evaluate multi-object
tracking errors, Figure \ref{fig:ospa_real_data} demonstrates the
$\text{OSPA}^{(2)}$ estimation errors \cite{schuhmacher2008,beard2020}
exhibited by these filters, while average runtime and processing speed
are given in Table \ref{tab:runtime}. Additional performance evaluation
with the CLEAR-MOT metric \cite{bernardin2008} and further details
regarding parameter sensitivity are provided in Subsections \ref{subsec:clear_mot}
and \ref{subsec:sensitivity_para} of the Supplementary Materials.

\begin{table}
	\centering
	\caption{Average runtime (ms) and processing speed (fps) reported from the standard GLMB filter, filtering density approximation, standard multi-scan GLMB, JointSFA-UA, and SFA-then-UA filters over the period of 65 time frames.}\label{tab:runtime}
	{\scriptsize{}%
	\begin{tabular}{|c||c|c||c|c||c|}
	\hline 
	\multicolumn{2}{|c|}{{\scriptsize\textbf{Methods}}} & \multicolumn{2}{c|}{{\scriptsize Runtime (ms)}} & \multicolumn{2}{c|}{{\scriptsize Speed (fps)}}\tabularnewline
	\hline 
	\multicolumn{2}{|c|}{{\scriptsize\textbf{Standard GLMB Filter}}} & \multicolumn{2}{c|}{{\scriptsize 7.5}} & \multicolumn{2}{c|}{{\scriptsize 133.3}}\tabularnewline
	\hline 
	\multicolumn{2}{|c|}{{\scriptsize\textbf{Filtering Density Approximation}}} & \multicolumn{2}{c|}{{\scriptsize 44.9}} & \multicolumn{2}{c|}{{\scriptsize 22.3}}\tabularnewline
	\hline 
	\multicolumn{2}{|c|}{{\scriptsize\textbf{Standard Multi-Scan GLMB}}} & \multicolumn{2}{c|}{{\scriptsize 126.0}} & \multicolumn{2}{c|}{{\scriptsize 7.9}}\tabularnewline
	\hline 
	\multicolumn{2}{|c|}{{\scriptsize\textbf{JointSFA-UA}}} & \multicolumn{2}{c|}{{\scriptsize 318.0}} & \multicolumn{2}{c|}{{\scriptsize 3.1}}\tabularnewline
	\hline 
	\multicolumn{2}{|c|}{{\scriptsize\textbf{SFA-then-UA}}} & \multicolumn{2}{c|}{{\scriptsize 336.0}} & \multicolumn{2}{c|}{{\scriptsize 3.0}}\tabularnewline
	\hline 
	\end{tabular}}{\scriptsize\par}
\end{table}

Figure \ref{fig:trajectory_ped_tracking} highlights the effectiveness
of integrating interaction modeling with posterior approximation by
comparing trajectory estimates across five filters. Under the standard
multi-object measurement model, filters that assume independent motion--specifically
the standard GLMB filter and standard multi-scan GLMB (Figures \ref{fig:estimates_peds_standard_GLMB_filter}
and \ref{fig:estimates_peds_standard_MS_GLMB}, respectively)--fail
to capture social force interactions, resulting in erroneous collisions
in Groups A and B, whereas the latter reduces track fragmentation
via full posterior computation. The filtering density approximation
(Figure \ref{fig:estimates_peds_filtering_approx}), while accurately
predicting trajectory estimates in Group A by incorporating the social
force model, still causes fragmentations and collisions in Group B
due to its lack of full posterior computation. In contrast, Figures
\ref{fig:estimates_peds_jointSFA_UA} and \ref{fig:estimates_peds_SFA_then_UA}
demonstrate that the proposed JointSFA-UA and SFA-then-UA methods
successfully address these limitations. By integrating the repulsive
social force model with the proposed posterior approximation, both
methods preserve trajectory continuity and prevent erroneous collisions,
effectively eliminating identity switching in close proximity for
all pedestrians. Figure \ref{fig:ospa_real_data} verifies these results
via the $\text{OSPA}^{(2)}$ (over a 10-scan moving window) errors
\cite{schuhmacher2008,beard2020}, confirming that the proposed interaction
modeling and posterior approximation outperform both standard independent-motion
filter/smoother and the scan-by-scan filtering density approximation.

Table \ref{tab:runtime} presents the average runtime and processing
speed for the five tracking methods. From the same set of measurements,
the standard GLMB filter is the fastest (7.5 ms, 133.3 fps), followed
by the filtering density approximation (44.9 ms, 22.3 fps) and the
standard multi-scan GLMB (126.0 ms, 7.9 fps). While the proposed JointSFA-UA
(318.0 ms, 3.1 fps) and SFA-then-UA (336.0 ms, 3.0 fps) are the slowest
due to higher computational costs for interaction modeling and full
posterior computation, trajectory estimates in Figure \ref{fig:trajectory_ped_tracking}
prove that this trade-off is necessary for accurate tracking results.
The processing speeds are obtained using an unoptimized MATLAB implementation,
which suggests that the proposed methods are potentially amenable
for real-world data. 

\subsubsection*{Monte Carlo Evaluation}

To demonstrate robustness of our proposed methods, 300 Monte Carlo
simulations are conducted by injecting measurement noise into the
six-pedestrian ETHZ ground truth. As detailed in Table \ref{tab:MC_simulation},
these simulations encompass three distinct scenarios designed to validate
performance consistency across the five algorithms. These scenarios
progressively increase in difficulty by varying the measurement noise
standard deviation $\sigma_{\epsilon}$, detection probability $P_{k,D}$,
and average Poisson clutter rate $\lambda_{c}$.

\begin{table}
	\centering
	\caption{Three distinct scenarios for generating noisy measurements with increasing difficulty levels across measurement noise standard deviation $\sigma_{\epsilon}$, detection probability $P_{k,D}$, and average Poisson clutter rate $\lambda_{c}$.}\label{tab:MC_simulation}
	{\footnotesize{}%
	\begin{tabular}{|c|c|c|c|}
	\hline 
	 & {\footnotesize$\sigma_{\epsilon}$} & {\footnotesize$P_{k,D}$} & {\footnotesize$\lambda_{c}$}\tabularnewline
	\hline 
	{\footnotesize\textbf{Scenario 1}} & {\footnotesize 0.15m} & {\footnotesize 0.90} & {\footnotesize 3}\tabularnewline
	\hline 
	{\footnotesize\textbf{Scenario 2}} & {\footnotesize 0.20m} & {\footnotesize 0.80} & {\footnotesize 10}\tabularnewline
	\hline 
	{\footnotesize\textbf{Scenario 3}} & {\footnotesize 0.25m} & {\footnotesize 0.70} & {\footnotesize 12}\tabularnewline
	\hline 
	\end{tabular}}{\scriptsize}{\scriptsize\par}
\end{table}

\begin{figure*}
	\centering
	\includegraphics[bb=2cm 4cm 40cm 173bp,clip,scale=0.35]{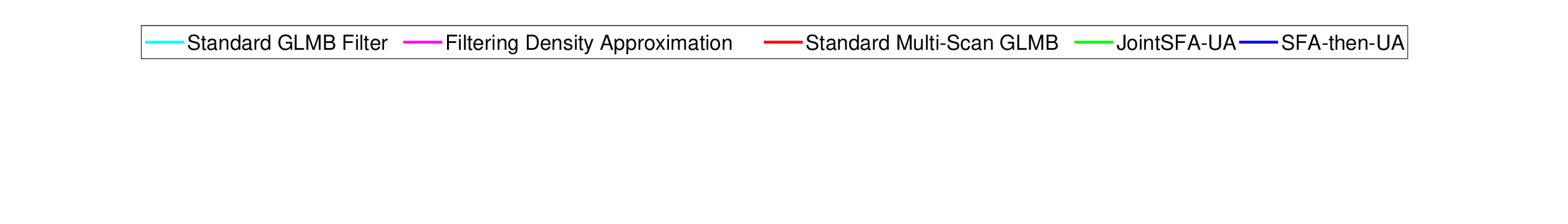}
	\hfil 
	\subfloat[]{\includegraphics[bb=0.8cm 5.5cm 14cm 11.5cm,clip,scale=0.42]{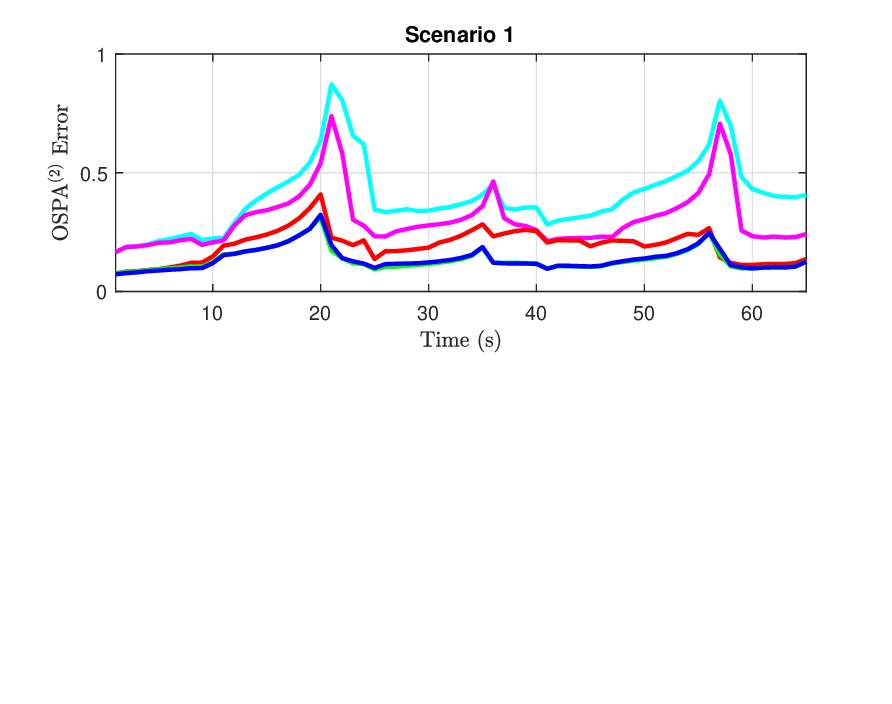}}
	\hfil 
	\subfloat[]{\includegraphics[bb=0.8cm 5.5cm 14cm 11.5cm,clip,scale=0.42]{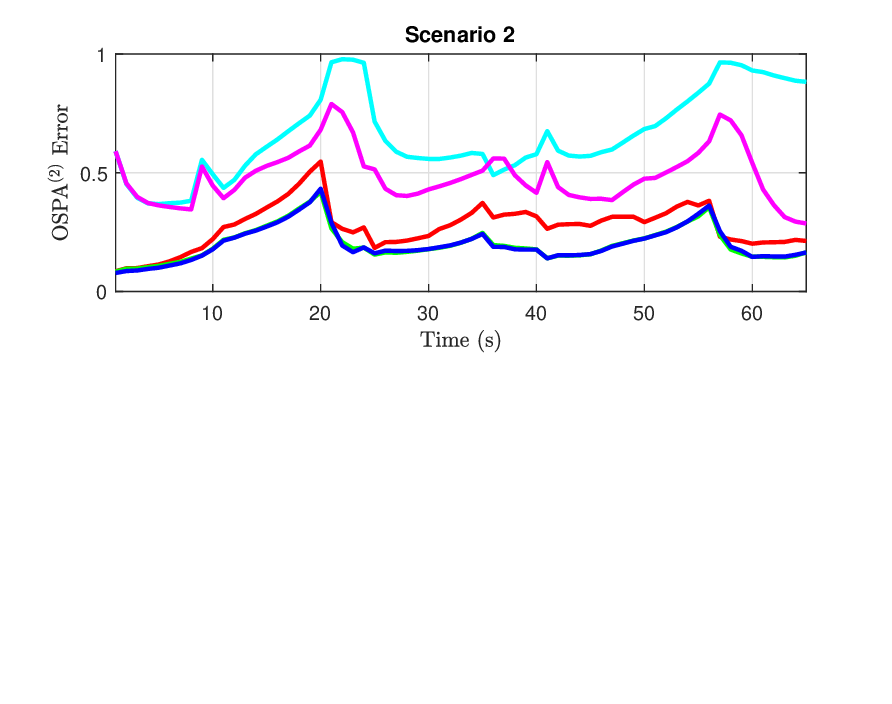}}
	\hfil 
	\subfloat[]{\includegraphics[bb=0.8cm 5.5cm 14cm 11.5cm,clip,scale=0.42]{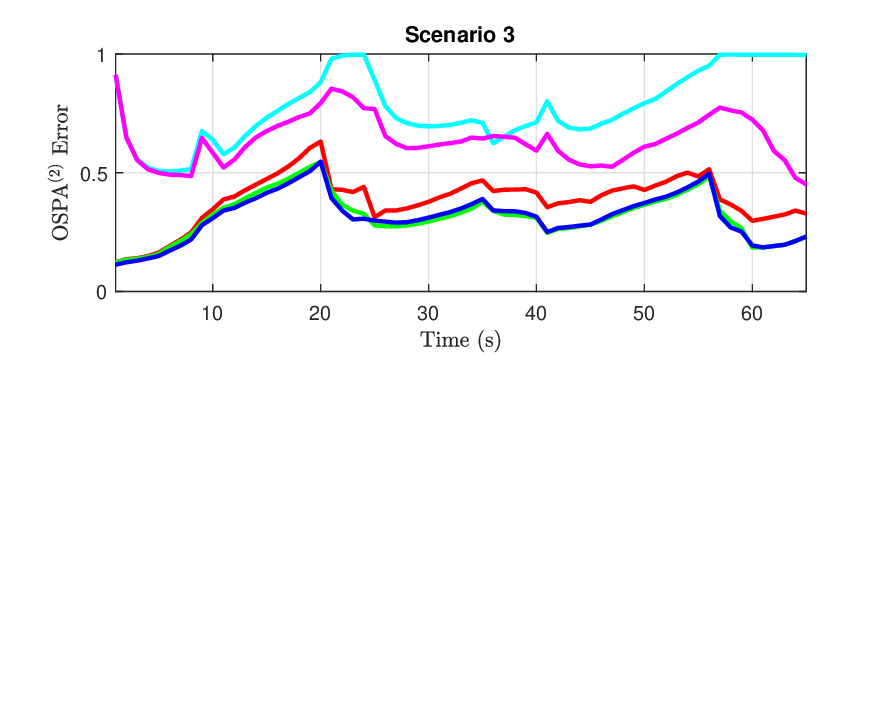}}
	
	\caption{$\text{OSPA}^{(2)}$ (cutoff $c=1m$; order $p=1$; over 10-scan window length) errors \cite{schuhmacher2008,beard2020}
	over 100 Monte Carlo simulations for (a) Scenario 1, (b) Scenario 2, and (c) Scenario 3. The subplots compare the final estimates
	from the standard GLMB filter, filtering density approximation, standard
    multi-scan GLMB, JointSFA-UA, and SFA-then-UA.}\label{fig:average_ospa_ospa2}
\end{figure*}

Figure \ref{fig:average_ospa_ospa2} depicts the average $\text{OSPA}^{(2)}$
(over a 10-scan moving window) errors \cite{schuhmacher2008,beard2020}
for the five tracking methods. While degraded detectability and higher
uncertainty increase errors for all methods as the scenarios become
more challenging, the proposed JointSFA-UA and SFA-then-UA consistently
maintain the lowest average errors. This demonstrates that integrating
the social force model with the proposed posterior approximation is
essential for robust tracking, effectively addressing the limitations
caused by the independent motion assumption in standard filters and
scan-by-scan approach of the filtering density approximation. 

\section{Conclusion}\label{sec:conclusion}

This work addresses the challenging problems of multi-object posterior
inference in non-standard SSMs, where exact posterior computation
is typically intractable. We introduced a tractable multi-scan GLMB
approximation, which preserves the trajectory cardinality distribution
of the labeled multi-object posterior density of interest. The proposed
approximation was shown to minimize the Kullback-Leibler divergence
over a special class of multi-scan GLMB model. Based on this result,
we developed a tractable algorithm, with constant per-step computational
complexity, to compute approximate multi-object posteriors over finite
windows. We validated the proposed solutions through numerical experiments
with object interactions through the social force model and merged
measurements that introduce additional inter-object correlations,
as well as real-world pedestrian tracking benchmarks. These experiments
confirmed the effectiveness of our approach, demonstrating that explicit
modeling of object interactions and prudent functional approximation
of the resulting intractable posterior successfully capture multi-object
correlations and improve estimation accuracy. 
\bibliographystyle{IEEEtran}
\bibliography{bibliography}

	\cleardoublepage{}

\setcounter{page}{1}

\setcounter{section}{0}

\noindent\begin{minipage}[t]{1\columnwidth}%
	{\large\textbf{Supplementary Materials\smallskip{}
	}}{\large\par}
	
	Tractable Approximation of Labeled Multi-Object
	
	Posterior Densities\medskip{}
	
	Thi Hong Thai Nguyen, Ba-Ngu Vo, and Ba-Tuong Vo%
\end{minipage}

\section{Mathematical Proofs}

\subsection{Proof of Proposition \ref{thm:MSapproximation}}\label{subsec:proof_approx}

Given the following labeled multi-object posterior density 
\[
\boldsymbol{\pi}_{j:k}(\boldsymbol{X}_{j:k})=w(\mathcal{L}(\boldsymbol{X}_{j:k}))p_{\mathcal{L}(\boldsymbol{X}_{j:k})}(\boldsymbol{X}_{j:k}),
\]
let $\hat{\boldsymbol{\pi}}_{j:k}$ be the multi-scan GLMB density
with the hypothesis weights $\hat{w}^{(\xi)}(I_{j:k})$, where $\xi\in\Xi$
and $I_{j:k}\in\cprod_{i=j}^{k}\mathcal{F}(\mathbb{L}_{i})$. For
each $I_{j:k}$, setting 
\[
\underset{\xi\in\Xi}{\sum}\hat{w}^{(\xi)}(I_{j:k})=\langle\boldsymbol{\pi}_{j:k}\rangle(I_{j:k}).
\]
\textit{\uline{Trajectory cardinality distribution:}} Following \cite{vo2024supp},
the trajectory cardinality distribution of the multi-scan GLMB $\hat{\boldsymbol{\pi}}_{j:k}$
is 
\[
\mathbb{P}_{\hat{\boldsymbol{\pi}}_{j:k}}(|\boldsymbol{X}_{j:k}|=n)=\sum_{\xi,I_{j:k}}\delta_{n}[|\cup_{i=j}^{k}I_{i}|]\hat{w}^{(\xi)}(I_{j:k}).
\]
Further, the trajectory cardinality distribution of $\boldsymbol{\pi}_{j:k}$
is 
\begin{align*}
	\rho(n) & =\mathbb{P}_{\boldsymbol{\pi}_{j:k}}(|\boldsymbol{X}_{j:k}|=n),\\
	& =\int\delta_{n}[|\boldsymbol{X}_{j:k}|]\boldsymbol{\pi}_{j:k}(\boldsymbol{X}_{j:k})\delta\boldsymbol{X}_{j:k},\\
	& =\int\delta_{n}[|\boldsymbol{X}_{j:k}|]w(\mathcal{L}(\boldsymbol{X}_{j:k}))p_{\mathcal{L}(\boldsymbol{X}_{j:k})}(\boldsymbol{X}_{j:k})\delta\boldsymbol{X}_{j:k},\\
	& =\sum_{I_{j:k}}\delta_{n}[|\cup_{i=j}^{k}I_{i}|]w(I_{j:k})\langle p_{I_{j:k}}\rangle(I_{j:k}),\\
	& =\sum_{I_{j:k}}\delta_{n}[|\cup_{i=j}^{k}I_{i}|]w(I_{j:k}),
\end{align*}
where derivations of the fourth equation come from the fact that $\delta_{n}[|\boldsymbol{X}_{j:k}|]=\delta_{n}[|\cup_{i=j}^{k}\mathcal{L}(\boldsymbol{X}_{i})|]$.
Since $w(I_{j:k})=\langle\boldsymbol{\pi}_{j:k}\rangle(I_{j:k})=\underset{\xi\in\Xi}{\sum}\hat{w}^{(\xi)}(I_{j:k})$,
$\boldsymbol{\pi}_{j:k}$ and $\hat{\boldsymbol{\pi}}_{j:k}$ have
the same trajectory cardinality distribution.

\textit{\uline{Kullback-Leibler divergence:}} Let $\mathring{\boldsymbol{\pi}}_{j:k}=\{\mathring{w}^{(I_{j:k})},\mathring{p}^{(I_{j:k})}:I_{j:k}\}$
be the M-GLMB density that matches the trajectory cardinality distribution
of $\boldsymbol{\pi}_{j:k}$. Hence, we have $\mathring{w}^{(I_{j:k})}=\langle\boldsymbol{\pi}_{j:k}\rangle(I_{j:k})$.
Given any multi-scan M-GLMB density of the form $\bar{\boldsymbol{\pi}}_{j:k}=\{\bar{w}^{(I_{j:k})},\bar{p}^{(I_{j:k})}:I_{j:k}\}$,
it is shown that $\bar{\boldsymbol{\pi}}_{j:k}$ can be rewritten
in one term with the sum over the sequence of label set $I_{j:k}$
is collapsed, i.e. 
\begin{align*}
	\bar{\boldsymbol{\pi}}_{j:k}(\boldsymbol{X}_{j:k}) & =\Delta(\boldsymbol{X}_{j:k})\sum_{I_{j:k}}\bar{w}^{(I_{j:k})}\delta_{I_{j:k}}[\mathcal{L}(\boldsymbol{X}_{j:k})]\left[\bar{p}^{(I_{j:k})}\right]^{\boldsymbol{X}_{j:k}}\\
	& =\bar{w}(\mathcal{L}(\boldsymbol{X}_{j:k})\left[\bar{p}^{(\mathcal{L}(\boldsymbol{X}_{j:k}))}\right]^{\boldsymbol{X}_{j:k}}.
\end{align*}

The Kullback-Leibler divergence of any multi-scan M-GLMB $\bar{\boldsymbol{\pi}}_{j:k}$
from $\boldsymbol{\pi}_{j:k}$ is  
\[
D_{\boldsymbol{\pi}_{j:k}}(\boldsymbol{\pi}_{j:k};\bar{\boldsymbol{\pi}}_{j:k})=D_{\boldsymbol{\pi}_{j:k}}(w_{j:k};\bar{w}_{j:k})+D_{\boldsymbol{\pi}_{j:k}}(p_{j:k};\bar{p}_{j:k}),
\]
where each of the terms is given by
\[
D_{\boldsymbol{\pi}_{j:k}}(\boldsymbol{\pi}_{j:k};\bar{\boldsymbol{\pi}}_{j:k})=\int\log\left(\frac{\boldsymbol{\pi}_{j:k}(\boldsymbol{X}_{j:k})}{\bar{\boldsymbol{\pi}}_{j:k}(\boldsymbol{X}_{j:k})}\right)\boldsymbol{\pi}_{j:k}(\boldsymbol{X}_{j:k})\delta\boldsymbol{X}_{j:k},
\]
\begin{align*}
	& D_{\boldsymbol{\pi}_{j:k}}(w_{j:k};\bar{w}_{j:k})=\\
	& \int\log\left(\frac{w(\mathcal{L}(\boldsymbol{X}_{j:k}))}{\bar{w}(\mathcal{L}(\boldsymbol{X}_{j:k}))}\right)w(\mathcal{L}(\boldsymbol{X}_{j:k}))p_{\mathcal{L}(\boldsymbol{X}_{j:k})}(\boldsymbol{X}_{j:k})\delta\boldsymbol{X}_{j:k},
\end{align*}
\begin{align*}
	& D_{\boldsymbol{\pi}_{j:k}}(p_{j:k};\bar{p}_{j:k})=\\
	& \int\log\left(\frac{p_{\mathcal{L}(\boldsymbol{X}_{j:k})}(\boldsymbol{X}_{j:k})}{\left[\bar{p}^{(\mathcal{L}(\boldsymbol{X}_{j:k}))}\right]^{\boldsymbol{X}_{j:k}}}\right)w(\mathcal{L}(\boldsymbol{X}_{j:k}))p_{\mathcal{L}(\boldsymbol{X}_{j:k})}(\boldsymbol{X}_{j:k})\delta\boldsymbol{X}_{j:k}.
\end{align*}

Evaluating each term, we obtain 
\begin{align*}
	D_{\boldsymbol{\pi}_{j:k}}(w_{j:k};\bar{w}_{j:k}) & =\sum_{I_{j:k}}\log\left(\frac{w(I_{j:k})}{\bar{w}(I_{j:k})}\right){\displaystyle w(I_{j:k})}\langle p_{I_{j:k}}\rangle(I_{j:k})\\
	& =\sum_{I_{j:k}}\log\left(\frac{w(I_{j:k})}{\bar{w}(I_{j:k})}\right){\displaystyle w(I_{j:k})}\\
	& =D_{w_{j:k}}(w_{j:k};\bar{w}_{j:k}),
\end{align*}
and
\begin{align*}
	& D_{\boldsymbol{\pi}_{j:k}}(p_{j:k};\bar{p}_{j:k})\\
	& \quad=\sum_{I_{j:k}}w(I_{j:k})\left\langle \log\left(\frac{p_{j:k}}{\underset{\ell\in\cup_{i=j}^{k}I_{i}}{\prod}\bar{p}_{j:k}}\right)p_{j:k}\right\rangle (I_{j:k})\\
	& \quad=\sum_{I_{j:k}}w(I_{j:k})D_{p_{j:k}}\left(p_{I_{j:k}};\underset{\ell\in\cup_{i=j}^{k}I_{i}}{\prod}\bar{p}^{(I_{j:k})}\right),
\end{align*}
where the second equation comes from the fact that
\begin{align*}
	& \left\langle \log\left(\frac{p_{j:k}}{\underset{\ell\in\cup_{i=j}^{k}I_{i}}{\prod}\bar{p}_{j:k}}\right)p_{j:k}\right\rangle (I_{j:k})\\
	& =\int\log\left(\frac{p_{I_{j:k}}(\boldsymbol{X}_{j:k})}{\underset{\ell\in\cup_{i=j}^{k}I_{i}}{\prod}\bar{p}^{(I_{j:k})}(\boldsymbol{x}_{T(\ell)})}\right)p_{I_{j:k}}(\boldsymbol{X}_{j:k})\delta\boldsymbol{X}_{j:k}\\
	& =D_{p_{j:k}}\left(p_{I_{j:k}};\underset{\ell\in\cup_{i=j}^{k}I_{i}}{\prod}\bar{p}^{(I_{j:k})}\right).
\end{align*}

Therefore, $D_{\boldsymbol{\pi}_{j:k}}(\boldsymbol{\pi}_{j:k},\bar{\boldsymbol{\pi}}_{j:k})$
is equivalent to 
\begin{align*}
	& D_{w_{j:k}}(w_{j:k};\bar{w}_{j:k})\\
	& +\sum_{I_{j:k}}w(I_{j:k})D_{p_{j:k}}\left(p_{I_{j:k}};\underset{\ell\in\cup_{i=j}^{k}I_{i}}{\prod}\bar{p}^{(I_{j:k})}\right).
\end{align*}

Setting $\bar{\boldsymbol{\pi}}_{j:k}=\mathring{\boldsymbol{\pi}}_{j:k}$
and since $\mathring{w}(I_{j:k})=w(I_{j:k})$, we obtain $D_{w_{j:k}}(w_{j:k};\mathring{w}_{j:k})=0$.
To minimize each Kullback-Leibler divergence of the above sum, for
each $I_{j:k}$ and each trajectory $\ell\in\cup_{i=j}^{k}I_{i}$,
we marginalize other labels from $p_{I_{j:k}}(\boldsymbol{X}_{j:k})$
to yield $\mathring{p}^{(I_{j:k})}(\boldsymbol{x}_{T(\ell)})$. Hence,
$D_{\boldsymbol{\pi}_{j:k}}(\boldsymbol{\pi}_{j:k};\bar{\boldsymbol{\pi}}_{j:k})$
is minimized over the class of multi-scan M-GLMB density.$\blacksquare$

\subsection{Proof of Proposition \ref{prop:smoothing_window_approx}}\label{subsec:proof_moving_window}

Given a labeled multi-object posterior density $\boldsymbol{\pi}_{j:k}$
on $\{j:k\}$, assume $\boldsymbol{\pi}_{j:k}(\boldsymbol{X}_{j:k})=w(\mathcal{L}(\boldsymbol{X}_{j:k}))p_{\mathcal{L}(\boldsymbol{X}_{j:k})}(\boldsymbol{X}_{j:k})$.
Since $\{j:k\}=\uplus_{i=1}^{N_{S}}\{j^{(i)}:k^{(i)}\}$, let $\check{\boldsymbol{\pi}}_{j^{(i)}:k^{(i)}}(\boldsymbol{X}_{j^{(i)}:k^{(i)}})=\check{w}(\mathcal{L}(\boldsymbol{X}_{j^{(i)}:k^{(i)}}))\check{p}_{\mathcal{L}(\boldsymbol{X}_{j^{(i)}:k^{(i)}})}(\boldsymbol{X}_{j^{(i)}:k^{(i)}})$
be the labeled multi-object posterior density on $\{j^{(i)}:k^{(i)}\}$
with $j^{(i)}\geq j$ and $k^{(i)}\leq k$, for all $i\in\{1:N_{S}\}$,
and let $\check{\boldsymbol{\pi}}_{j:k}(\boldsymbol{X}_{j:k})=\prod_{i=1}^{N_{S}}\check{\boldsymbol{\pi}}_{j^{(i)}:k^{(i)}}(\boldsymbol{X}_{j^{(i)}:k^{(i)}})$. 

The Kullback-Leibler divergence of $\check{\boldsymbol{\pi}}_{j:k}$
from $\boldsymbol{\pi}_{j:k}$ is 
\begin{align}
	& D_{\boldsymbol{\pi}_{j:k}}\left(\boldsymbol{\pi}_{j:k};\check{\boldsymbol{\pi}}_{j:k}\right)\label{eq:DKL-SW}\\
	& =D_{\boldsymbol{\pi}_{j:k}}\left(w_{j:k};\prod_{i=1}^{N_{S}}\check{w}_{j^{(i)}:k^{(i)}}\right)+D_{\boldsymbol{\pi}_{j:k}}\left(p_{j:k};\prod_{i=1}^{N_{S}}\check{p}_{j^{(i)}:k^{(i)}}\right).\nonumber 
\end{align}

Since $I_{j:k}=\left[I_{j^{(1)}:k^{(1)}},...,I_{j^{(N_{S})}:k^{(N_{S})}}\right]\in\cprod_{i=j}^{k}\mathcal{F}(\mathbb{L}_{i})$,
evaluating each term of \eqref{eq:DKL-SW}, we obtain 
\begin{align*}
	& D_{\boldsymbol{\pi}_{j:k}}\left(w_{j:k};\prod_{i=1}^{N_{S}}\check{w}_{j^{(i)}:k^{(i)}}\right)\\
	& =\sum_{I_{j:k}}w(I_{j:k})\log\left(\frac{w(I_{j:k})}{\prod_{i=1}^{N_{S}}\check{w}(I_{j^{(i)}:k^{(i)}})}\right)\langle p_{I_{j:k}}\rangle(I_{j:k}),\\
	& =\sum_{I_{j:k}}w(I_{j:k})\log\left(\frac{w(I_{j:k})}{\prod_{i=1}^{N_{S}}\check{w}(I_{j^{(i)}:k^{(i)}})}\right),\\
	& =D_{w_{j:k}}\left(w_{j:k};\prod_{i=1}^{N_{S}}\check{w}_{j^{(i)}:k^{(i)}}\right),
\end{align*}
and 
\begin{align*}
	& D_{\boldsymbol{\pi}_{j:k}}\left(p_{j:k};\prod_{i=1}^{N_{S}}\check{p}_{j^{(i)}:k^{(i)}}\right)\\
	& =\sum_{I_{j:k}}w(I_{j:k})\left\langle \log\left(\frac{p_{j:k}}{\prod_{i=1}^{N_{S}}\check{p}_{j^{(i)}:k^{(i)}}}\right)p_{j:k}\right\rangle (I_{j:k}),\\
	& =\sum_{I_{j:k}}w(I_{j:k})D_{p_{j:k}}\left(p_{I_{j:k}};\prod_{i=1}^{N_{S}}\check{p}_{I_{j^{(i)}:k^{(i)}}}\right).
\end{align*}

Denote $\{\bar{j}^{(i)}:\bar{k}^{(i)}\}=\{j:k\}\setminus\{j^{(i)}:k^{(i)}\}$,
choosing 
\begin{align*}
	\check{w}(I_{j^{(i)}:k^{(i)}}) & =\sum_{I_{\bar{j}^{(i)}:\bar{k}^{(i)}}}w(I_{j:k}),\\
	\check{p}_{\mathcal{L}(\boldsymbol{X}_{j^{(i)}:k^{(i)}})}(\boldsymbol{X}_{j^{(i)}:k^{(i)}}) & =\int p_{\mathcal{L}(\boldsymbol{X}_{j:k})}(\boldsymbol{X}_{j:k})\delta\boldsymbol{X}_{\bar{j}^{(i)}:\bar{k}^{(i)}},
\end{align*}
minimizes the Kullback-Leibler divergence \eqref{eq:DKL-SW}.$\blacksquare$

\subsection{Proof of Proposition \ref{prop:jointpredictionprop}}\label{subsec:proof_pred}

The multi-object trajectory $\boldsymbol{X}_{0:k-1}$ at time $k-1$
can be decomposed as $\boldsymbol{X}_{0:k-1}=\boldsymbol{S}_{0:k-1}\uplus\boldsymbol{D}_{0:k-1}$,
where $\boldsymbol{S}_{0:k-1}=\{\boldsymbol{x}_{T(\ell)}\in\boldsymbol{X}_{0:k-1}:\ell\in\mathcal{L}(\boldsymbol{X}_{0:k-1})\cap\mathcal{L}(\boldsymbol{X}_{k})\}$
is the set of surviving trajectories at time $k$, and $\boldsymbol{D}_{0:k-1}=\{\boldsymbol{x}_{T(\ell)}\in\boldsymbol{X}_{0:k-1}:\ell\in\mathcal{L}(\boldsymbol{X}_{0:k-1})-\mathcal{L}(\boldsymbol{X}_{k})\}$
is the set of trajectories that have either just disappeared at time
$k$ or were previously terminated. Thus, the joint probability density
$p_{-}^{(\xi)}(\boldsymbol{X}_{0:k-1})$ can be rewritten as 
\[
p_{-}^{(\xi)}(\boldsymbol{S}_{0:k-1}\uplus\boldsymbol{D}_{0:k-1})=p_{-}^{(\xi)}(\boldsymbol{S}_{0:k-1}|\boldsymbol{D}_{0:k-1})p_{-}^{(\xi)}(\boldsymbol{D}_{0:k-1}).
\]

Multiplying the multi-object Markov transition kernel $\boldsymbol{f}_{k}(\boldsymbol{X}_{k}|\boldsymbol{X}_{k-1})$
with the multi-object posterior density $\boldsymbol{\pi}_{0:k-1}=\{w_{-}^{(\xi)}(I_{0:k-1}),p_{-}^{(\xi)}:(\xi,I_{0:k-1})\}$
gives 
\begin{align}
	& \boldsymbol{\pi}_{0:k}(\boldsymbol{X}_{0:k})\nonumber \\
	& =\boldsymbol{\pi}_{0:k-1}(\boldsymbol{X}_{0:k-1})\boldsymbol{f}_{k}(\boldsymbol{X}_{k}|\boldsymbol{X}_{k-1}),\nonumber \\
	& =\Delta(\boldsymbol{X}_{0:k})\sum\limits_{\xi,I_{0:k}}w_{-}^{(\xi)}(I_{0:k-1})\delta_{I_{0:k}}[\mathcal{L}(\boldsymbol{X}_{0:k})]\nonumber \\
	& \times\left[\boldsymbol{f}_{k,B}(\boldsymbol{B}_{k})\boldsymbol{\Phi}_{k,S}(\boldsymbol{S}_{k}|\boldsymbol{X}_{k-1})p_{-}^{(\xi)}(\boldsymbol{X}_{0:k-1})\right],\nonumber \\
	& =\{(w^{(\xi)}(I_{0:k}),p^{(\xi)}(\boldsymbol{X}_{0:k}):(\xi,I_{0:k})\},\label{eq:pred_density}
\end{align}
where for each $\xi\in\Xi$, $I_{0:k}=(I_{0:k-1},I_{k})$, 
\begin{align*}
	w^{(\xi)}(I_{0:k}) & =\boldsymbol{1}_{I_{k-1}}^{I_{k}-\mathbb{B}_{k}}\eta_{k}^{(I_{k-1},I_{k})}w_{-}^{(\xi)}(I_{0:k-1}),\\
	\boldsymbol{1}_{I_{k-1}}^{I_{k}-\mathbb{B}_{k}} & =\prod_{\ell\in I_{k}-\mathbb{B}_{k}}\boldsymbol{1}_{I_{k-1}}(\ell),\\
	\eta_{k}^{(I_{k-1},I_{k})} & =w_{k,B}(\mathbb{B}_{k}\cap I_{k})w_{k,S}(I_{k}-\mathbb{B}_{k}),\\
	w_{k,B}(\mathbb{B}_{k}\cap I_{k}) & =[Q_{k,B}]^{\mathbb{B}_{k}-(\mathbb{B}_{k}\cap I_{k})}[P_{k,B}]^{I_{k}\cap\mathbb{B}_{k}},\\
	w_{k,S}(I_{k}-\mathbb{B}_{k}) & =[Q_{k,S}]^{I_{k-1}-I_{k}-\mathbb{B}_{k}}[P_{k,S}]^{I_{k}-\mathbb{B}_{k}},\\
	p^{(\xi)}(\boldsymbol{X}_{0:k}) & =\left[p_{k,B}\right]^{\boldsymbol{B}_{k}}p_{k,S}^{(\xi)}(\boldsymbol{S}_{0:k}\uplus\boldsymbol{D}_{0:k-1}),\\
	p_{k,S}^{(\xi)}(\boldsymbol{S}_{0:k}\uplus\boldsymbol{D}_{0:k-1}) & =p_{S,k}^{(\xi)}(\boldsymbol{S}_{0:k}|\boldsymbol{D}_{0:k-1})p_{-}^{(\xi)}(\boldsymbol{D}_{0:k-1}),\\
	p_{k,S}^{(\xi)}(\boldsymbol{S}_{0:k}|\boldsymbol{D}_{0:k-1}) & =\boldsymbol{f}_{k,S}(\boldsymbol{S}_{k}|\boldsymbol{X}_{k-1})p_{-}^{(\xi)}(\boldsymbol{S}_{0:k-1}|\boldsymbol{D}_{0:k-1}).\blacksquare
\end{align*}

\subsection{Proof of Proposition \ref{prop:jointupdateprop}}\label{subsec:proof_update}

Since $\boldsymbol{X}_{0:k}=\boldsymbol{B}_{k}\uplus\boldsymbol{S}_{0:k}\uplus\boldsymbol{D}_{0:k-1}$
and the disappearing trajectories $\boldsymbol{D}_{0:k-1}$ are not
updated with measurements, the standard multi-object likelihood can
be rewritten as 
\begin{equation}
	[\psi_{k,Z}^{(\theta_{k}\circ\mathcal{L}(\cdot))}]^{\boldsymbol{X}_{0:k}}=[\psi_{k,Z}^{(\theta_{k}\circ\mathcal{L}(\cdot))}]^{\boldsymbol{B}_{k}}[\psi_{k,Z}^{(\theta_{k}\circ\mathcal{L}(\cdot))}]^{\boldsymbol{S}_{0:k}},\label{eq:meas_likelihood}
\end{equation}

Multiplying the measurement likelihood \eqref{eq:meas_likelihood}
with the multi-object prediction density \eqref{eq:pred_density},
we obtain 
\begin{align*}
	& \boldsymbol{\pi}_{0:k}(\boldsymbol{X}_{0:k}|Z_{0:k})\\
	& \propto\Delta(\boldsymbol{X}_{0:k})\sum\limits_{\xi,I_{0:k},\theta_{k}}\boldsymbol{1}_{\Theta_{k}(I_{k})}(\theta_{k})w^{(\xi)}(I_{0:k})\delta_{I_{0:k}}[\mathcal{L}(\boldsymbol{X}_{0:k})]\\
	& \qquad\qquad\qquad\times[\psi_{k,Z}^{(\theta_{k}\circ\mathcal{L}(\cdot))}]^{\boldsymbol{B}_{k}}[\psi_{k,Z}^{(\theta_{k}\circ\mathcal{L}(\cdot))}]^{\boldsymbol{S}_{0:k}}p^{(\xi)}(\boldsymbol{X}_{0:k}),\\
	& =\{(w_{Z}^{(\xi,\theta_{k})}(I_{0:k}),p_{Z}^{(\xi,\theta_{k})}(\boldsymbol{X}_{0:k})):(\xi,I_{0:k},\theta_{k})\},
\end{align*}
where for each $\xi\in\Xi$, $I_{0:k}=(I_{0:k-1},I_{k})$, $\theta_{k}\in\Theta_{k}$,
\begin{align*}
	w_{Z}^{(\xi,\theta_{k})}(I_{0:k}) & =\boldsymbol{1}_{\Theta_{k}(I_{k})}(\theta_{k})\mu_{Z}^{(\xi,I_{k},\theta_{k})}w^{(\xi)}(I_{0:k}),\\
	\mu_{Z}^{(\xi,I_{k},\theta_{k})} & =[\mu_{B,Z}^{(\theta_{k})}]^{I_{k}\cap\mathbb{B}_{k}}\mu_{S,Z}^{(\xi,\theta_{k})}(I_{k}-\mathbb{B}_{k}),\\
	\mu_{B,Z}^{(\theta_{k})}(\ell) & =\left\langle p_{k,B}(\cdot,\ell),\psi_{k,Z}^{(\theta_{k}\circ\mathcal{L}(\cdot))}(\cdot,\ell)\right\rangle ,\\
	\mu_{S,Z}^{(\xi,\theta_{k})}(L) & =\left\langle p_{k,S}^{(\xi)}(\cdot),\left[\psi_{k,Z}^{(\theta_{k}\circ\mathcal{L}(\cdot))}\right]^{(\cdot)}\right\rangle (L),\\
	p_{Z}^{(\xi,\theta_{k})}(\boldsymbol{X}_{0:k}) & \propto[p_{B,Z}^{(\theta_{k})}]^{\boldsymbol{B}_{k}}p_{S,Z}^{(\xi,\theta_{k})}(\boldsymbol{S}_{0:k}\uplus\boldsymbol{D}_{0:k-1}),\\{}
	[p_{B,Z}^{(\theta_{k})}]^{\boldsymbol{B}_{k}} & =[p_{k,B}\psi_{k,Z}^{(\theta_{k}\circ\mathcal{L}(\cdot))}]^{\boldsymbol{B}_{k}},\\
	p_{S,Z}^{(\xi,\theta_{k})}(\boldsymbol{S}_{0:k}\uplus\boldsymbol{D}_{0:k-1}) & =p_{S,Z}^{(\xi,\theta_{k})}(\boldsymbol{S}_{0:k}|\boldsymbol{D}_{0:k-1})p_{-}^{(\xi)}(\boldsymbol{D}_{0:k-1}),\\
	p_{S,Z}^{(\xi,\theta_{k})}(\boldsymbol{S}_{0:k}|\boldsymbol{D}_{0:k-1}) & =p_{k,S}^{(\xi)}(\boldsymbol{S}_{0:k}|\boldsymbol{D}_{0:k-1})[\psi_{k,Z}^{(\theta_{k}\circ\mathcal{L}(\cdot))}]^{\boldsymbol{S}_{0:k}}.\blacksquare
\end{align*}

\section{Additional Performance Evaluations on the ETHZ Walking Pedestrian
	Dataset}

\subsection{Evaluation with CLEAR-MOT}\label{subsec:clear_mot}

To further assess the tracking performance on the ETHZ walking pedestrian
dataset \cite{pellegrini2009-supp}, Table \ref{tab:MOTA_table} shows
a detailed breakdown using standard CLEAR-MOT metrics \cite{bernadin2008-supp}
evaluated across the five tracking methods: standard GLMB filter,
filtering density approximation, standard multi-scan GLMB, and the
proposed JointSFA-UA and SFA-then-UA. The results are presented by
MOTA (\%), MOTP (\%), ID Switches (Total), and False Positives (Total). 

\begin{table*}
	\centering
	\caption{Tracking performance metrics for the
		ETHZ walking pedestrian dataset based on the CLEAR formulations \cite{bernadin2008-supp}.
		Results for MOTA and MOTP are reported as percentages (\%), while
		ID Switches and False Positives are recorded as the aggregate sum.}\label{tab:MOTA_table}
	\begin{tabular}{|c|c|c|c|c|}
		\hline 
		\multirow{2}{*}{\textbf{Methods}} & \multirow{2}{*}{\textbf{MOTA (\%) $\uparrow$}} & \multirow{2}{*}{\textbf{MOTP (\%) $\uparrow$}} & \multirow{2}{*}{\begin{varwidth}[t]{3cm}
				\centering
				\textbf{ID Switches}\\
				\textbf{(Total) $\downarrow$}
		\end{varwidth}} & \multirow{2}{*}{\begin{varwidth}[t]{3cm}
				\centering
				\textbf{False Positives}\\
				\textbf{(Total) $\downarrow$}
		\end{varwidth}}\tabularnewline
		&  &  &  & \tabularnewline
		\hline 
		\textbf{Standard GLMB Filter} & -31.53 & 71.83 & 2 & 73\tabularnewline
		\hline 
		\textbf{Filtering Density Approximation} & 81.08 & 81.48 & 1 & 11\tabularnewline
		\hline 
		\textbf{Standard Multi-Scan GLMB} & 87.39 & 84.72 & 2 & 6\tabularnewline
		\hline 
		\textbf{JointSFA-UA} & 96.40 & 88.48 & 0 & 2\tabularnewline
		\hline 
		\textbf{SFA-then-UA} & 96.40 & 88.11 & 0 & 2\tabularnewline
		\hline 
	\end{tabular}
\end{table*}

As presented in Table \ref{tab:MOTA_table}, the proposed JointSFA-UA
and SFA-then-UA methods outperform the three baselines. Consistent
with the $\text{OSPA}^{(2)}$ errors presented by Figure \ref{fig:ospa_real_data}
in Subsection VI-D, the proposed methods achieve significant improvements
in tracking performance by incorporating the repulsive social force
model into the pedestrian dynamics. Specifically, MOTA score significantly
increases from -31.53\% (standard GLMB filter) to 96.40\% (proposed),
and ID Switches are entirely eliminated from 2 to 0. Further, the
improvements in MOTP score and the reduction in False Positives demonstrate
the effectiveness of the proposed functional approximation in achieving
accurate spatial localization and successfully rejecting clutter.
These results confirm that the proposed object interaction modeling
achieves accurate trajectory estimates for multi-object posteriors
compared to the standard GLMB filter/standard multi-scan GLMB which
assumes independent motion and the filtering density approximation
which applies scan-by-scan approximation technique.

\subsection{Sensitivity of Algorithm Parameters}\label{subsec:sensitivity_para}

To examine the sensitivity of the parameters of the proposed JointSFA-UA
and SFA-then-UA, we summarize, in Table \ref{tab:sensitivity_testing},
the output metrics with respect to the number of Markov chain iterations
$T$, the component weight truncation, and the sub-window size/length
of overlap, respectively, based on the ETHZ dataset \cite{pellegrini2009-supp}.

In this paper, these parameters are chosen to achieve a balance between
estimation accuracy and computation tractability. These settings are
presented by a maximum of 1000 components for the multi-object posterior
computation, $T=10$ iterations of the Markov chain for the multi-scan
Gibbs sampler, and component weight truncation threshold of $10^{-5}$
for computational tractability. Due to intractability over long duration,
the proposed windowing approximation technique is employed on a set
of $10$-scan sub-windows with an overlap of length 5. In the multi-object
tracking scenario with complex inter-object interactions, such as
repulsive social forces combined with merged measurement in Subsection
\ref{subsec:merged_meas}, the size of sub-window/length of overlap
can be adaptively modified to reduce the computational load of capturing
these interactions in the posterior recursion.

\begin{sidewaystable*}
	\centering
	\caption{Tracking performance and computational
		cost of the proposed JointSFA-UA and SFA-then-UA under various parameter
		settings. The results show the impact of varying the number of Markov
		chain iterations $T$ (top), the component weight truncation thresholds
		(middle), and the window size/length of overlap (bottom). All of the
		multi-object posteriors are computed with a maximum of 1000 components.}\label{tab:sensitivity_testing}
	\subfloat[]{%
		\begin{tabular}{|c|c|c|c|c|c|c|c|c|c|c|c|c|}
			\hline 
			\textbf{$T$} & \multicolumn{3}{c|}{\textbf{2}} & \multicolumn{3}{c|}{\textbf{5}} & \multicolumn{3}{c|}{\textbf{10}} & \multicolumn{3}{c|}{\textbf{20}}\tabularnewline
			\hline 
			& \begin{varwidth}[t]{3cm}
				\centering
				Runtime\\
				(ms)
			\end{varwidth} & \begin{varwidth}[t]{3cm}
				\centering
				Memory\\
				(MB)
			\end{varwidth} & \begin{varwidth}[t]{3cm}
				\centering
				$\text{OSPA}^{(2)}$\\
				(m)
			\end{varwidth} & \begin{varwidth}[t]{3cm}
				\centering
				Runtime\\
				(ms)
			\end{varwidth} & \begin{varwidth}[t]{3cm}
				\centering
				Memory\\
				(MB)
			\end{varwidth} & \begin{varwidth}[t]{3cm}
				\centering
				$\text{OSPA}^{(2)}$\\
				(m)
			\end{varwidth} & \begin{varwidth}[t]{3cm}
				\centering
				Runtime\\
				(ms)
			\end{varwidth} & \begin{varwidth}[t]{3cm}
				\centering
				Memory\\
				(MB)
			\end{varwidth} & \begin{varwidth}[t]{3cm}
				\centering
				$\text{OSPA}^{(2)}$\\
				(m)
			\end{varwidth} & \begin{varwidth}[t]{3cm}
				\centering
				Runtime\\
				(ms)
			\end{varwidth} & \begin{varwidth}[t]{3cm}
				\centering
				Memory\\
				(MB)
			\end{varwidth} & \begin{varwidth}[t]{3cm}
				\centering
				$\text{OSPA}^{(2)}$\\
				(m)
			\end{varwidth}\tabularnewline
			\hline 
			\textbf{JointSFA-UA} & 108 & 12.67 & 0.0601 & 156 & 12.67 & 0.0601 & 318 & 12.67 & 0.0601 & 606 & 12.67 & 0.0601\tabularnewline
			\hline 
			\textbf{SFA-then-UA} & 96 & 12.67 & 0.0600 & 174 & 12.67 & 0.0600 & 336 & 12.67 & 0.0600 & 636 & 12.67 & 0.0600\tabularnewline
			\hline 
		\end{tabular}
		
	}
	
	\subfloat[]{%
		\begin{tabular}{|c|c|c|c|c|c|c|c|c|c|c|c|c|}
			\hline 
			\textbf{Threshold} & \multicolumn{3}{c|}{\textbf{$10^{-3}$}} & \multicolumn{3}{c|}{\textbf{$10^{-5}$}} & \multicolumn{3}{c|}{\textbf{$10^{-7}$}} & \multicolumn{3}{c|}{\textbf{$10^{-10}$}}\tabularnewline
			\hline 
			& \begin{varwidth}[t]{3cm}
				\centering
				Runtime\\
				(ms)
			\end{varwidth} & \begin{varwidth}[t]{3cm}
				\centering
				Memory\\
				(MB)
			\end{varwidth} & \begin{varwidth}[t]{3cm}
				\centering
				$\text{OSPA}^{(2)}$\\
				(m)
			\end{varwidth} & \begin{varwidth}[t]{3cm}
				\centering
				Runtime\\
				(ms)
			\end{varwidth} & \begin{varwidth}[t]{3cm}
				\centering
				Memory\\
				(MB)
			\end{varwidth} & \begin{varwidth}[t]{3cm}
				\centering
				$\text{OSPA}^{(2)}$\\
				(m)
			\end{varwidth} & \begin{varwidth}[t]{3cm}
				\centering
				Runtime\\
				(ms)
			\end{varwidth} & \begin{varwidth}[t]{3cm}
				\centering
				Memory\\
				(MB)
			\end{varwidth} & \begin{varwidth}[t]{3cm}
				\centering
				$\text{OSPA}^{(2)}$\\
				(m)
			\end{varwidth} & \begin{varwidth}[t]{3cm}
				\centering
				Runtime\\
				(ms)
			\end{varwidth} & \begin{varwidth}[t]{3cm}
				\centering
				Memory\\
				(MB)
			\end{varwidth} & \begin{varwidth}[t]{3cm}
				\centering
				$\text{OSPA}^{(2)}$\\
				(m)
			\end{varwidth}\tabularnewline
			\hline 
			\textbf{JointSFA-UA} & 282 & 12.67 & 0.0601 & 318 & 12.67 & 0.0601 & 360 & 12.67 & 0.0601 & 510 & 12.67 & 0.0601\tabularnewline
			\hline 
			\textbf{SFA-then-UA} & 300 & 12.67 & 0.0600 & 336 & 12.67 & 0.0600 & 396 & 12.67 & 0.0600 & 504 & 12.67 & 0.0600\tabularnewline
			\hline 
		\end{tabular}
		
	}
	
	\subfloat[]{%
		\begin{tabular}{|c|c|c|c|c|c|c|}
			\hline 
			\textbf{Window size/overlap} & \multicolumn{3}{c|}{\textbf{5/5}} & \multicolumn{3}{c|}{\textbf{10/5}}\tabularnewline
			\hline 
			& \begin{varwidth}[t]{3cm}
				\centering
				Runtime\\
				(ms)
			\end{varwidth} & \begin{varwidth}[t]{3cm}
				\centering
				Memory\\
				(MB)
			\end{varwidth} & \begin{varwidth}[t]{3cm}
				\centering
				$\text{OSPA}^{(2)}$\\
				(m)
			\end{varwidth} & \begin{varwidth}[t]{3cm}
				\centering
				Runtime\\
				(ms)
			\end{varwidth} & \begin{varwidth}[t]{3cm}
				\centering
				Memory\\
				(MB)
			\end{varwidth} & \begin{varwidth}[t]{3cm}
				\centering
				$\text{OSPA}^{(2)}$\\
				(m)
			\end{varwidth}\tabularnewline
			\hline 
			\textbf{JointSFA-UA} & 192 & 12.67 & 0.0641 & 318 & 12.67 & 0.0601\tabularnewline
			\hline 
			\textbf{SFA-then-UA} & 210 & 12.67 & 0.0640 & 336 & 12.67 & 0.0600\tabularnewline
			\hline 
		\end{tabular}
	}
\end{sidewaystable*}

\end{document}